\documentclass[aps,prc,preprint,amsmath,amssymb,preprintnumbers,superscriptaddress]{revtex4} % twocolumn/preprint

\usepackage{CJK}
\usepackage{amsmath}
\usepackage{graphicx}% Include figure files
\usepackage{bm}% bold math
\usepackage{hyperref}
\usepackage{braket}
\usepackage{multirow}
\usepackage{mathrsfs}
\usepackage{longtable}
\usepackage{pdflscape}
\usepackage[dvipsnames,svgnames]{xcolor}

\allowdisplaybreaks[4]

\begin{document}
\begin{CJK*}{UTF8}{}

\title{  Charge radii and their deformation correlation for even-$Z$ nuclei in deformed relativistic Hartree-Bogoliubov theory in continuum 
}

\author{Cong Pan \CJKfamily{gbsn}(潘琮)}
\affiliation{Department of Physics, Anhui Normal University, Wuhu 241000, China}

\author{Jie Meng \CJKfamily{gbsn}(孟杰)}
\email{mengj@pku.edu.cn}
\affiliation{State Key Laboratory of Nuclear Physics and Technology, School of Physics, Peking University, Beijing 100871, China}
\affiliation{Center for Theoretical Physics, China Institute of Atomic Energy, Beijing 102413, China}

\begin{abstract}

The systematics are investigated for the charge radii of the even-$Z$ nuclei with $8 \leqslant Z \leqslant 120$ calculated by the deformed relativistic Hartree-Bogoliubov theory in continuum (DRHBc) with the functional PC-PK1, and their deformation correlation is explored. 
The available data of the charge radius are reproduced with a root-mean-square deviation $\sigma=0.033$ fm. 
In particular, for the nuclei between the closed shells, the descriptions of the charge radii are remarkably improved by including the deformation. 
Taking molybdenum isotopes as examples, both the evolutions of the charge radius and deformation are well reproduced. 
It is found that while the charge radius typically increases with the deformation, there also exist different cases. 
For example, in $^{346}$Sg, the charge radius of the deformed ground state is smaller than the one of the spherical state, and the largest binding energy does not necessarily correspond to the smallest charge radius. 
The increase or decrease of the charge radii with deformation is related to specific shell structures, particularly the key single-particle levels near the Fermi energy.

\end{abstract}

\date{\today}

\maketitle

\section{Introduction}

The nuclear charge radius is one of the fundamental physical observables characterizing an atomic nucleus. 
Among about ten thousand nuclei predicted theoretically \cite{Erler2012Nat,Xia2018ADNDT,Zhang2022ADNDT,Guo2024ADNDT}, 3359 nuclei have been experimentally observed \cite{nndc}, and the charge radii of more than 1039 nuclei have been measured
\cite{Angeli2013ADNDT,Li2021ADNDT}. 
The nuclei with the charge radii measured mainly distribute along or near the $\beta$-stability line. 
For the exotic nuclei far away from the $\beta$-stability line, the charge radius data are extremely scarce, even though considerable experimental progress has been achieved in recent years \cite{Zhao2024PLB}. 
The charge radii of exotic nuclei are highly important to provide information for nuclear structures, such as the shell evolutions and new magic numbers \cite{Angeli2013ADNDT,Wang2013PRC,Angeli2015JPG}, halo nuclei \cite{Geithner2008PRL,Nortershauser2009PRL}, and isomers \cite{Yordanov2016PRL}. 
Therefore, the predictions on the charge radii from the robust theoretical models are highly desired to provide references for experiments. 

The relativistic density functional theory (RDFT) has been proved to be a powerful tool in nuclear physics by its successful applications in describing many nuclear phenomena \cite{Meng2016book}. 
Due to its compelling advantages, such as the automatic inclusion of the spin-orbit coupling \cite{Koepf1991ZPA,Ren2020PRC} and the reasonable description of its isospin dependence \cite{Sharma1995PRL}, the new saturation mechanism by the competition between the scalar and vector densities \cite{Walecka1974AP}, the natural explanation of the pseudospin symmetry in the nucleon spectrum \cite{Ginocchio1997PRL,Meng1998PRC,Liang2015PR} and the spin symmetry in antinucleon spectrum \cite{Liang2015PR,Zhou2003PRL,He2006EPJA}, and the self-consistent treatment of the nuclear magnetism \cite{Koepf1989NPA,Koepf1990NPA}, the RDFT has continuously attracted worldwide attentions in recent decades \cite{Ring1996PPNP,Vretenar2005PR,Meng2006PPNP,Niksic2011PPNP,Meng2013FoP,Meng2015JPG,Zhou2016PS,Meng2016book,Shen2019PPNP,Meng2021AAAPS}. 
Based on the RDFT, efforts have been made for the theoretical descriptions of the charge radii and the improvement of the accuracy \cite{An2020PRC,Perera2021PRC,An2022PRC,An2024PRC}. 
The descriptions of the charge radius by the macroscopic-microscopic model and the nonrelativistic density functional theory can be found in Refs.~\cite{Wang2013PRC,Gaidarov2014PRC,Bassem2015IJMPE,Sarriguren2019PRC,Reinhard2021PRC,Scamps2021EPJA}. 

In exotic nuclei, the Fermi energy lies close to the continuum threshold, enabling the pairing interaction to scatter nucleons from bound states to the resonant states within the continuum. 
This could potentially result in a more diffuse density, thereby influencing ground-state properties including binding energy and charge radius, and even the dripline location, which is the so-called continuum effect \cite{Meng2006PPNP}.
Based on the RDFT and taking into account the effects of pairing correlations and continuum, the relativistic continuum Hartree-Bogoliubov (RCHB) theory for spherical nuclei was developed \cite{Meng1996PRL,Meng1998NPA}. 
The RCHB theory has achieved great success in the descriptions for both stable and exotic nuclei \cite{Meng1998PRL,Meng1998PLB,Meng2002PLB,Meng2002PRC,Zhang2002CPL,Lv2003EPJA,Zhang2005NPA,Meng2006PPNP,Lim2016PRC,Zhang2016CPC,Kuang2023EPJA,Guo2024PRC,Wu2024PRC_KRR}. 
Utilizing the RCHB calculations, the charge radii for the nuclei with $A\geqslant 40$ were systematically investigated, and a new isospin-dependent $Z^{1/3}$ formula for the nuclear charge radii was proposed \cite{Zhang2002EPJA}. 
In Ref.~\cite{Xia2018ADNDT}, the first nuclear mass table that incorporates continuum effects was constructed based on the RCHB theory, and the continuum effects on the limits of the nuclear landscape were studied.

Except for doubly-magic nuclei which exhibit spherical shape, most atomic nuclei are deformed. 
Based on the RDFT and treating the effects of deformation, pairing correlations and continuum simultaneously, 
the deformed relativistic Hartree-Bogoliubov theory in continuum (DRHBc) was developed \cite{Zhou2010PRC,Li2012PRC}. 
Inheriting the advantages of the RCHB theory and including the deformation degrees of freedom, the DRHBc theory has been successfully applied in investigating the halo structures in B \cite{Yang2021PRL,Sun2021PRC}, C \cite{Sun2018PLB,Sun2020NPA,Wang2024EPJA}, Ne \cite{Zhong2022SCP,Pan2024PLB}, Na \cite{Zhang2023PRC_Na}, Mg \cite{Zhang2023PLB,An2024PLB}, and Al \cite{Zhang2024PRC_Al} isotopes, dripline locations \cite{In2021IJMPE,Zhang2021PRC,Pan2021PRC,He2021CPC,He2024PRC}, nuclear shapes and sizes \cite{Pan2019IJMPE,Choi2022PRC,Kim2022PRC,Guo2023PRC,Mun2023PLB}, nuclear stability and radioactive decays \cite{Xiao2023PLB,Choi2024PRC,Zhang2024CPC}, as well as other applications and extensions  \cite{Zhang2019PRC,Sun2021SciB,Sun2021PRC_AMP,Sun2022CPC,Sun2022PRC,Zhang2022PRC,Zhang2023PRC_2DCH,Zhao2023PLB,Mun2024PRC,Zhang2024PRC_shell,Pan2025arXiv}. 
Based on the DRHBc theory and one of the most successful relativistic density functionals PC-PK1 \cite{Zhao2010PRC}, a nuclear mass table including both deformation and continuum effects is under construction \cite{Zhang2020PRC,Pan2022PRC}. 
Up to now, the even-$Z$ part of the DRHBc mass table has been completed, and the ground-state properties, including the binding energy, charge radius and quadrupole deformation, have been published \cite{Zhang2022ADNDT,Guo2024ADNDT}. 

In this work, the systematics of the charge radii for even-$Z$ nuclei based on the DRHBc theory with the relativistic density functional PC-PK1 \cite{Zhao2010PRC} are investigated.
In Sec.~\ref{theory}, a brief framework of the DRHBc theory is introduced. 
The numerical details in the DRHBc mass table calculations are provided in Sec.~\ref{num}. 
The systematics of the charge radii and their deformation correlation are discussed in Sec.~\ref{results}. 
A summary is given in Sec.~\ref{summary}.

\section{Theoretical framework} \label{theory}

The details of the DRHBc theory can be found in Refs.~\cite{Li2012PRC,Zhang2020PRC}. 
Here a very brief theoretical framework is presented. 

In the DRHBc theory, the mean fields and pairing correlations are self-consistently treated by the relativistic Hartree-Bogoliubov (RHB) equations, 
\begin{equation}
	\label{RHB}
	\begin{pmatrix} \hat{h}_D - \lambda & \hat{\Delta} \\ -\hat{\Delta}^* & -\hat{h}_D^* + \lambda \end{pmatrix} 
	\begin{pmatrix} U_k \\ V_k \end{pmatrix} = E_k \begin{pmatrix} U_k \\ V_k \end{pmatrix} , 
\end{equation}
where $\hat{h}_D$ is the Dirac Hamiltonian, $\lambda$ is the Fermi energy, $\hat{\Delta}$ is the pairing potential, $E_k$ is the quasiparticle energy, and $U_k$ and $V_k$ are the quasiparticle wavefunctions. 

For an axially deformed nucleus with spatial reflection symmetry, the densities and potentials are expanded in terms of the Legendre polynomials, 
\begin{equation}
	\label{mlb}
	f(\bm{r}) = \sum_\lambda f_\lambda(r) P_\lambda(\cos\theta) , \qquad \lambda=0,2,4,\dots,\lambda_{\max}
\end{equation}
where only even numbers are taken for $\lambda$ due to spatial reflection symmetry. 
In order to take into account the continuum effect in exotic nuclei, the RHB equations \eqref{RHB} are solved in a spherical Dirac Woods-Saxon (DWS) basis \cite{Zhou2003PRC,Zhang2022PRC}, whose radial wavefunctions have a proper asymptotic behavior at large $r$. 

For an odd-$A$ nucleus, the blocking effect of the unpaired nucleon needs to be considered \cite{Ring1980NMBP}. 
In the DRHBc theory, the blocking effect is included with the equal filling approximation \cite{Li2012CPL,Pan2022PRC}. 

After self-consistently solving the RHB equations \eqref{RHB}, the nucleon densities are obtained and the physical observables, such as binding energy, charge radius and quadrupole deformation, can be calculated \cite{Zhang2020PRC,Pan2022PRC}. 
The charge radius is calculated by 
\begin{equation}
	\label{Rch}
	R_{\text{ch}} = \sqrt{\braket{r^2}_p + 0.64 ~ \text{fm}^2} , 
\end{equation}
where $\braket{r^2}_p$ is the root-mean-square (rms) radius of proton. 
Since Eq.~\eqref{Rch} has been adopted in the fitting of the functional, the neutron contribution and the Darwin-Foldy correction \cite{Ekstrom2015PRC} are not taken into account here. 

\section{Numerical details} \label{num}

The numerical details in the DRHBc mass table calculations have been introduced in Refs.~\cite{Zhang2022ADNDT,Guo2024ADNDT}, and they are briefly listed as follows: 
For the particle-hole channel, the relativistic density functional PC-PK1 \cite{Zhao2010PRC} is employed. 
For the particle-particle channel, a density-dependent zero-range pairing force is adopted, with the pairing strength $V_0 = -325 ~ \mathrm{MeV ~ fm}^3$, the saturation density $\rho_{\text{sat}} = 0.152 ~ \mathrm{fm}^{-3}$, and a pairing window 100 MeV. 
The box size is $R_{\text{box}} = 20$ fm and the mesh size is $\Delta r=0.1$ fm. 
For the Dirac Woods-Saxon basis, the angular momentum cutoff is $J_{\max} = 23/2~\hbar$, the energy cutoff is $E_{\text{cut}}^+ = 300$ MeV, and the number of the basis states in the Dirac sea is the same as that in the
Fermi sea \cite{Zhou2003PRC}. 
The Legendre expansion truncation for potentials and densities in Eq.~\eqref{mlb} is $\lambda_{\max} = 6$, 8 and 10 for the nuclei with $8 \leqslant Z \leqslant 70$, $71\leqslant Z \leqslant 100$ and $101 \leqslant Z \leqslant 120$, respectively.  
The convergence for the above numerical conditions has been examined in Refs.~\cite{Zhang2020PRC,Pan2022PRC}. 

\section{Results and discussions} \label{results}

\subsection{Systematics of the charge radii } \label{global}

\begin{figure*}[htbp]
\centering
\includegraphics[width=.75\linewidth]{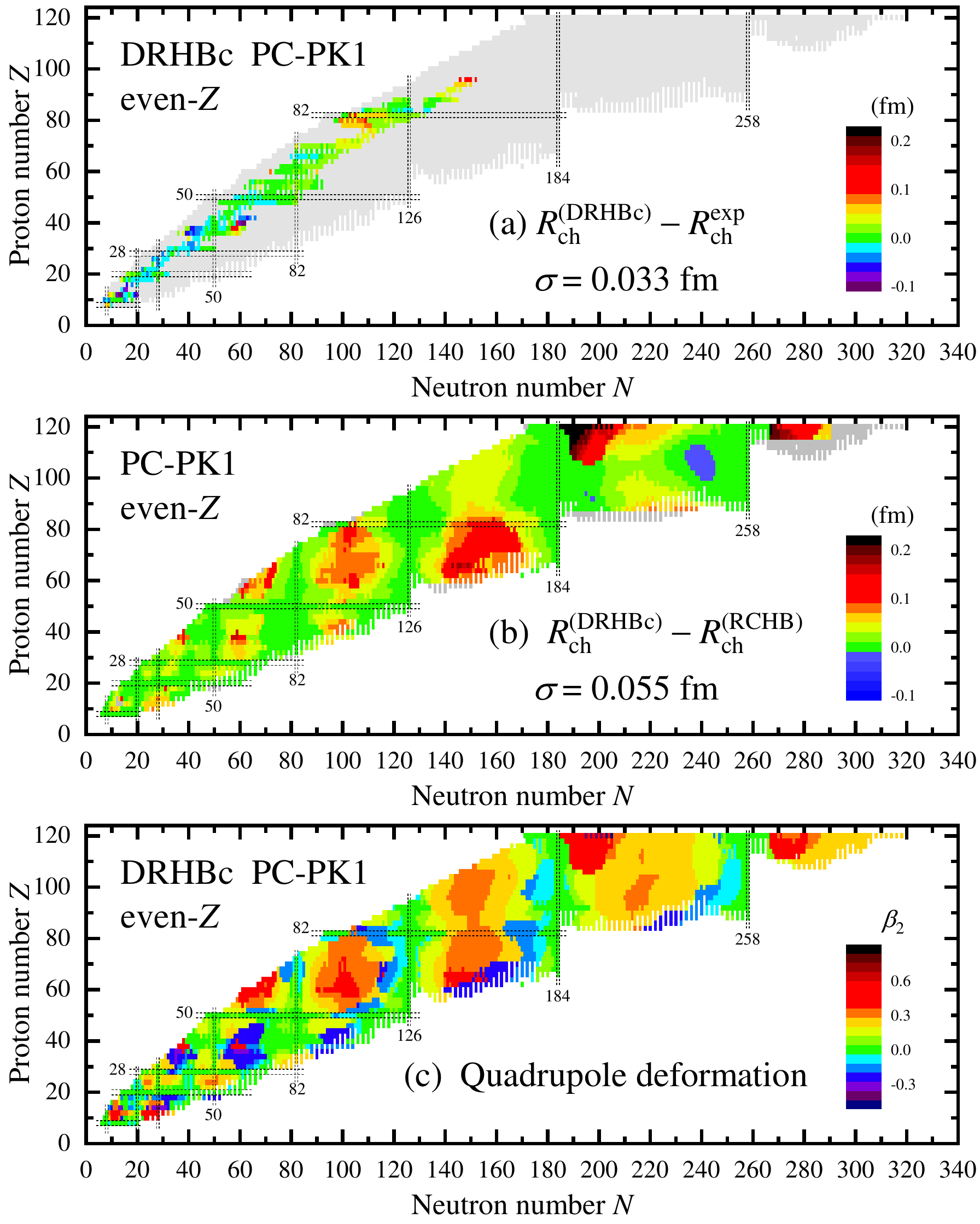}
\caption{ (Color online) For even-$Z$ nuclei with $8 \leqslant Z \leqslant 120$, (a) the differences between the charge radii $R_{\text{ch}}$ in DRHBc calculations \cite{Guo2024ADNDT} with PC-PK1 and the 620 data available \cite{Angeli2013ADNDT,Li2021ADNDT},
(b) the differences of $R_{\text{ch}}$ calculated in DRHBc and RCHB \cite{Xia2018ADNDT} with PC-PK1, and (c) the calculated quadrupole deformations $\beta_2$ in DRHBc with PC-PK1, scaled by colors.  }
\label{fig_lscp}
\end{figure*}

In order to study the systematics of the charge radii $R_{\text{ch}}$ and their correlation with the deformation, the comparisons for the DRHBc charge radii respectively with the data and with the RCHB results, as well as the systematics of the DRHBc quadrupole deformation $\beta_2$, are shown in Fig.~\ref{fig_lscp}. 

The charge radius differences between the DRHBc calculations \cite{Guo2024ADNDT} and the data available \cite{Angeli2013ADNDT,Li2021ADNDT}, $R_{\text{ch}}^{\text{(DRHBc)}} - R_{\text{ch}}^{\text{exp}}$, for even-$Z$ nuclei are shown in Fig.~\ref{fig_lscp}(a). 
As shown in the figure, the data for most nuclei are reproduced within 0.05 fm. 
The rms deviation for the total 620 nuclei is $\sigma=0.033$ fm, with $\sigma=0.032$ fm for even-even nuclei \cite{Zhang2022ADNDT} and $\sigma=0.034$ fm for even-odd nuclei. 
In comparison with $\sigma=0.036$ fm in RCHB calculations \cite{Xia2018ADNDT}, it is demonstrated that the deformation degree of freedom improves the description. 

Figure \ref{fig_lscp}(b) shows the charge radius differences between the DRHBc \cite{Guo2024ADNDT} and RCHB \cite{Xia2018ADNDT} calculations, $R_{\text{ch}}^{\text{(DRHBc)}} - R_{\text{ch}}^{\text{(RCHB)}}$, for even-$Z$ nuclei. 
The rms difference is $\sigma=0.055$ fm. 
For nuclei near the regions with either proton or neutron magic numbers, the differences are negligible. 
Remarkable differences appear in the middle of the shell, for example, the regions with $N$ between 50 and 82, 82 and 126, 126 and 184, as well as 184 and 258. 
Among 4609 nuclei, there are 2808 nuclei with $R_{\text{ch}}^{\text{(DRHBc)}} - R_{\text{ch}}^{\text{(RCHB)}} \geqslant  0.01$ fm and only 67 nuclei with $R_{\text{ch}}^{\text{(DRHBc)}} - R_{\text{ch}}^{\text{(RCHB)}} \leqslant -0.01$ fm. 
This indicates that the deformation typically leads to a larger radius. 
Such effects are closely related to the shell structures. 
The details of the underlying mechanisms are discussed in Section \ref{defor}. 

Figure \ref{fig_lscp}(c) shows the calculated quadrupole deformations $\beta_2$ in DRHBc. 
It is clearly seen that spherical or near-spherical shapes mainly occur near the regions with either proton or neutron magic numbers, corresponding to the negligible differences shown in Fig.~\ref{fig_lscp}(b). 
The nuclei with remarkable quadrupole deformations, either prolate or oblate, normally occur in the middle of the shell, roughly corresponding to the regions with large $R_{\text{ch}}^{\text{(DRHBc)}} - R_{\text{ch}}^{\text{(RCHB)}}$ in Fig.~\ref{fig_lscp}(b). 
These results further demonstrate the correlation between the radius and deformation. 

	\begin{figure}[htbp]
	\centering
	\includegraphics[width=0.5\linewidth]{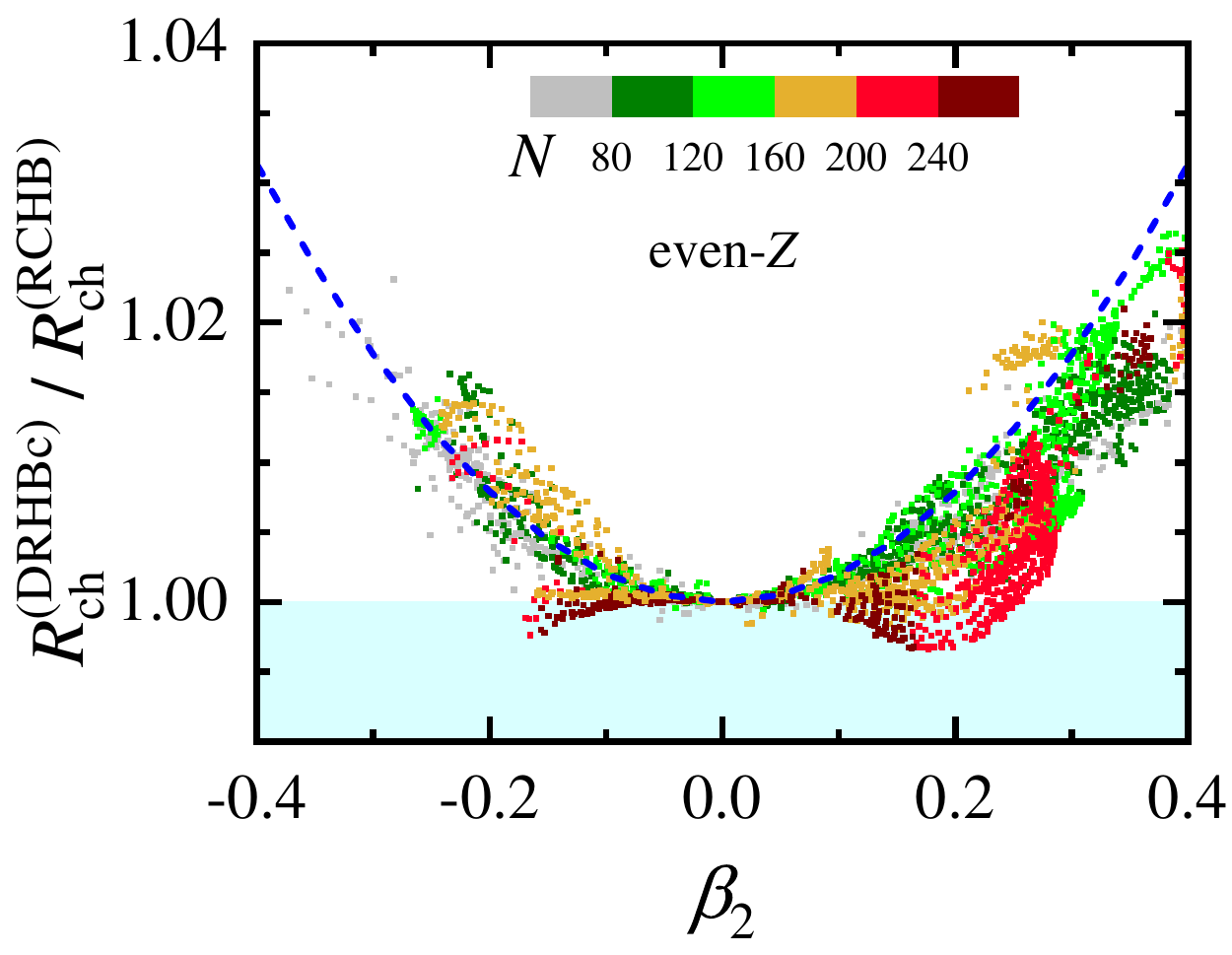} 
	\caption{ The ratio of $R_{\mathrm{ch}}$ calculated in DRHBc to that in RCHB as a function of $\beta_2$. 
	The neutron numbers are scaled by colors. 
	The empirical formula $R_{\mathrm{ch}}(\beta_2) / R_{\mathrm{ch}}(0) = (1+ 5\beta_2^2 / 4\pi)^{1/2}$ is shown by the blue dashed line.  }
	\label{Rch_beta}
	\end{figure}

It would be interesting to systematically compare the charge radii between the prolate and oblate cases. 
In Fig.~\ref{Rch_beta}, the ratio of $R_{\mathrm{ch}}$ calculated in DRHBc to that in RCHB as a function of $\beta_2$ is shown. 
The blue dashed line corresponds to the empirical formula $R_{\text{ch}}(\beta_2) = \left( 1+\frac{5}{4\pi}\beta_2^2 \right)^{1/2} R_{\text{ch}}(0)$. 
At $\beta_2 \approx 0$, the ratios in the prolate and oblate sides exhibit similar behaviors, both close to the empirical curve. 
With the increase of $|\beta_2|$, the number of prolate nuclei is significantly larger than that of oblate nuclei. 
One cannot simply conclude that the radius is larger in the prolate or oblate side, as it varies significantly from case to case.

\subsection{Evolution of the charge radius with the neutron number} \label{evolution}
% and the deformation dependence

\begin{figure}[htbp]
  \centering
  \includegraphics[width=0.5\linewidth]{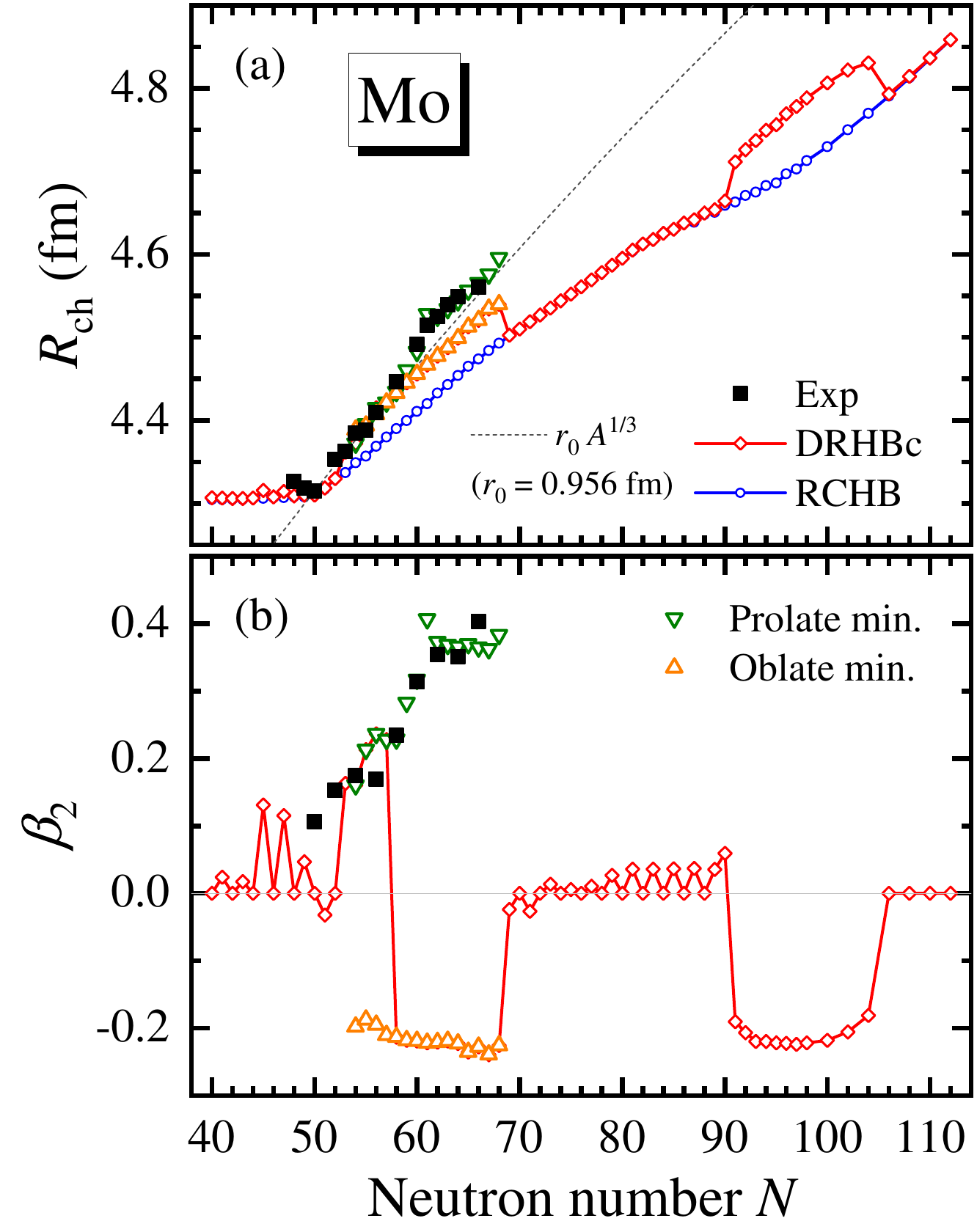} 
  \caption{ (a) Charge radius and (b) quadrupole deformation as functions of the neutron number in the DRHBc \cite{Guo2024ADNDT} calculations with PC-PK1 for Mo isotopes. 
  The RCHB charge radii \cite{Xia2018ADNDT}, the data available \cite{Angeli2013ADNDT,Li2021ADNDT}, and the empirical formula $r_0 A^{1/3}$ with $r_0=0.956$ fm determined by data at $N=50$, are shown for comparison in panel (a). 
  The available experimental deformation \cite{Pritychenko2016ADNDT} is shown in panel (b). 
  For the isotopes with significant prolate and oblate minima simultaneously, both minima are shown with green triangles and orange triangles, respectively, for comparison in both panels. }
  \label{fig_Z42}
\end{figure}

To investigate the evolution of the charge radius with respect to the neutron number and the roles played by the deformation in detail, the Mo ($Z=42$) isotopes, which exhibit rich information on shape evolution as shown in Fig.~\ref{fig_lscp}(c), are taken as examples. 
The charge radii of Mo isotopes calculated by the DRHBc theory from the proton drip line at $N=40$ to the neutron drip line at $N=112$ are shown in Fig.~\ref{fig_Z42}(a) with red diamonds. 
For the isotopes with $54 \leqslant N \leqslant 68$, two minima at the prolate and oblate sides are obtained with the energy differences ranging down to 0.1 MeV. 
Such distinct shapes with similar energies in the same nucleus suggest possible shape coexistence. 
Therefore, both the prolate and oblate solutions in the isotopes with $54 \leqslant N \leqslant 68$ are presented with green and orange triangles, respectively. 
For comparison, the RCHB results \cite{Xia2018ADNDT} are shown with blue circles, the data available \cite{Angeli2013ADNDT,Li2021ADNDT} are shown with black squares, and the empirical formula $r_0 A^{1/3}$ is shown with the dashed line, where $r_0=0.956$ fm is determined by the data at $N=50$. 
% RCHB, DRHBc (sigma: empirical 0.021; DRHBc 0.030; RCHB 0.062 fm)
In Fig.~\ref{fig_Z42}(a), the empirical formula roughly reproduces the general trend of the data. 
For the isotopes near $N=50$ with spherical or near-spherical shapes, both the RCHB and DRHBc calculations reproduce the data well, including the kink at $N=50$ corresponding to the shell closure. 
For the isotopes with $N>52$, the data are underestimated by RCHB but well reproduced by DRHBc. 
For the isotopes with shape coexistence, $58 \leqslant N \leqslant 66$, the charge radii corresponding to the prolate minima in the DRHBc calculations better reproduce the data. 
The rms deviation from the data is $\sigma=0.030$ fm for DRHBc, in comparison with $\sigma=0.062$ fm for RCHB. 
In short, taking into account the deformation effect remarkably improves the description of charge radii for Mo isotopes. 

The quadrupole deformations of Mo isotopes calculated by the DRHBc theory from the proton drip line at $N=40$ to the neutron drip line at $N=112$ are shown in Fig.~\ref{fig_Z42}(b) with red diamonds. 
For the isotopes with $54 \leqslant N \leqslant 68$ exhibiting shape coexistence, the prolate and oblate minima are also presented with green and orange triangles, respectively. 
The data available \cite{Pritychenko2016ADNDT} are shown with black squares for comparison. 
It should be mentioned that the data are determined from the observed $B(E2, 0_1^+\to 2_1^+)$ by assuming that the nucleus is a rigid rotor, and are only available in the absolute values for even-even nuclei. 
The isotopes near magic numbers $N=50$ and 82 exhibit spherical or near-spherical shapes, and the charge radii in DRHBc are the same or close to the RCHB results. 
For the isotopes with pronounced deformations, the charge radii in DRHBc are significantly larger than the RCHB results. 
For the isotopes with shape coexistence at $58 \leqslant N \leqslant 66$, the prolate minima reproduce better the data than the oblate ones, corresponding to the better descriptions for charge radii. 
At $N = 69$ and 91, the transitions between well-deformed and (near-)spherical shapes lead to the abrupt changes in charge radii. 
More data are desired to further examine the DRHBc predictions. 

\begin{figure}[htbp]
  \centering
  \includegraphics[width=0.5\linewidth]{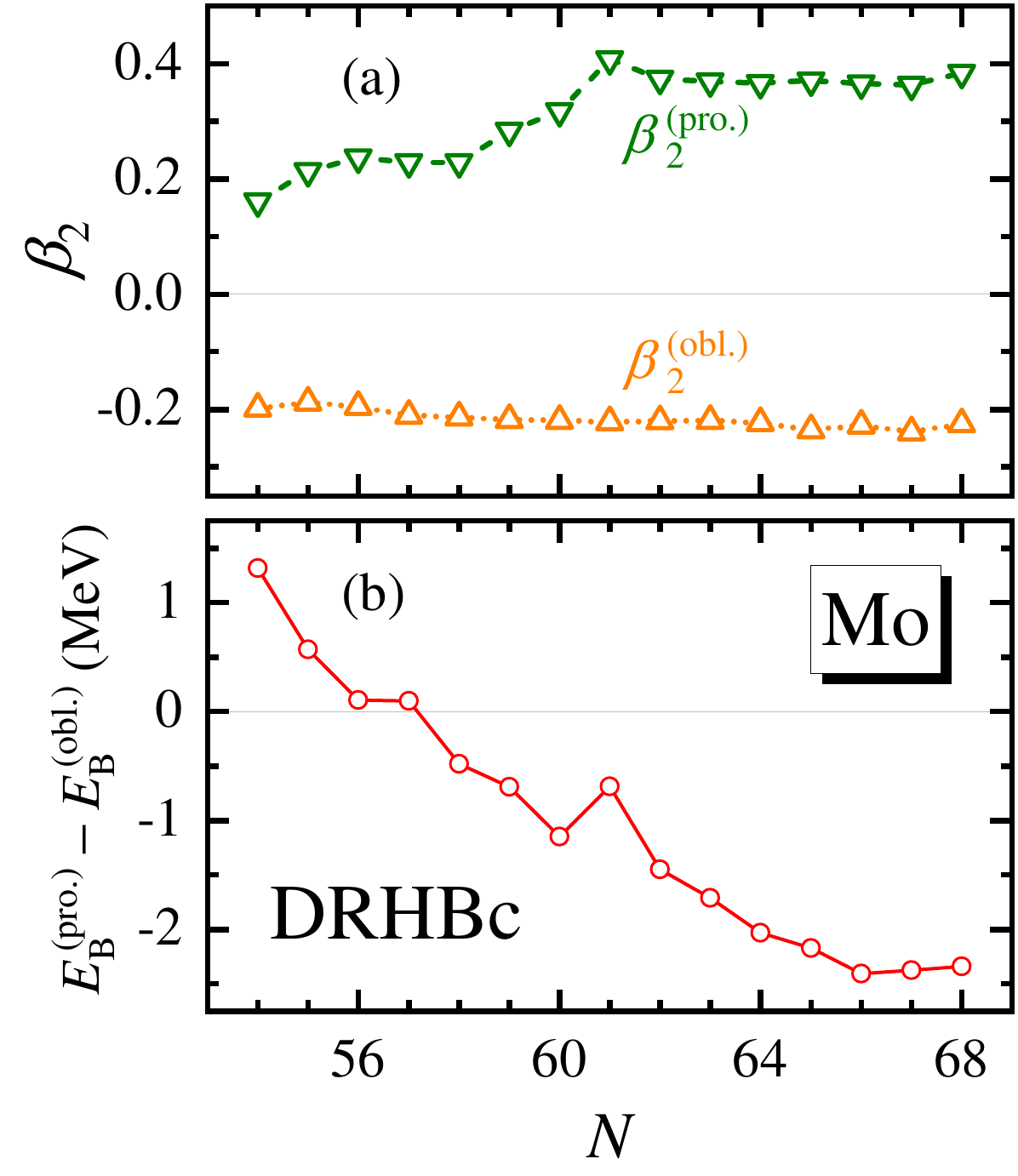} 
  \caption{ Quadrupole deformation parameters $\beta_2$ of the prolate (green triangles) and oblate (orange triangles) minima, as well as the binding energy difference (purple squares) between them, $E_{\text{B}}^{\text{(pro.)}} - E_{\text{B}}^{\text{(obl.)}}$, as functions of the neutron number for Mo isotopes with $54 \leqslant N \leqslant 68$ in the DRHBc calculations with PC-PK1. 
   }
  \label{fig_MTeZ_Z42Mo_coex_1}
\end{figure}

For the isotopes with shape coexistence, it is interesting to see whether the prolate or oblate shape wins in energy. 
In Fig.~\ref{fig_MTeZ_Z42Mo_coex_1}(a), the quadrupole deformation parameters $\beta_2$ of the prolate and oblate minima for the Mo isotopes with $54 \leqslant N \leqslant 68$ are shown.  
The deformation parameters $\beta_2$ of the oblate minima are around $-0.2$. 
The $\beta_2$ values of the prolate minima are around $0.2$ for $53\leqslant N \leqslant 58$. 
For $59\leqslant N \leqslant 61$, the $\beta_2$ values gradually increase to 0.41. 
For $62 \leqslant N \leqslant 68$, the $\beta_2$ values are around 0.4. 

In Fig.~\ref{fig_MTeZ_Z42Mo_coex_1}(b), the energy difference between the prolate and oblate minima, i.e., $E_{\text{B}}^{\text{(pro.)}} - E_{\text{B}}^{\text{(obl.)}}$, for the Mo isotopes with $54 \leqslant N \leqslant 68$ are shown. 
The absolute values of $E_{\text{B}}^{\text{(pro.)}} - E_{\text{B}}^{\text{(obl.)}}$ vary from 0.1 to 2.5 MeV, and the remarkable differences of the deformation parameters $\beta_2$ are around 0.4, indicating the strong competition between the prolate and oblate shapes. 
For $54 \leqslant N \leqslant 57$, the positive values suggest that the prolate minima are the ground states. 
For $58 \leqslant N \leqslant 68$, the negative values suggest that the oblate minima are the ground states. 

\subsection{Deformation dependence of the charge radius} \label{defor}

\begin{figure}[htbp]
\centering
\includegraphics[width=.9\linewidth]{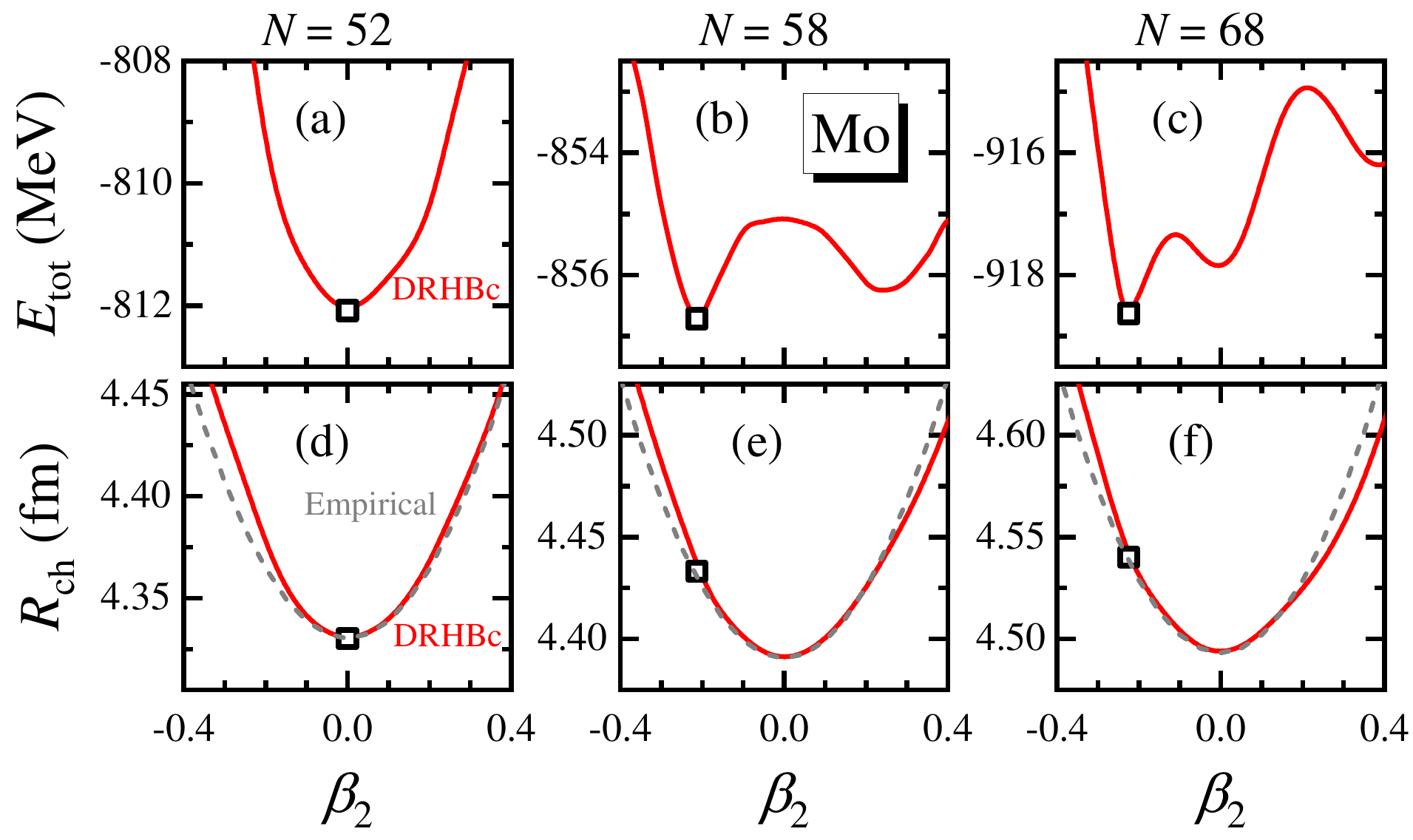}
\caption{ (Color online) Total energies $E_{\text{tot}}$ [(a), (b) and (c)] and charge radii $R_{\text{ch}}$ [(d), (e) and (f)] as functions of quadrupole deformation $\beta_2$ of the Mo isotopes with $N=52,58,68$ from the constrained DRHBc calculations with PC-PK1. 
The ground-state deformations are denoted by squares. 
The empirical formula for charge radius, $R_{\text{ch}}(\beta_2) = (1+5\beta_2^2/4\pi)^{1/2} R_{\text{ch}}(0)$, is shown by the gray dashed line. }
\label{fig_Z42PEC}
\end{figure}

To further investigate the deformation dependence of the charge radius, 
the shape coexistence and the competition between the prolate and oblate minima, taking the Mo isotopes with $N=52, 58$ and 68 as examples, the potential energy curves (PECs), as well as the evolutions of the charge radii with the quadrupole deformation are shown in Fig.~\ref{fig_Z42PEC}. 
The empirical formula for the charge radius, $R_{\text{ch}}(\beta_2) = \left( 1+\frac{5}{4\pi}\beta_2^2 \right)^{1/2} R_{\text{ch}}(0)$ is also shown in Fig.~\ref{fig_Z42PEC}.

For the isotope with $N=52$, there is only one spherical minimum, as shown in Fig.~\ref{fig_Z42PEC}(a). 
The charge radius increases monotonically with the absolute value of the quadrupole deformation parameter, as shown in Fig.~\ref{fig_Z42PEC}(d). 

For the isotope with $N=58$, the ground state locates at $\beta_2=-0.21$, and another minimum locates at $\beta_2=0.23$, with the energy difference 0.48 MeV, as shown in Fig.~\ref{fig_Z42PEC}(b). 
The charge radius increases monotonically with the absolute value of the quadrupole deformation parameter, as shown in Fig.~\ref{fig_Z42PEC}(e). 

For the isotope with $N=68$, the ground state corresponds to the oblate minimum with $\beta_2=-0.23$, and there is a spherical minimum with the excitation energy 0.70 MeV, as shown in Fig.~\ref{fig_Z42PEC}(c). 
For the neighboring isotopes with $N=69$ and 70, the ground states will move to the near-spherical and spherical shapes, respectively. 
The charge radius increases monotonically with the absolute value of the quadrupole deformation parameter, as shown in Fig.~\ref{fig_Z42PEC}(f). 
It seems that the charge radius always increases with the deformation. 
The dependence of the charge radius on the deformation might provide a way to characterize the nuclear shape by a radius measurement, as has been discussed for triaxial nuclei in Ref.~\cite{Maass2025arXiv}. 

\begin{figure}[htbp]
\centering
\includegraphics[width=.5\linewidth]{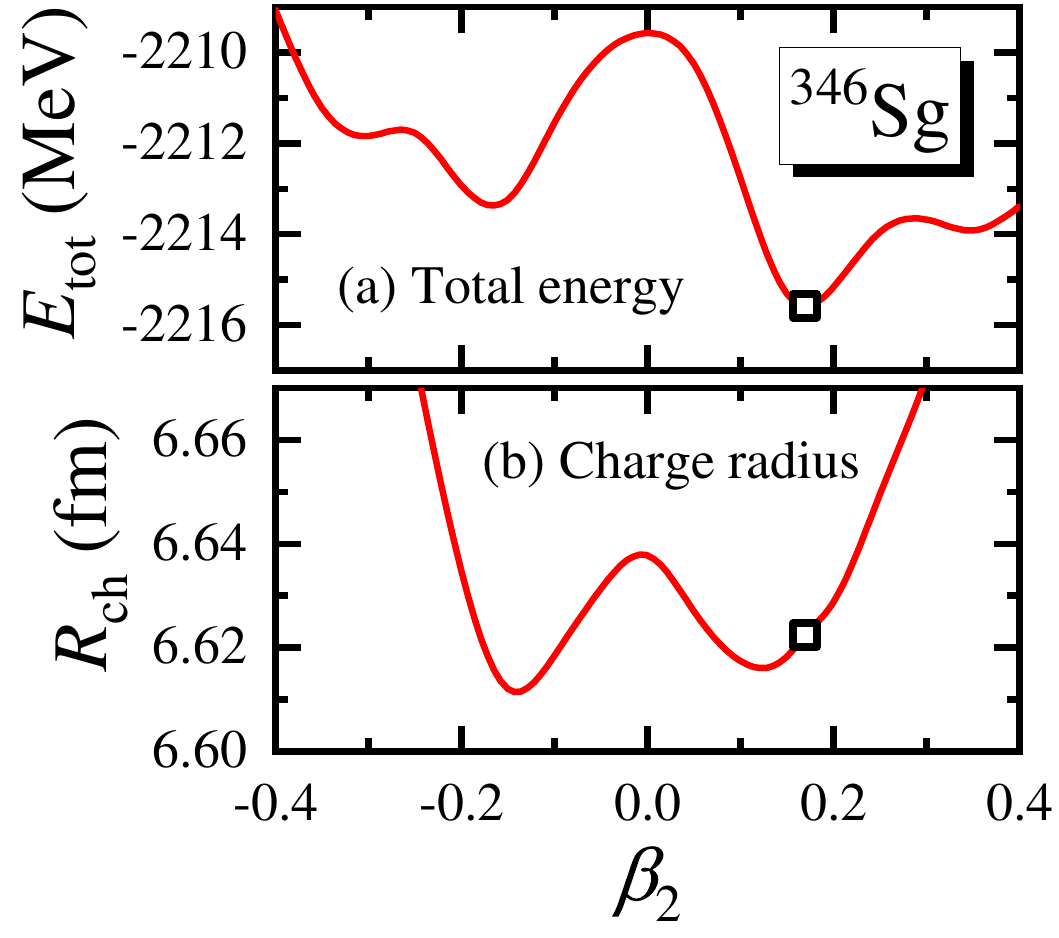}
\caption{ (Color online) Total energy $E_{\text{tot}}$ (a) and charge radius $R_{\text{ch}}$ as functions of quadrupole deformation $\beta_2$ of $^{346}$Sg from the constrained DRHBc calculations with PC-PK1. 
The ground-state deformation is denoted by the square. }
\label{fig_Z106PEC}
\end{figure}

It is interesting to check whether the charge radii may decrease with the deformation $\beta_2$. 
In Fig.~\ref{fig_Z106PEC}, the PEC, as well as the evolution of the charge radius with the quadrupole deformation are shown for $^{346}$Sg. 
In Fig.~\ref{fig_Z106PEC}(a), there are two minima of the energy with $\beta_2=-0.15$ and $0.17$, and the ground state corresponds to the prolate one. 
The spherical result is a saddle point on the PEC, and its energy is 6.00 MeV higher than the ground state. 
In Fig.~~\ref{fig_Z106PEC}(b), there are two minima of the charge radius with $\beta_2=-0.15$ and $0.10$, which do not correspond to the minima in Fig.~\ref{fig_Z106PEC}(a). 
Between these two minima of the charge radius, the charge radius increases with the decrease of the absolute value of the quadrupole deformation parameter.
These results demonstrate that one cannot simply conclude the charge radius as a monotonically increasing function of the quadrupole deformation parameter, and the largest binding energy does not necessarily correspond to the lowest charge radius. 

\begin{figure}[htbp]
\centering
\includegraphics[width=1\linewidth]{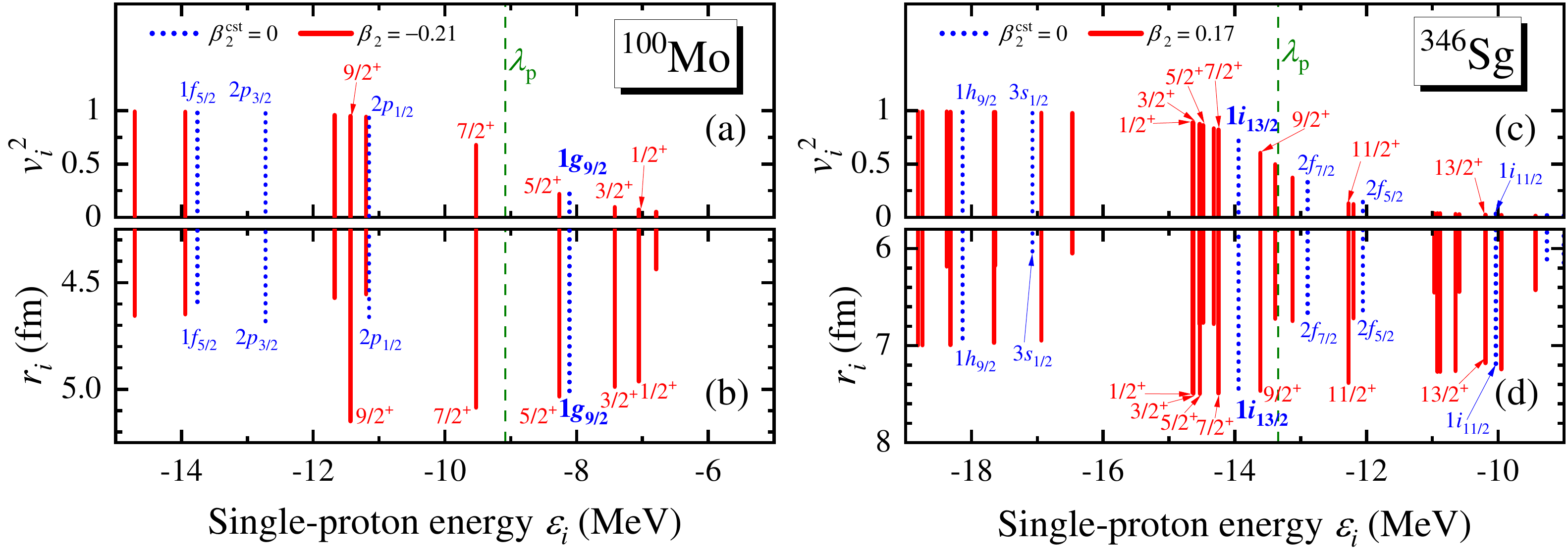}
\caption{ (Color online) The occupation probability $v_i^2$ and rms radius $r_i$ versus the single-proton energy $\epsilon_i$ near the Fermi energy $\lambda_p$ in the canonical basis for the deformed ground states of $^{100}$Mo [(a) and (b)] and $^{346}$Sg [(c) and (d)] with red solid lines. 
The corresponding spherical results are shown with blue dotted lines for comparison. 
In the spherical case, the levels are labeled by the spherical quantum numbers. % $(n,l,j)$. 
In the deformed case, the levels stemming from $1g_{9/2}$ in $^{100}$Mo and $1i_{13/2}$ in $^{346}$Sg are labeled by the quantum numbers $m^\pi$. 
The proton Fermi energies $\lambda_p$ of the ground states are shown with green dashed lines. }
\label{Rch_spl}
\end{figure}

To understand the underlying structural cause of the deformation effects on the charge radius, the composition of the rms radii for proton will be discussed. 
In Fig.~\ref{Rch_spl}, taking the deformed ground states of $^{100}$Mo and $^{346}$Sg as examples, the occupation probability $v_i^2$ and rms radius $r_i$ for the  single-proton levels near the Fermi energy $\lambda_p$ in the canonical basis are shown with red solid lines. 
The corresponding spherical results are shown with blue dotted lines for comparison. 

For $^{100}$Mo shown in Fig.~\ref{Rch_spl}(a) and (b), the inclusion of deformation leads to the splitting of single-proton levels, which significantly impacts the $v_i^2$ near $\lambda_p$ in Fig.~\ref{Rch_spl}(a), but the influence on the corresponding rms radius is small in Fig.~\ref{Rch_spl}(b). 
For the spherical case, the level $1g_{9/2}$ above the Fermi energy has the largest $r_i$, but small $v_i^2$. 
For the deformed case, this level $1g_{9/2}$ splits into $9/2^+$ and $7/2^+$ below the Fermi energy and $5/2^+$, $3/2^+$ and $1/2^+$ above the Fermi energy. 
The levels $9/2^+$ and $7/2^+$ have similar $r_i$ but large $v_i^2$, resulting in the larger $R_{\mathrm{ch}}$ for the deformed case in $^{100}$Mo. 

For $^{346}$Sg shown in Fig.~\ref{Rch_spl}(c) and (d), the inclusion of deformation leads to the splitting of single-proton levels. 
For the spherical case, the level $1i_{13/2}$ below the Fermi energy has the largest $r_i$ and large $v_i^2$. 
For the deformed case, this level $1i_{13/2}$ splits into $1/2^+$, $3/2^+$, $5/2^+$, $7/2^+$ and $9/2^+$ below the Fermi energy and $11/2^+$ and $13/2^+$ above the Fermi energy. 
The levels $11/2^+$ and $13/2^+$ have similar $r_i$ but small $v_i^2$, resulting in the smaller $R_{\mathrm{ch}}$ for the deformed case in $^{346}$Sg. 

The deformation dependences of the charge radii in $^{100}$Mo and $^{346}$Sg show that the increase or decrease of the charge radii with deformation is related to specific shell structures, particularly the key single-particle levels near the Fermi energy.

\section{Summary} \label{summary}

In summary, the systematics of the charge radii for even-$Z$ nuclei with $8 \leqslant Z \leqslant 120$ are investigated based on the DRHBc theory with PC-PK1, and their deformation correlation is explored. 
The available data are well reproduced with the rms deviation $\sigma=0.033$ fm, improved from $\sigma=0.036$ fm in the RCHB calculations \cite{Xia2018ADNDT}. 
In particular, for the nuclei between the closed shells, the descriptions of the charge radii are remarkably improved by the deformation degree of freedom. 
Taking Mo isotopes as examples, the rms deviation from the data is reduced from $0.062$ fm to $0.030$ fm. 
It is found that the charge radius typically increases with the deformation. 
However, this is not always true. 
For example, in $^{346}$Sg, the charge radius for the ground state at $\beta_2=0.17$ is 6.622 fm, which is smaller than that of the state with spherical shape, 6.638 fm. 

In the present, only the results of even-$Z$ nuclei are available \cite{Guo2024ADNDT}. 
A complete DRHBc mass table including odd-$Z$ nuclei will appear soon, enabling a more comprehensive investigation on the systematics of the charge radii and their deformation correlation. 
More experimental data, including charge radii and deformations,  will deepen our understanding on nuclear structures, especially the data distinguishing prolate from oblate shapes, which will provide valuable constraints in examining existing nuclear models. 
It would also be interesting to use DRHBc to improve the isospin-dependent formula for the charge radius in Ref.~\cite{Zhang2002EPJA} by incorporating the deformation effect. 
As the DRHBc theory is based on the axial symmetry approximation, the effects of the triaxial deformation \cite{Zhang2023PRC_TRHBc} and the deformations with higher orders on the charge radius are issues worthy of further investigation.

\begin{acknowledgments}
Helpful discussions with members of the DRHBc Mass Table Collaboration are highly appreciated. 
This work was partly supported by 
the National Natural Science Foundation of China (Grants No. 11935003, No. 12435006), 
the State Key Laboratory of Nuclear Physics and Technology, Peking University (Grant No. NPT2023ZX03), 
and the High-performance Computing Platform of Peking University.
\end{acknowledgments}

\end{CJK*}

%\bibliography{refcpan}

\begin{thebibliography}{105}%
\makeatletter
\providecommand \@ifxundefined [1]{%
 \@ifx{#1\undefined}
}%
\providecommand \@ifnum [1]{%
 \ifnum #1\expandafter \@firstoftwo
 \else \expandafter \@secondoftwo
 \fi
}%
\providecommand \@ifx [1]{%
 \ifx #1\expandafter \@firstoftwo
 \else \expandafter \@secondoftwo
 \fi
}%
\providecommand \natexlab [1]{#1}%
\providecommand \enquote  [1]{``#1''}%
\providecommand \bibnamefont  [1]{#1}%
\providecommand \bibfnamefont [1]{#1}%
\providecommand \citenamefont [1]{#1}%
\providecommand \href@noop [0]{\@secondoftwo}%
\providecommand \href [0]{\begingroup \@sanitize@url \@href}%
\providecommand \@href[1]{\@@startlink{#1}\@@href}%
\providecommand \@@href[1]{\endgroup#1\@@endlink}%
\providecommand \@sanitize@url [0]{\catcode `\\12\catcode `\$12\catcode
  `\&12\catcode `\#12\catcode `\^12\catcode `\_12\catcode `\%12\relax}%
\providecommand \@@startlink[1]{}%
\providecommand \@@endlink[0]{}%
\providecommand \url  [0]{\begingroup\@sanitize@url \@url }%
\providecommand \@url [1]{\endgroup\@href {#1}{\urlprefix }}%
\providecommand \urlprefix  [0]{URL }%
\providecommand \Eprint [0]{\href }%
\providecommand \doibase [0]{http://dx.doi.org/}%
\providecommand \selectlanguage [0]{\@gobble}%
\providecommand \bibinfo  [0]{\@secondoftwo}%
\providecommand \bibfield  [0]{\@secondoftwo}%
\providecommand \translation [1]{[#1]}%
\providecommand \BibitemOpen [0]{}%
\providecommand \bibitemStop [0]{}%
\providecommand \bibitemNoStop [0]{.\EOS\space}%
\providecommand \EOS [0]{\spacefactor3000\relax}%
\providecommand \BibitemShut  [1]{\csname bibitem#1\endcsname}%
\let\auto@bib@innerbib\@empty
%</preamble>
\bibitem [{\citenamefont {Erler}\ \emph {et~al.}(2012)\citenamefont {Erler},
  \citenamefont {Birge}, \citenamefont {Kortelainen}, \citenamefont
  {Nazarewicz}, \citenamefont {Olsen}, \citenamefont {Perhac},\ and\
  \citenamefont {Stoitsov}}]{Erler2012Nat}%
  \BibitemOpen
  \bibfield  {author} {\bibinfo {author} {\bibfnamefont {J.}~\bibnamefont
  {Erler}}, \bibinfo {author} {\bibfnamefont {N.}~\bibnamefont {Birge}},
  \bibinfo {author} {\bibfnamefont {M.}~\bibnamefont {Kortelainen}}, \bibinfo
  {author} {\bibfnamefont {W.}~\bibnamefont {Nazarewicz}}, \bibinfo {author}
  {\bibfnamefont {E.}~\bibnamefont {Olsen}}, \bibinfo {author} {\bibfnamefont
  {A.~M.}\ \bibnamefont {Perhac}}, \ and\ \bibinfo {author} {\bibfnamefont
  {M.}~\bibnamefont {Stoitsov}},\ }\href {\doibase 10.1038/nature11188}
  {\bibfield  {journal} {\bibinfo  {journal} {Nature}\ }\textbf {\bibinfo
  {volume} {486}},\ \bibinfo {pages} {509} (\bibinfo {year}
  {2012})}\BibitemShut {NoStop}%
\bibitem [{\citenamefont {Xia}\ \emph {et~al.}(2018)\citenamefont {Xia},
  \citenamefont {Lim}, \citenamefont {Zhao}, \citenamefont {Liang},
  \citenamefont {Qu}, \citenamefont {Chen}, \citenamefont {Liu}, \citenamefont
  {Zhang}, \citenamefont {Zhang}, \citenamefont {Kim},\ and\ \citenamefont
  {Meng}}]{Xia2018ADNDT}%
  \BibitemOpen
  \bibfield  {author} {\bibinfo {author} {\bibfnamefont {X.~W.}\ \bibnamefont
  {Xia}}, \bibinfo {author} {\bibfnamefont {Y.}~\bibnamefont {Lim}}, \bibinfo
  {author} {\bibfnamefont {P.~W.}\ \bibnamefont {Zhao}}, \bibinfo {author}
  {\bibfnamefont {H.~Z.}\ \bibnamefont {Liang}}, \bibinfo {author}
  {\bibfnamefont {X.~Y.}\ \bibnamefont {Qu}}, \bibinfo {author} {\bibfnamefont
  {Y.}~\bibnamefont {Chen}}, \bibinfo {author} {\bibfnamefont {H.}~\bibnamefont
  {Liu}}, \bibinfo {author} {\bibfnamefont {L.~F.}\ \bibnamefont {Zhang}},
  \bibinfo {author} {\bibfnamefont {S.~Q.}\ \bibnamefont {Zhang}}, \bibinfo
  {author} {\bibfnamefont {Y.}~\bibnamefont {Kim}}, \ and\ \bibinfo {author}
  {\bibfnamefont {J.}~\bibnamefont {Meng}},\ }\href {\doibase
  10.1016/j.adt.2017.09.001} {\bibfield  {journal} {\bibinfo  {journal} {At.
  Data Nucl. Data Tables}\ }\textbf {\bibinfo {volume} {121-122}},\ \bibinfo
  {pages} {1 } (\bibinfo {year} {2018})}\BibitemShut {NoStop}%
\bibitem [{\citenamefont {Zhang}\ \emph
  {et~al.}(2022{\natexlab{a}})\citenamefont {Zhang}, \citenamefont {Cheoun},
  \citenamefont {Choi}, \citenamefont {Chong}, \citenamefont {Dong},
  \citenamefont {Dong}, \citenamefont {Du}, \citenamefont {Geng}, \citenamefont
  {Ha}, \citenamefont {He}, \citenamefont {Heo}, \citenamefont {Ho},
  \citenamefont {In}, \citenamefont {Kim}, \citenamefont {Kim}, \citenamefont
  {Lee}, \citenamefont {Lee}, \citenamefont {Li}, \citenamefont {Li},
  \citenamefont {Luo}, \citenamefont {Meng}, \citenamefont {Mun}, \citenamefont
  {Niu}, \citenamefont {Pan}, \citenamefont {Papakonstantinou}, \citenamefont
  {Shang}, \citenamefont {Shen}, \citenamefont {Shen}, \citenamefont {Sun},
  \citenamefont {Sun}, \citenamefont {Tam}, \citenamefont {Thaivayongnou},
  \citenamefont {Wang}, \citenamefont {Wang}, \citenamefont {Wong},
  \citenamefont {Wu}, \citenamefont {Wu}, \citenamefont {Xia}, \citenamefont
  {Yan}, \citenamefont {Yeung}, \citenamefont {Yiu}, \citenamefont {Zhang},
  \citenamefont {Zhang}, \citenamefont {Zhang}, \citenamefont {Zhao},\ and\
  \citenamefont {Zhou}}]{Zhang2022ADNDT}%
  \BibitemOpen
  \bibfield  {author} {\bibinfo {author} {\bibfnamefont {K.}~\bibnamefont
  {Zhang}}, \bibinfo {author} {\bibfnamefont {M.-K.}\ \bibnamefont {Cheoun}},
  \bibinfo {author} {\bibfnamefont {Y.-B.}\ \bibnamefont {Choi}}, \bibinfo
  {author} {\bibfnamefont {P.~S.}\ \bibnamefont {Chong}}, \bibinfo {author}
  {\bibfnamefont {J.}~\bibnamefont {Dong}}, \bibinfo {author} {\bibfnamefont
  {Z.}~\bibnamefont {Dong}}, \bibinfo {author} {\bibfnamefont {X.}~\bibnamefont
  {Du}}, \bibinfo {author} {\bibfnamefont {L.}~\bibnamefont {Geng}}, \bibinfo
  {author} {\bibfnamefont {E.}~\bibnamefont {Ha}}, \bibinfo {author}
  {\bibfnamefont {X.-T.}\ \bibnamefont {He}}, \bibinfo {author} {\bibfnamefont
  {C.}~\bibnamefont {Heo}}, \bibinfo {author} {\bibfnamefont {M.~C.}\
  \bibnamefont {Ho}}, \bibinfo {author} {\bibfnamefont {E.~J.}\ \bibnamefont
  {In}}, \bibinfo {author} {\bibfnamefont {S.}~\bibnamefont {Kim}}, \bibinfo
  {author} {\bibfnamefont {Y.}~\bibnamefont {Kim}}, \bibinfo {author}
  {\bibfnamefont {C.-H.}\ \bibnamefont {Lee}}, \bibinfo {author} {\bibfnamefont
  {J.}~\bibnamefont {Lee}}, \bibinfo {author} {\bibfnamefont {H.}~\bibnamefont
  {Li}}, \bibinfo {author} {\bibfnamefont {Z.}~\bibnamefont {Li}}, \bibinfo
  {author} {\bibfnamefont {T.}~\bibnamefont {Luo}}, \bibinfo {author}
  {\bibfnamefont {J.}~\bibnamefont {Meng}}, \bibinfo {author} {\bibfnamefont
  {M.-H.}\ \bibnamefont {Mun}}, \bibinfo {author} {\bibfnamefont
  {Z.}~\bibnamefont {Niu}}, \bibinfo {author} {\bibfnamefont {C.}~\bibnamefont
  {Pan}}, \bibinfo {author} {\bibfnamefont {P.}~\bibnamefont
  {Papakonstantinou}}, \bibinfo {author} {\bibfnamefont {X.}~\bibnamefont
  {Shang}}, \bibinfo {author} {\bibfnamefont {C.}~\bibnamefont {Shen}},
  \bibinfo {author} {\bibfnamefont {G.}~\bibnamefont {Shen}}, \bibinfo {author}
  {\bibfnamefont {W.}~\bibnamefont {Sun}}, \bibinfo {author} {\bibfnamefont
  {X.-X.}\ \bibnamefont {Sun}}, \bibinfo {author} {\bibfnamefont {C.~K.}\
  \bibnamefont {Tam}}, \bibinfo {author} {\bibnamefont {Thaivayongnou}},
  \bibinfo {author} {\bibfnamefont {C.}~\bibnamefont {Wang}}, \bibinfo {author}
  {\bibfnamefont {X.}~\bibnamefont {Wang}}, \bibinfo {author} {\bibfnamefont
  {S.~H.}\ \bibnamefont {Wong}}, \bibinfo {author} {\bibfnamefont
  {J.}~\bibnamefont {Wu}}, \bibinfo {author} {\bibfnamefont {X.}~\bibnamefont
  {Wu}}, \bibinfo {author} {\bibfnamefont {X.}~\bibnamefont {Xia}}, \bibinfo
  {author} {\bibfnamefont {Y.}~\bibnamefont {Yan}}, \bibinfo {author}
  {\bibfnamefont {R.~W.-Y.}\ \bibnamefont {Yeung}}, \bibinfo {author}
  {\bibfnamefont {T.~C.}\ \bibnamefont {Yiu}}, \bibinfo {author} {\bibfnamefont
  {S.}~\bibnamefont {Zhang}}, \bibinfo {author} {\bibfnamefont
  {W.}~\bibnamefont {Zhang}}, \bibinfo {author} {\bibfnamefont
  {X.}~\bibnamefont {Zhang}}, \bibinfo {author} {\bibfnamefont
  {Q.}~\bibnamefont {Zhao}}, \ and\ \bibinfo {author} {\bibfnamefont {S.-G.}\
  \bibnamefont {Zhou}} (\bibinfo {collaboration} {DRHBc Mass Table
  Collaboration}),\ }\href {\doibase 10.1016/j.adt.2022.101488} {\bibfield
  {journal} {\bibinfo  {journal} {Atom. Data Nucl. Data Tabl.}\ }\textbf
  {\bibinfo {volume} {144}},\ \bibinfo {pages} {101488} (\bibinfo {year}
  {2022}{\natexlab{a}})}\BibitemShut {NoStop}%
\bibitem [{\citenamefont {Guo}\ \emph {et~al.}(2024{\natexlab{a}})\citenamefont
  {Guo}, \citenamefont {Cao}, \citenamefont {Chen}, \citenamefont {Chen},
  \citenamefont {Cheoun}, \citenamefont {Choi}, \citenamefont {Lam},
  \citenamefont {Deng}, \citenamefont {Dong}, \citenamefont {Du}, \citenamefont
  {Du}, \citenamefont {Duan}, \citenamefont {Fan}, \citenamefont {Gao},
  \citenamefont {Geng}, \citenamefont {Ha}, \citenamefont {He}, \citenamefont
  {Hu}, \citenamefont {Huang}, \citenamefont {Huang}, \citenamefont {Huang},
  \citenamefont {Huang}, \citenamefont {Hyung}, \citenamefont {Chan},
  \citenamefont {Jiang}, \citenamefont {Kim}, \citenamefont {Kim},
  \citenamefont {Lee}, \citenamefont {Lee}, \citenamefont {Li}, \citenamefont
  {Li}, \citenamefont {Li}, \citenamefont {Li}, \citenamefont {Lian},
  \citenamefont {Liang}, \citenamefont {Liu}, \citenamefont {Lu}, \citenamefont
  {Liu}, \citenamefont {Meng}, \citenamefont {Meng}, \citenamefont {Mun},
  \citenamefont {Niu}, \citenamefont {Niu}, \citenamefont {Pan}, \citenamefont
  {Peng}, \citenamefont {Qu}, \citenamefont {Papakonstantinou}, \citenamefont
  {Shang}, \citenamefont {Shang}, \citenamefont {Shen}, \citenamefont {Shen},
  \citenamefont {Sun}, \citenamefont {Sun}, \citenamefont {Wang}, \citenamefont
  {Wang}, \citenamefont {Wang}, \citenamefont {Wang}, \citenamefont {Wu},
  \citenamefont {Wu}, \citenamefont {Wu}, \citenamefont {Xia}, \citenamefont
  {Xie}, \citenamefont {Yao}, \citenamefont {Ip}, \citenamefont {Yiu},
  \citenamefont {Yu}, \citenamefont {Yu}, \citenamefont {Zhang}, \citenamefont
  {Zhang}, \citenamefont {Zhang}, \citenamefont {Zhang}, \citenamefont {Zhang},
  \citenamefont {Zhang}, \citenamefont {Zhang}, \citenamefont {Zhang},
  \citenamefont {Zhang}, \citenamefont {Zhao}, \citenamefont {Zhao},
  \citenamefont {Zheng}, \citenamefont {Zhou}, \citenamefont {Zhou},\ and\
  \citenamefont {Zou}}]{Guo2024ADNDT}%
  \BibitemOpen
  \bibfield  {author} {\bibinfo {author} {\bibfnamefont {P.}~\bibnamefont
  {Guo}}, \bibinfo {author} {\bibfnamefont {X.}~\bibnamefont {Cao}}, \bibinfo
  {author} {\bibfnamefont {K.}~\bibnamefont {Chen}}, \bibinfo {author}
  {\bibfnamefont {Z.}~\bibnamefont {Chen}}, \bibinfo {author} {\bibfnamefont
  {M.-K.}\ \bibnamefont {Cheoun}}, \bibinfo {author} {\bibfnamefont {Y.-B.}\
  \bibnamefont {Choi}}, \bibinfo {author} {\bibfnamefont {P.~C.}\ \bibnamefont
  {Lam}}, \bibinfo {author} {\bibfnamefont {W.}~\bibnamefont {Deng}}, \bibinfo
  {author} {\bibfnamefont {J.}~\bibnamefont {Dong}}, \bibinfo {author}
  {\bibfnamefont {P.}~\bibnamefont {Du}}, \bibinfo {author} {\bibfnamefont
  {X.}~\bibnamefont {Du}}, \bibinfo {author} {\bibfnamefont {K.}~\bibnamefont
  {Duan}}, \bibinfo {author} {\bibfnamefont {X.}~\bibnamefont {Fan}}, \bibinfo
  {author} {\bibfnamefont {W.}~\bibnamefont {Gao}}, \bibinfo {author}
  {\bibfnamefont {L.}~\bibnamefont {Geng}}, \bibinfo {author} {\bibfnamefont
  {E.}~\bibnamefont {Ha}}, \bibinfo {author} {\bibfnamefont {X.-T.}\
  \bibnamefont {He}}, \bibinfo {author} {\bibfnamefont {J.}~\bibnamefont {Hu}},
  \bibinfo {author} {\bibfnamefont {J.}~\bibnamefont {Huang}}, \bibinfo
  {author} {\bibfnamefont {K.}~\bibnamefont {Huang}}, \bibinfo {author}
  {\bibfnamefont {Y.}~\bibnamefont {Huang}}, \bibinfo {author} {\bibfnamefont
  {Z.}~\bibnamefont {Huang}}, \bibinfo {author} {\bibfnamefont {K.~D.}\
  \bibnamefont {Hyung}}, \bibinfo {author} {\bibfnamefont {H.~Y.}\ \bibnamefont
  {Chan}}, \bibinfo {author} {\bibfnamefont {X.}~\bibnamefont {Jiang}},
  \bibinfo {author} {\bibfnamefont {S.}~\bibnamefont {Kim}}, \bibinfo {author}
  {\bibfnamefont {Y.}~\bibnamefont {Kim}}, \bibinfo {author} {\bibfnamefont
  {C.-H.}\ \bibnamefont {Lee}}, \bibinfo {author} {\bibfnamefont
  {J.}~\bibnamefont {Lee}}, \bibinfo {author} {\bibfnamefont {J.}~\bibnamefont
  {Li}}, \bibinfo {author} {\bibfnamefont {M.}~\bibnamefont {Li}}, \bibinfo
  {author} {\bibfnamefont {Z.}~\bibnamefont {Li}}, \bibinfo {author}
  {\bibfnamefont {Z.}~\bibnamefont {Li}}, \bibinfo {author} {\bibfnamefont
  {Z.}~\bibnamefont {Lian}}, \bibinfo {author} {\bibfnamefont {H.}~\bibnamefont
  {Liang}}, \bibinfo {author} {\bibfnamefont {L.}~\bibnamefont {Liu}}, \bibinfo
  {author} {\bibfnamefont {X.}~\bibnamefont {Lu}}, \bibinfo {author}
  {\bibfnamefont {Z.-R.}\ \bibnamefont {Liu}}, \bibinfo {author} {\bibfnamefont
  {J.}~\bibnamefont {Meng}}, \bibinfo {author} {\bibfnamefont {Z.}~\bibnamefont
  {Meng}}, \bibinfo {author} {\bibfnamefont {M.-H.}\ \bibnamefont {Mun}},
  \bibinfo {author} {\bibfnamefont {Y.}~\bibnamefont {Niu}}, \bibinfo {author}
  {\bibfnamefont {Z.}~\bibnamefont {Niu}}, \bibinfo {author} {\bibfnamefont
  {C.}~\bibnamefont {Pan}}, \bibinfo {author} {\bibfnamefont {J.}~\bibnamefont
  {Peng}}, \bibinfo {author} {\bibfnamefont {X.}~\bibnamefont {Qu}}, \bibinfo
  {author} {\bibfnamefont {P.}~\bibnamefont {Papakonstantinou}}, \bibinfo
  {author} {\bibfnamefont {T.}~\bibnamefont {Shang}}, \bibinfo {author}
  {\bibfnamefont {X.}~\bibnamefont {Shang}}, \bibinfo {author} {\bibfnamefont
  {C.}~\bibnamefont {Shen}}, \bibinfo {author} {\bibfnamefont {G.}~\bibnamefont
  {Shen}}, \bibinfo {author} {\bibfnamefont {T.}~\bibnamefont {Sun}}, \bibinfo
  {author} {\bibfnamefont {X.-X.}\ \bibnamefont {Sun}}, \bibinfo {author}
  {\bibfnamefont {S.}~\bibnamefont {Wang}}, \bibinfo {author} {\bibfnamefont
  {T.}~\bibnamefont {Wang}}, \bibinfo {author} {\bibfnamefont {Y.}~\bibnamefont
  {Wang}}, \bibinfo {author} {\bibfnamefont {Y.}~\bibnamefont {Wang}}, \bibinfo
  {author} {\bibfnamefont {J.}~\bibnamefont {Wu}}, \bibinfo {author}
  {\bibfnamefont {L.}~\bibnamefont {Wu}}, \bibinfo {author} {\bibfnamefont
  {X.}~\bibnamefont {Wu}}, \bibinfo {author} {\bibfnamefont {X.}~\bibnamefont
  {Xia}}, \bibinfo {author} {\bibfnamefont {H.}~\bibnamefont {Xie}}, \bibinfo
  {author} {\bibfnamefont {J.}~\bibnamefont {Yao}}, \bibinfo {author}
  {\bibfnamefont {K.~Y.}\ \bibnamefont {Ip}}, \bibinfo {author} {\bibfnamefont
  {T.~C.}\ \bibnamefont {Yiu}}, \bibinfo {author} {\bibfnamefont
  {J.}~\bibnamefont {Yu}}, \bibinfo {author} {\bibfnamefont {Y.}~\bibnamefont
  {Yu}}, \bibinfo {author} {\bibfnamefont {K.}~\bibnamefont {Zhang}}, \bibinfo
  {author} {\bibfnamefont {S.}~\bibnamefont {Zhang}}, \bibinfo {author}
  {\bibfnamefont {S.}~\bibnamefont {Zhang}}, \bibinfo {author} {\bibfnamefont
  {W.}~\bibnamefont {Zhang}}, \bibinfo {author} {\bibfnamefont
  {X.}~\bibnamefont {Zhang}}, \bibinfo {author} {\bibfnamefont
  {Y.}~\bibnamefont {Zhang}}, \bibinfo {author} {\bibfnamefont
  {Y.}~\bibnamefont {Zhang}}, \bibinfo {author} {\bibfnamefont
  {Y.}~\bibnamefont {Zhang}}, \bibinfo {author} {\bibfnamefont
  {Z.}~\bibnamefont {Zhang}}, \bibinfo {author} {\bibfnamefont
  {Q.}~\bibnamefont {Zhao}}, \bibinfo {author} {\bibfnamefont {Y.}~\bibnamefont
  {Zhao}}, \bibinfo {author} {\bibfnamefont {R.}~\bibnamefont {Zheng}},
  \bibinfo {author} {\bibfnamefont {C.}~\bibnamefont {Zhou}}, \bibinfo {author}
  {\bibfnamefont {S.-G.}\ \bibnamefont {Zhou}}, \ and\ \bibinfo {author}
  {\bibfnamefont {L.}~\bibnamefont {Zou}} (\bibinfo {collaboration} {DRHBc Mass
  Table Collaboration}),\ }\href {\doibase
  https://doi.org/10.1016/j.adt.2024.101661} {\bibfield  {journal} {\bibinfo
  {journal} {Atomic Data and Nuclear Data Tables}\ }\textbf {\bibinfo {volume}
  {158}},\ \bibinfo {pages} {101661} (\bibinfo {year}
  {2024}{\natexlab{a}})}\BibitemShut {NoStop}%
\bibitem [{nnd()}]{nndc}%
  \BibitemOpen
  \href {https://www.nndc.bnl.gov/} {\enquote {\bibinfo {title} {{National
  Nuclear Data Center (NNDC), https://www.nndc.bnl.gov/}},}\ }\BibitemShut
  {NoStop}%
\bibitem [{\citenamefont {Angeli}\ and\ \citenamefont
  {Marinova}(2013)}]{Angeli2013ADNDT}%
  \BibitemOpen
  \bibfield  {author} {\bibinfo {author} {\bibfnamefont {I.}~\bibnamefont
  {Angeli}}\ and\ \bibinfo {author} {\bibfnamefont {K.~P.}\ \bibnamefont
  {Marinova}},\ }\href {\doibase 10.1016/j.adt.2011.12.006} {\bibfield
  {journal} {\bibinfo  {journal} {At. Data Nucl. Data Tables}\ }\textbf
  {\bibinfo {volume} {99}},\ \bibinfo {pages} {69} (\bibinfo {year}
  {2013})}\BibitemShut {NoStop}%
\bibitem [{\citenamefont {Li}\ \emph {et~al.}(2021)\citenamefont {Li},
  \citenamefont {Luo},\ and\ \citenamefont {Wang}}]{Li2021ADNDT}%
  \BibitemOpen
  \bibfield  {author} {\bibinfo {author} {\bibfnamefont {T.}~\bibnamefont
  {Li}}, \bibinfo {author} {\bibfnamefont {Y.}~\bibnamefont {Luo}}, \ and\
  \bibinfo {author} {\bibfnamefont {N.}~\bibnamefont {Wang}},\ }\href {\doibase
  https://doi.org/10.1016/j.adt.2021.101440} {\bibfield  {journal} {\bibinfo
  {journal} {Atomic Data and Nuclear Data Tables}\ }\textbf {\bibinfo {volume}
  {140}},\ \bibinfo {pages} {101440} (\bibinfo {year} {2021})}\BibitemShut
  {NoStop}%
\bibitem [{\citenamefont {Zhao}\ \emph {et~al.}(2024)\citenamefont {Zhao},
  \citenamefont {Sun}, \citenamefont {Tanihata}, \citenamefont {Xu},
  \citenamefont {Zhang}, \citenamefont {Prochazka}, \citenamefont {Zhu},
  \citenamefont {Terashima}, \citenamefont {Meng}, \citenamefont {He},
  \citenamefont {Liu}, \citenamefont {Li}, \citenamefont {Lu}, \citenamefont
  {Lin}, \citenamefont {Lin}, \citenamefont {Liu}, \citenamefont {Ren},
  \citenamefont {Sun}, \citenamefont {Wang}, \citenamefont {Wang},
  \citenamefont {Wang}, \citenamefont {Wang}, \citenamefont {Wei},
  \citenamefont {Xu}, \citenamefont {Zhang}, \citenamefont {Zhang},\ and\
  \citenamefont {Zhang}}]{Zhao2024PLB}%
  \BibitemOpen
  \bibfield  {author} {\bibinfo {author} {\bibfnamefont {J.}~\bibnamefont
  {Zhao}}, \bibinfo {author} {\bibfnamefont {B.-H.}\ \bibnamefont {Sun}},
  \bibinfo {author} {\bibfnamefont {I.}~\bibnamefont {Tanihata}}, \bibinfo
  {author} {\bibfnamefont {J.}~\bibnamefont {Xu}}, \bibinfo {author}
  {\bibfnamefont {K.}~\bibnamefont {Zhang}}, \bibinfo {author} {\bibfnamefont
  {A.}~\bibnamefont {Prochazka}}, \bibinfo {author} {\bibfnamefont
  {L.}~\bibnamefont {Zhu}}, \bibinfo {author} {\bibfnamefont {S.}~\bibnamefont
  {Terashima}}, \bibinfo {author} {\bibfnamefont {J.}~\bibnamefont {Meng}},
  \bibinfo {author} {\bibfnamefont {L.}~\bibnamefont {He}}, \bibinfo {author}
  {\bibfnamefont {C.}~\bibnamefont {Liu}}, \bibinfo {author} {\bibfnamefont
  {G.}~\bibnamefont {Li}}, \bibinfo {author} {\bibfnamefont {C.}~\bibnamefont
  {Lu}}, \bibinfo {author} {\bibfnamefont {W.}~\bibnamefont {Lin}}, \bibinfo
  {author} {\bibfnamefont {W.}~\bibnamefont {Lin}}, \bibinfo {author}
  {\bibfnamefont {Z.}~\bibnamefont {Liu}}, \bibinfo {author} {\bibfnamefont
  {P.}~\bibnamefont {Ren}}, \bibinfo {author} {\bibfnamefont {Z.}~\bibnamefont
  {Sun}}, \bibinfo {author} {\bibfnamefont {F.}~\bibnamefont {Wang}}, \bibinfo
  {author} {\bibfnamefont {J.}~\bibnamefont {Wang}}, \bibinfo {author}
  {\bibfnamefont {M.}~\bibnamefont {Wang}}, \bibinfo {author} {\bibfnamefont
  {S.}~\bibnamefont {Wang}}, \bibinfo {author} {\bibfnamefont {X.}~\bibnamefont
  {Wei}}, \bibinfo {author} {\bibfnamefont {X.}~\bibnamefont {Xu}}, \bibinfo
  {author} {\bibfnamefont {J.}~\bibnamefont {Zhang}}, \bibinfo {author}
  {\bibfnamefont {M.}~\bibnamefont {Zhang}}, \ and\ \bibinfo {author}
  {\bibfnamefont {X.}~\bibnamefont {Zhang}},\ }\href {\doibase
  https://doi.org/10.1016/j.physletb.2024.139082} {\bibfield  {journal}
  {\bibinfo  {journal} {Phys. Lett. B}\ }\textbf {\bibinfo {volume} {858}},\
  \bibinfo {pages} {139082} (\bibinfo {year} {2024})}\BibitemShut {NoStop}%
\bibitem [{\citenamefont {Wang}\ and\ \citenamefont {Li}(2013)}]{Wang2013PRC}%
  \BibitemOpen
  \bibfield  {author} {\bibinfo {author} {\bibfnamefont {N.}~\bibnamefont
  {Wang}}\ and\ \bibinfo {author} {\bibfnamefont {T.}~\bibnamefont {Li}},\
  }\href {\doibase 10.1103/PhysRevC.88.011301} {\bibfield  {journal} {\bibinfo
  {journal} {Phys. Rev. C}\ }\textbf {\bibinfo {volume} {88}},\ \bibinfo
  {pages} {011301} (\bibinfo {year} {2013})}\BibitemShut {NoStop}%
\bibitem [{\citenamefont {Angeli}\ and\ \citenamefont
  {Marinova}(2015)}]{Angeli2015JPG}%
  \BibitemOpen
  \bibfield  {author} {\bibinfo {author} {\bibfnamefont {I.}~\bibnamefont
  {Angeli}}\ and\ \bibinfo {author} {\bibfnamefont {K.~P.}\ \bibnamefont
  {Marinova}},\ }\href {\doibase 10.1088/0954-3899/42/5/055108} {\bibfield
  {journal} {\bibinfo  {journal} {J. Phys. G}\ }\textbf {\bibinfo {volume}
  {42}},\ \bibinfo {pages} {055108} (\bibinfo {year} {2015})}\BibitemShut
  {NoStop}%
\bibitem [{\citenamefont {Geithner}\ \emph {et~al.}(2008)\citenamefont
  {Geithner}, \citenamefont {Neff}, \citenamefont {Audi}, \citenamefont
  {Blaum}, \citenamefont {Delahaye}, \citenamefont {Feldmeier}, \citenamefont
  {George}, \citenamefont {Gu\'enaut}, \citenamefont {Herfurth}, \citenamefont
  {Herlert}, \citenamefont {Kappertz}, \citenamefont {Keim}, \citenamefont
  {Kellerbauer}, \citenamefont {Kluge}, \citenamefont {Kowalska}, \citenamefont
  {Lievens}, \citenamefont {Lunney}, \citenamefont {Marinova}, \citenamefont
  {Neugart}, \citenamefont {Schweikhard}, \citenamefont {Wilbert},\ and\
  \citenamefont {Yazidjian}}]{Geithner2008PRL}%
  \BibitemOpen
  \bibfield  {author} {\bibinfo {author} {\bibfnamefont {W.}~\bibnamefont
  {Geithner}}, \bibinfo {author} {\bibfnamefont {T.}~\bibnamefont {Neff}},
  \bibinfo {author} {\bibfnamefont {G.}~\bibnamefont {Audi}}, \bibinfo {author}
  {\bibfnamefont {K.}~\bibnamefont {Blaum}}, \bibinfo {author} {\bibfnamefont
  {P.}~\bibnamefont {Delahaye}}, \bibinfo {author} {\bibfnamefont
  {H.}~\bibnamefont {Feldmeier}}, \bibinfo {author} {\bibfnamefont
  {S.}~\bibnamefont {George}}, \bibinfo {author} {\bibfnamefont
  {C.}~\bibnamefont {Gu\'enaut}}, \bibinfo {author} {\bibfnamefont
  {F.}~\bibnamefont {Herfurth}}, \bibinfo {author} {\bibfnamefont
  {A.}~\bibnamefont {Herlert}}, \bibinfo {author} {\bibfnamefont
  {S.}~\bibnamefont {Kappertz}}, \bibinfo {author} {\bibfnamefont
  {M.}~\bibnamefont {Keim}}, \bibinfo {author} {\bibfnamefont {A.}~\bibnamefont
  {Kellerbauer}}, \bibinfo {author} {\bibfnamefont {H.-J.}\ \bibnamefont
  {Kluge}}, \bibinfo {author} {\bibfnamefont {M.}~\bibnamefont {Kowalska}},
  \bibinfo {author} {\bibfnamefont {P.}~\bibnamefont {Lievens}}, \bibinfo
  {author} {\bibfnamefont {D.}~\bibnamefont {Lunney}}, \bibinfo {author}
  {\bibfnamefont {K.}~\bibnamefont {Marinova}}, \bibinfo {author}
  {\bibfnamefont {R.}~\bibnamefont {Neugart}}, \bibinfo {author} {\bibfnamefont
  {L.}~\bibnamefont {Schweikhard}}, \bibinfo {author} {\bibfnamefont
  {S.}~\bibnamefont {Wilbert}}, \ and\ \bibinfo {author} {\bibfnamefont
  {C.}~\bibnamefont {Yazidjian}},\ }\href {\doibase
  10.1103/PhysRevLett.101.252502} {\bibfield  {journal} {\bibinfo  {journal}
  {Phys. Rev. Lett.}\ }\textbf {\bibinfo {volume} {101}},\ \bibinfo {pages}
  {252502} (\bibinfo {year} {2008})}\BibitemShut {NoStop}%
\bibitem [{\citenamefont {N\"ortersh\"auser}\ \emph {et~al.}(2009)\citenamefont
  {N\"ortersh\"auser}, \citenamefont {Tiedemann}, \citenamefont
  {\ifmmode~\check{Z}\else \v{Z}\fi{}\'akov\'a}, \citenamefont {Andjelkovic},
  \citenamefont {Blaum}, \citenamefont {Bissell}, \citenamefont {Cazan},
  \citenamefont {Drake}, \citenamefont {Geppert}, \citenamefont {Kowalska},
  \citenamefont {Kr\"amer}, \citenamefont {Krieger}, \citenamefont {Neugart},
  \citenamefont {S\'anchez}, \citenamefont {Schmidt-Kaler}, \citenamefont
  {Yan}, \citenamefont {Yordanov},\ and\ \citenamefont
  {Zimmermann}}]{Nortershauser2009PRL}%
  \BibitemOpen
  \bibfield  {author} {\bibinfo {author} {\bibfnamefont {W.}~\bibnamefont
  {N\"ortersh\"auser}}, \bibinfo {author} {\bibfnamefont {D.}~\bibnamefont
  {Tiedemann}}, \bibinfo {author} {\bibfnamefont {M.}~\bibnamefont
  {\ifmmode~\check{Z}\else \v{Z}\fi{}\'akov\'a}}, \bibinfo {author}
  {\bibfnamefont {Z.}~\bibnamefont {Andjelkovic}}, \bibinfo {author}
  {\bibfnamefont {K.}~\bibnamefont {Blaum}}, \bibinfo {author} {\bibfnamefont
  {M.~L.}\ \bibnamefont {Bissell}}, \bibinfo {author} {\bibfnamefont
  {R.}~\bibnamefont {Cazan}}, \bibinfo {author} {\bibfnamefont {G.~W.~F.}\
  \bibnamefont {Drake}}, \bibinfo {author} {\bibfnamefont {C.}~\bibnamefont
  {Geppert}}, \bibinfo {author} {\bibfnamefont {M.}~\bibnamefont {Kowalska}},
  \bibinfo {author} {\bibfnamefont {J.}~\bibnamefont {Kr\"amer}}, \bibinfo
  {author} {\bibfnamefont {A.}~\bibnamefont {Krieger}}, \bibinfo {author}
  {\bibfnamefont {R.}~\bibnamefont {Neugart}}, \bibinfo {author} {\bibfnamefont
  {R.}~\bibnamefont {S\'anchez}}, \bibinfo {author} {\bibfnamefont
  {F.}~\bibnamefont {Schmidt-Kaler}}, \bibinfo {author} {\bibfnamefont {Z.-C.}\
  \bibnamefont {Yan}}, \bibinfo {author} {\bibfnamefont {D.~T.}\ \bibnamefont
  {Yordanov}}, \ and\ \bibinfo {author} {\bibfnamefont {C.}~\bibnamefont
  {Zimmermann}},\ }\href {\doibase 10.1103/PhysRevLett.102.062503} {\bibfield
  {journal} {\bibinfo  {journal} {Phys. Rev. Lett.}\ }\textbf {\bibinfo
  {volume} {102}},\ \bibinfo {pages} {062503} (\bibinfo {year}
  {2009})}\BibitemShut {NoStop}%
\bibitem [{\citenamefont {Yordanov}\ \emph {et~al.}(2016)\citenamefont
  {Yordanov}, \citenamefont {Balabanski}, \citenamefont {Bissell},
  \citenamefont {Blaum}, \citenamefont {Budin\ifmmode \check{c}\else
  \v{c}\fi{}evi\ifmmode~\acute{c}\else \'{c}\fi{}}, \citenamefont {Cheal},
  \citenamefont {Flanagan}, \citenamefont {Fr\"ommgen}, \citenamefont
  {Georgiev}, \citenamefont {Geppert}, \citenamefont {Hammen}, \citenamefont
  {Kowalska}, \citenamefont {Kreim}, \citenamefont {Krieger}, \citenamefont
  {Meng}, \citenamefont {Neugart}, \citenamefont {Neyens}, \citenamefont
  {N\"ortersh\"auser}, \citenamefont {Rajabali}, \citenamefont {Papuga},
  \citenamefont {Schmidt},\ and\ \citenamefont {Zhao}}]{Yordanov2016PRL}%
  \BibitemOpen
  \bibfield  {author} {\bibinfo {author} {\bibfnamefont {D.~T.}\ \bibnamefont
  {Yordanov}}, \bibinfo {author} {\bibfnamefont {D.~L.}\ \bibnamefont
  {Balabanski}}, \bibinfo {author} {\bibfnamefont {M.~L.}\ \bibnamefont
  {Bissell}}, \bibinfo {author} {\bibfnamefont {K.}~\bibnamefont {Blaum}},
  \bibinfo {author} {\bibfnamefont {I.}~\bibnamefont {Budin\ifmmode
  \check{c}\else \v{c}\fi{}evi\ifmmode~\acute{c}\else \'{c}\fi{}}}, \bibinfo
  {author} {\bibfnamefont {B.}~\bibnamefont {Cheal}}, \bibinfo {author}
  {\bibfnamefont {K.}~\bibnamefont {Flanagan}}, \bibinfo {author}
  {\bibfnamefont {N.}~\bibnamefont {Fr\"ommgen}}, \bibinfo {author}
  {\bibfnamefont {G.}~\bibnamefont {Georgiev}}, \bibinfo {author}
  {\bibfnamefont {C.}~\bibnamefont {Geppert}}, \bibinfo {author} {\bibfnamefont
  {M.}~\bibnamefont {Hammen}}, \bibinfo {author} {\bibfnamefont
  {M.}~\bibnamefont {Kowalska}}, \bibinfo {author} {\bibfnamefont
  {K.}~\bibnamefont {Kreim}}, \bibinfo {author} {\bibfnamefont
  {A.}~\bibnamefont {Krieger}}, \bibinfo {author} {\bibfnamefont
  {J.}~\bibnamefont {Meng}}, \bibinfo {author} {\bibfnamefont {R.}~\bibnamefont
  {Neugart}}, \bibinfo {author} {\bibfnamefont {G.}~\bibnamefont {Neyens}},
  \bibinfo {author} {\bibfnamefont {W.}~\bibnamefont {N\"ortersh\"auser}},
  \bibinfo {author} {\bibfnamefont {M.~M.}\ \bibnamefont {Rajabali}}, \bibinfo
  {author} {\bibfnamefont {J.}~\bibnamefont {Papuga}}, \bibinfo {author}
  {\bibfnamefont {S.}~\bibnamefont {Schmidt}}, \ and\ \bibinfo {author}
  {\bibfnamefont {P.~W.}\ \bibnamefont {Zhao}},\ }\href {\doibase
  10.1103/PhysRevLett.116.032501} {\bibfield  {journal} {\bibinfo  {journal}
  {Phys. Rev. Lett.}\ }\textbf {\bibinfo {volume} {116}},\ \bibinfo {pages}
  {032501} (\bibinfo {year} {2016})}\BibitemShut {NoStop}%
\bibitem [{\citenamefont {Meng}(2016)}]{Meng2016book}%
  \BibitemOpen
  \bibinfo {editor} {\bibfnamefont {J.}~\bibnamefont {Meng}},\ ed.,\ \href
  {\doibase 10.1142/9872} {\emph {\bibinfo {title} {{Relativistic Density
  Functional for Nuclear Structure}}}}\ (\bibinfo  {publisher} {World
  Scientific},\ \bibinfo {year} {2016})\BibitemShut {NoStop}%
\bibitem [{\citenamefont {Koepf}\ and\ \citenamefont
  {Ring}(1991)}]{Koepf1991ZPA}%
  \BibitemOpen
  \bibfield  {author} {\bibinfo {author} {\bibfnamefont {W.}~\bibnamefont
  {Koepf}}\ and\ \bibinfo {author} {\bibfnamefont {P.}~\bibnamefont {Ring}},\
  }\href {\doibase 10.1007/BF01282936} {\bibfield  {journal} {\bibinfo
  {journal} {Z. Phys. A}\ }\textbf {\bibinfo {volume} {339}},\ \bibinfo {pages}
  {81} (\bibinfo {year} {1991})}\BibitemShut {NoStop}%
\bibitem [{\citenamefont {Ren}\ and\ \citenamefont {Zhao}(2020)}]{Ren2020PRC}%
  \BibitemOpen
  \bibfield  {author} {\bibinfo {author} {\bibfnamefont {Z.~X.}\ \bibnamefont
  {Ren}}\ and\ \bibinfo {author} {\bibfnamefont {P.~W.}\ \bibnamefont {Zhao}},\
  }\href {\doibase 10.1103/PhysRevC.102.021301} {\bibfield  {journal} {\bibinfo
   {journal} {Phys. Rev. C}\ }\textbf {\bibinfo {volume} {102}},\ \bibinfo
  {pages} {021301} (\bibinfo {year} {2020})}\BibitemShut {NoStop}%
\bibitem [{\citenamefont {Sharma}\ \emph {et~al.}(1995)\citenamefont {Sharma},
  \citenamefont {Lalazissis}, \citenamefont {K\"onig},\ and\ \citenamefont
  {Ring}}]{Sharma1995PRL}%
  \BibitemOpen
  \bibfield  {author} {\bibinfo {author} {\bibfnamefont {M.~M.}\ \bibnamefont
  {Sharma}}, \bibinfo {author} {\bibfnamefont {G.}~\bibnamefont {Lalazissis}},
  \bibinfo {author} {\bibfnamefont {J.}~\bibnamefont {K\"onig}}, \ and\
  \bibinfo {author} {\bibfnamefont {P.}~\bibnamefont {Ring}},\ }\href {\doibase
  10.1103/PhysRevLett.74.3744} {\bibfield  {journal} {\bibinfo  {journal}
  {Phys. Rev. Lett.}\ }\textbf {\bibinfo {volume} {74}},\ \bibinfo {pages}
  {3744} (\bibinfo {year} {1995})}\BibitemShut {NoStop}%
\bibitem [{\citenamefont {Walecka}(1974)}]{Walecka1974AP}%
  \BibitemOpen
  \bibfield  {author} {\bibinfo {author} {\bibfnamefont {J.~D.}\ \bibnamefont
  {Walecka}},\ }\href {\doibase https://doi.org/10.1016/0003-4916(74)90208-5}
  {\bibfield  {journal} {\bibinfo  {journal} {Ann. Phys.}\ }\textbf {\bibinfo
  {volume} {83}},\ \bibinfo {pages} {491} (\bibinfo {year} {1974})}\BibitemShut
  {NoStop}%
\bibitem [{\citenamefont {Ginocchio}(1997)}]{Ginocchio1997PRL}%
  \BibitemOpen
  \bibfield  {author} {\bibinfo {author} {\bibfnamefont {J.~N.}\ \bibnamefont
  {Ginocchio}},\ }\href {\doibase 10.1103/PhysRevLett.78.436} {\bibfield
  {journal} {\bibinfo  {journal} {Phys. Rev. Lett.}\ }\textbf {\bibinfo
  {volume} {78}},\ \bibinfo {pages} {436} (\bibinfo {year} {1997})}\BibitemShut
  {NoStop}%
\bibitem [{\citenamefont {Meng}\ \emph
  {et~al.}(1998{\natexlab{a}})\citenamefont {Meng}, \citenamefont
  {Sugawara-Tanabe}, \citenamefont {Yamaji}, \citenamefont {Ring},\ and\
  \citenamefont {Arima}}]{Meng1998PRC}%
  \BibitemOpen
  \bibfield  {author} {\bibinfo {author} {\bibfnamefont {J.}~\bibnamefont
  {Meng}}, \bibinfo {author} {\bibfnamefont {K.}~\bibnamefont
  {Sugawara-Tanabe}}, \bibinfo {author} {\bibfnamefont {S.}~\bibnamefont
  {Yamaji}}, \bibinfo {author} {\bibfnamefont {P.}~\bibnamefont {Ring}}, \ and\
  \bibinfo {author} {\bibfnamefont {A.}~\bibnamefont {Arima}},\ }\href
  {\doibase 10.1103/PhysRevC.58.R628} {\bibfield  {journal} {\bibinfo
  {journal} {Phys. Rev. C}\ }\textbf {\bibinfo {volume} {58}},\ \bibinfo
  {pages} {R628} (\bibinfo {year} {1998}{\natexlab{a}})}\BibitemShut {NoStop}%
\bibitem [{\citenamefont {Liang}\ \emph {et~al.}(2015)\citenamefont {Liang},
  \citenamefont {Meng},\ and\ \citenamefont {Zhou}}]{Liang2015PR}%
  \BibitemOpen
  \bibfield  {author} {\bibinfo {author} {\bibfnamefont {H.}~\bibnamefont
  {Liang}}, \bibinfo {author} {\bibfnamefont {J.}~\bibnamefont {Meng}}, \ and\
  \bibinfo {author} {\bibfnamefont {S.-G.}\ \bibnamefont {Zhou}},\ }\href
  {\doibase 10.1016/j.physrep.2014.12.005} {\bibfield  {journal} {\bibinfo
  {journal} {Phys. Rep.}\ }\textbf {\bibinfo {volume} {570}},\ \bibinfo {pages}
  {1 } (\bibinfo {year} {2015})}\BibitemShut {NoStop}%
\bibitem [{\citenamefont {Zhou}\ \emph
  {et~al.}(2003{\natexlab{a}})\citenamefont {Zhou}, \citenamefont {Meng},\ and\
  \citenamefont {Ring}}]{Zhou2003PRL}%
  \BibitemOpen
  \bibfield  {author} {\bibinfo {author} {\bibfnamefont {S.-G.}\ \bibnamefont
  {Zhou}}, \bibinfo {author} {\bibfnamefont {J.}~\bibnamefont {Meng}}, \ and\
  \bibinfo {author} {\bibfnamefont {P.}~\bibnamefont {Ring}},\ }\href {\doibase
  10.1103/PhysRevLett.91.262501} {\bibfield  {journal} {\bibinfo  {journal}
  {Phys. Rev. Lett.}\ }\textbf {\bibinfo {volume} {91}},\ \bibinfo {pages}
  {262501} (\bibinfo {year} {2003}{\natexlab{a}})}\BibitemShut {NoStop}%
\bibitem [{\citenamefont {He}\ \emph {et~al.}(2006)\citenamefont {He},
  \citenamefont {Zhou}, \citenamefont {Meng}, \citenamefont {Zhao},\ and\
  \citenamefont {Scheid}}]{He2006EPJA}%
  \BibitemOpen
  \bibfield  {author} {\bibinfo {author} {\bibfnamefont {X.~T.}\ \bibnamefont
  {He}}, \bibinfo {author} {\bibfnamefont {S.~G.}\ \bibnamefont {Zhou}},
  \bibinfo {author} {\bibfnamefont {J.}~\bibnamefont {Meng}}, \bibinfo {author}
  {\bibfnamefont {E.~G.}\ \bibnamefont {Zhao}}, \ and\ \bibinfo {author}
  {\bibfnamefont {W.}~\bibnamefont {Scheid}},\ }\href {\doibase
  10.1140/epja/i2006-10066-0} {\bibfield  {journal} {\bibinfo  {journal} {Eur.
  Phys. J. A}\ }\textbf {\bibinfo {volume} {28}},\ \bibinfo {pages} {265}
  (\bibinfo {year} {2006})}\BibitemShut {NoStop}%
\bibitem [{\citenamefont {Koepf}\ and\ \citenamefont
  {Ring}(1989)}]{Koepf1989NPA}%
  \BibitemOpen
  \bibfield  {author} {\bibinfo {author} {\bibfnamefont {W.}~\bibnamefont
  {Koepf}}\ and\ \bibinfo {author} {\bibfnamefont {P.}~\bibnamefont {Ring}},\
  }\href {\doibase 10.1016/0375-9474(89)90532-0} {\bibfield  {journal}
  {\bibinfo  {journal} {Nucl. Phys. A}\ }\textbf {\bibinfo {volume} {493}},\
  \bibinfo {pages} {61 } (\bibinfo {year} {1989})}\BibitemShut {NoStop}%
\bibitem [{\citenamefont {Koepf}\ and\ \citenamefont
  {Ring}(1990)}]{Koepf1990NPA}%
  \BibitemOpen
  \bibfield  {author} {\bibinfo {author} {\bibfnamefont {W.}~\bibnamefont
  {Koepf}}\ and\ \bibinfo {author} {\bibfnamefont {P.}~\bibnamefont {Ring}},\
  }\href {\doibase https://doi.org/10.1016/0375-9474(90)90160-N} {\bibfield
  {journal} {\bibinfo  {journal} {Nucl. Phys. A}\ }\textbf {\bibinfo {volume}
  {511}},\ \bibinfo {pages} {279} (\bibinfo {year} {1990})}\BibitemShut
  {NoStop}%
\bibitem [{\citenamefont {Ring}(1996)}]{Ring1996PPNP}%
  \BibitemOpen
  \bibfield  {author} {\bibinfo {author} {\bibfnamefont {P.}~\bibnamefont
  {Ring}},\ }\href {\doibase 10.1016/0146-6410(96)00054-3} {\bibfield
  {journal} {\bibinfo  {journal} {Prog. Part. Nucl. Phys.}\ }\textbf {\bibinfo
  {volume} {37}},\ \bibinfo {pages} {193 } (\bibinfo {year}
  {1996})}\BibitemShut {NoStop}%
\bibitem [{\citenamefont {Vretenar}\ \emph {et~al.}(2005)\citenamefont
  {Vretenar}, \citenamefont {Afanasjev}, \citenamefont {Lalazissis},\ and\
  \citenamefont {Ring}}]{Vretenar2005PR}%
  \BibitemOpen
  \bibfield  {author} {\bibinfo {author} {\bibfnamefont {D.}~\bibnamefont
  {Vretenar}}, \bibinfo {author} {\bibfnamefont {A.~V.}\ \bibnamefont
  {Afanasjev}}, \bibinfo {author} {\bibfnamefont {G.~A.}\ \bibnamefont
  {Lalazissis}}, \ and\ \bibinfo {author} {\bibfnamefont {R.}~\bibnamefont
  {Ring}},\ }\href {\doibase 10.1016/j.physrep.2004.10.001} {\bibfield
  {journal} {\bibinfo  {journal} {Phys. Rep.}\ }\textbf {\bibinfo {volume}
  {409}},\ \bibinfo {pages} {101 } (\bibinfo {year} {2005})}\BibitemShut
  {NoStop}%
\bibitem [{\citenamefont {Meng}\ \emph {et~al.}(2006)\citenamefont {Meng},
  \citenamefont {Toki}, \citenamefont {Zhou}, \citenamefont {Zhang},
  \citenamefont {Long},\ and\ \citenamefont {Geng}}]{Meng2006PPNP}%
  \BibitemOpen
  \bibfield  {author} {\bibinfo {author} {\bibfnamefont {J.}~\bibnamefont
  {Meng}}, \bibinfo {author} {\bibfnamefont {H.}~\bibnamefont {Toki}}, \bibinfo
  {author} {\bibfnamefont {S.~G.}\ \bibnamefont {Zhou}}, \bibinfo {author}
  {\bibfnamefont {S.~Q.}\ \bibnamefont {Zhang}}, \bibinfo {author}
  {\bibfnamefont {W.~H.}\ \bibnamefont {Long}}, \ and\ \bibinfo {author}
  {\bibfnamefont {L.~S.}\ \bibnamefont {Geng}},\ }\href {\doibase
  10.1016/j.ppnp.2005.06.001} {\bibfield  {journal} {\bibinfo  {journal} {Prog.
  Part. Nucl. Phys.}\ }\textbf {\bibinfo {volume} {57}},\ \bibinfo {pages} {470
  } (\bibinfo {year} {2006})}\BibitemShut {NoStop}%
\bibitem [{\citenamefont {Nik\v{s}i\'{c}}\ \emph {et~al.}(2011)\citenamefont
  {Nik\v{s}i\'{c}}, \citenamefont {Vretenar},\ and\ \citenamefont
  {Ring}}]{Niksic2011PPNP}%
  \BibitemOpen
  \bibfield  {author} {\bibinfo {author} {\bibfnamefont {T.}~\bibnamefont
  {Nik\v{s}i\'{c}}}, \bibinfo {author} {\bibfnamefont {D.}~\bibnamefont
  {Vretenar}}, \ and\ \bibinfo {author} {\bibfnamefont {P.}~\bibnamefont
  {Ring}},\ }\href {\doibase 10.1016/j.ppnp.2011.01.055} {\bibfield  {journal}
  {\bibinfo  {journal} {Prog. Part. Nucl. Phys.}\ }\textbf {\bibinfo {volume}
  {66}},\ \bibinfo {pages} {519 } (\bibinfo {year} {2011})}\BibitemShut
  {NoStop}%
\bibitem [{\citenamefont {Meng}\ \emph {et~al.}(2013)\citenamefont {Meng},
  \citenamefont {Peng}, \citenamefont {Zhang},\ and\ \citenamefont
  {Zhao}}]{Meng2013FoP}%
  \BibitemOpen
  \bibfield  {author} {\bibinfo {author} {\bibfnamefont {J.}~\bibnamefont
  {Meng}}, \bibinfo {author} {\bibfnamefont {J.}~\bibnamefont {Peng}}, \bibinfo
  {author} {\bibfnamefont {S.-Q.}\ \bibnamefont {Zhang}}, \ and\ \bibinfo
  {author} {\bibfnamefont {P.-W.}\ \bibnamefont {Zhao}},\ }\href {\doibase
  10.1007/s11467-013-0287-y} {\bibfield  {journal} {\bibinfo  {journal} {Front.
  Phys.}\ }\textbf {\bibinfo {volume} {8}},\ \bibinfo {pages} {55} (\bibinfo
  {year} {2013})}\BibitemShut {NoStop}%
\bibitem [{\citenamefont {Meng}\ and\ \citenamefont
  {Zhou}(2015)}]{Meng2015JPG}%
  \BibitemOpen
  \bibfield  {author} {\bibinfo {author} {\bibfnamefont {J.}~\bibnamefont
  {Meng}}\ and\ \bibinfo {author} {\bibfnamefont {S.-G.}\ \bibnamefont
  {Zhou}},\ }\href {\doibase 10.1088/0954-3899/42/9/093101} {\bibfield
  {journal} {\bibinfo  {journal} {J. Phys. G}\ }\textbf {\bibinfo {volume}
  {42}},\ \bibinfo {pages} {093101} (\bibinfo {year} {2015})}\BibitemShut
  {NoStop}%
\bibitem [{\citenamefont {Zhou}(2016)}]{Zhou2016PS}%
  \BibitemOpen
  \bibfield  {author} {\bibinfo {author} {\bibfnamefont {S.-G.}\ \bibnamefont
  {Zhou}},\ }\href {\doibase 10.1088/0031-8949/91/6/063008} {\bibfield
  {journal} {\bibinfo  {journal} {Phys. Scr.}\ }\textbf {\bibinfo {volume}
  {91}},\ \bibinfo {pages} {063008} (\bibinfo {year} {2016})}\BibitemShut
  {NoStop}%
\bibitem [{\citenamefont {Shen}\ \emph {et~al.}(2019)\citenamefont {Shen},
  \citenamefont {Liang}, \citenamefont {Long}, \citenamefont {Meng},\ and\
  \citenamefont {Ring}}]{Shen2019PPNP}%
  \BibitemOpen
  \bibfield  {author} {\bibinfo {author} {\bibfnamefont {S.}~\bibnamefont
  {Shen}}, \bibinfo {author} {\bibfnamefont {H.}~\bibnamefont {Liang}},
  \bibinfo {author} {\bibfnamefont {W.~H.}\ \bibnamefont {Long}}, \bibinfo
  {author} {\bibfnamefont {J.}~\bibnamefont {Meng}}, \ and\ \bibinfo {author}
  {\bibfnamefont {P.}~\bibnamefont {Ring}},\ }\href {\doibase
  10.1016/j.ppnp.2019.103713} {\bibfield  {journal} {\bibinfo  {journal} {Prog.
  Part. Nucl. Phys.}\ }\textbf {\bibinfo {volume} {109}},\ \bibinfo {pages}
  {103713} (\bibinfo {year} {2019})}\BibitemShut {NoStop}%
\bibitem [{\citenamefont {Meng}\ and\ \citenamefont
  {Zhao}(2021)}]{Meng2021AAAPS}%
  \BibitemOpen
  \bibfield  {author} {\bibinfo {author} {\bibfnamefont {J.}~\bibnamefont
  {Meng}}\ and\ \bibinfo {author} {\bibfnamefont {P.}~\bibnamefont {Zhao}},\
  }\href {\doibase 10.1007/s43673-021-00001-8} {\bibfield  {journal} {\bibinfo
  {journal} {AAAPS Bulletin}\ }\textbf {\bibinfo {volume} {31}},\ \bibinfo
  {pages} {2} (\bibinfo {year} {2021})}\BibitemShut {NoStop}%
\bibitem [{\citenamefont {An}\ \emph {et~al.}(2020)\citenamefont {An},
  \citenamefont {Geng},\ and\ \citenamefont {Zhang}}]{An2020PRC}%
  \BibitemOpen
  \bibfield  {author} {\bibinfo {author} {\bibfnamefont {R.}~\bibnamefont
  {An}}, \bibinfo {author} {\bibfnamefont {L.-S.}\ \bibnamefont {Geng}}, \ and\
  \bibinfo {author} {\bibfnamefont {S.-S.}\ \bibnamefont {Zhang}},\ }\href
  {\doibase 10.1103/PhysRevC.102.024307} {\bibfield  {journal} {\bibinfo
  {journal} {Phys. Rev. C}\ }\textbf {\bibinfo {volume} {102}},\ \bibinfo
  {pages} {024307} (\bibinfo {year} {2020})}\BibitemShut {NoStop}%
\bibitem [{\citenamefont {Perera}\ \emph {et~al.}(2021)\citenamefont {Perera},
  \citenamefont {Afanasjev},\ and\ \citenamefont {Ring}}]{Perera2021PRC}%
  \BibitemOpen
  \bibfield  {author} {\bibinfo {author} {\bibfnamefont {U.~C.}\ \bibnamefont
  {Perera}}, \bibinfo {author} {\bibfnamefont {A.~V.}\ \bibnamefont
  {Afanasjev}}, \ and\ \bibinfo {author} {\bibfnamefont {P.}~\bibnamefont
  {Ring}},\ }\href {\doibase 10.1103/PhysRevC.104.064313} {\bibfield  {journal}
  {\bibinfo  {journal} {Phys. Rev. C}\ }\textbf {\bibinfo {volume} {104}},\
  \bibinfo {pages} {064313} (\bibinfo {year} {2021})}\BibitemShut {NoStop}%
\bibitem [{\citenamefont {An}\ \emph {et~al.}(2022)\citenamefont {An},
  \citenamefont {Jiang}, \citenamefont {Cao},\ and\ \citenamefont
  {Zhang}}]{An2022PRC}%
  \BibitemOpen
  \bibfield  {author} {\bibinfo {author} {\bibfnamefont {R.}~\bibnamefont
  {An}}, \bibinfo {author} {\bibfnamefont {X.}~\bibnamefont {Jiang}}, \bibinfo
  {author} {\bibfnamefont {L.-G.}\ \bibnamefont {Cao}}, \ and\ \bibinfo
  {author} {\bibfnamefont {F.-S.}\ \bibnamefont {Zhang}},\ }\href {\doibase
  10.1103/PhysRevC.105.014325} {\bibfield  {journal} {\bibinfo  {journal}
  {Phys. Rev. C}\ }\textbf {\bibinfo {volume} {105}},\ \bibinfo {pages}
  {014325} (\bibinfo {year} {2022})}\BibitemShut {NoStop}%
\bibitem [{\citenamefont {An}\ \emph {et~al.}(2024{\natexlab{a}})\citenamefont
  {An}, \citenamefont {Jiang}, \citenamefont {Tang}, \citenamefont {Cao},\ and\
  \citenamefont {Zhang}}]{An2024PRC}%
  \BibitemOpen
  \bibfield  {author} {\bibinfo {author} {\bibfnamefont {R.}~\bibnamefont
  {An}}, \bibinfo {author} {\bibfnamefont {X.}~\bibnamefont {Jiang}}, \bibinfo
  {author} {\bibfnamefont {N.}~\bibnamefont {Tang}}, \bibinfo {author}
  {\bibfnamefont {L.-G.}\ \bibnamefont {Cao}}, \ and\ \bibinfo {author}
  {\bibfnamefont {F.-S.}\ \bibnamefont {Zhang}},\ }\href {\doibase
  10.1103/PhysRevC.109.064302} {\bibfield  {journal} {\bibinfo  {journal}
  {Phys. Rev. C}\ }\textbf {\bibinfo {volume} {109}},\ \bibinfo {pages}
  {064302} (\bibinfo {year} {2024}{\natexlab{a}})}\BibitemShut {NoStop}%
\bibitem [{\citenamefont {Gaidarov}\ \emph {et~al.}(2014)\citenamefont
  {Gaidarov}, \citenamefont {Sarriguren}, \citenamefont {Antonov},\ and\
  \citenamefont {Moya~de Guerra}}]{Gaidarov2014PRC}%
  \BibitemOpen
  \bibfield  {author} {\bibinfo {author} {\bibfnamefont {M.~K.}\ \bibnamefont
  {Gaidarov}}, \bibinfo {author} {\bibfnamefont {P.}~\bibnamefont
  {Sarriguren}}, \bibinfo {author} {\bibfnamefont {A.~N.}\ \bibnamefont
  {Antonov}}, \ and\ \bibinfo {author} {\bibfnamefont {E.}~\bibnamefont
  {Moya~de Guerra}},\ }\href {\doibase 10.1103/PhysRevC.89.064301} {\bibfield
  {journal} {\bibinfo  {journal} {Phys. Rev. C}\ }\textbf {\bibinfo {volume}
  {89}},\ \bibinfo {pages} {064301} (\bibinfo {year} {2014})}\BibitemShut
  {NoStop}%
\bibitem [{\citenamefont {Bassem}\ and\ \citenamefont
  {Oulne}(2015)}]{Bassem2015IJMPE}%
  \BibitemOpen
  \bibfield  {author} {\bibinfo {author} {\bibfnamefont {Y.~E.}\ \bibnamefont
  {Bassem}}\ and\ \bibinfo {author} {\bibfnamefont {M.}~\bibnamefont {Oulne}},\
  }\href {\doibase 10.1142/S0218301315500731} {\bibfield  {journal} {\bibinfo
  {journal} {Int. J. Mod. Phys. E}\ }\textbf {\bibinfo {volume} {24}},\
  \bibinfo {pages} {1550073} (\bibinfo {year} {2015})}\BibitemShut {NoStop}%
\bibitem [{\citenamefont {Sarriguren}(2019)}]{Sarriguren2019PRC}%
  \BibitemOpen
  \bibfield  {author} {\bibinfo {author} {\bibfnamefont {P.}~\bibnamefont
  {Sarriguren}},\ }\href {\doibase 10.1103/PhysRevC.100.054306} {\bibfield
  {journal} {\bibinfo  {journal} {Phys. Rev. C}\ }\textbf {\bibinfo {volume}
  {100}},\ \bibinfo {pages} {054306} (\bibinfo {year} {2019})}\BibitemShut
  {NoStop}%
\bibitem [{\citenamefont {Reinhard}\ and\ \citenamefont
  {Nazarewicz}(2021)}]{Reinhard2021PRC}%
  \BibitemOpen
  \bibfield  {author} {\bibinfo {author} {\bibfnamefont {P.-G.}\ \bibnamefont
  {Reinhard}}\ and\ \bibinfo {author} {\bibfnamefont {W.}~\bibnamefont
  {Nazarewicz}},\ }\href {\doibase 10.1103/PhysRevC.103.054310} {\bibfield
  {journal} {\bibinfo  {journal} {Phys. Rev. C}\ }\textbf {\bibinfo {volume}
  {103}},\ \bibinfo {pages} {054310} (\bibinfo {year} {2021})}\BibitemShut
  {NoStop}%
\bibitem [{\citenamefont {Scamps}\ \emph {et~al.}(2021)\citenamefont {Scamps},
  \citenamefont {Goriely}, \citenamefont {Olsen}, \citenamefont {Bender},\ and\
  \citenamefont {Ryssens}}]{Scamps2021EPJA}%
  \BibitemOpen
  \bibfield  {author} {\bibinfo {author} {\bibfnamefont {G.}~\bibnamefont
  {Scamps}}, \bibinfo {author} {\bibfnamefont {S.}~\bibnamefont {Goriely}},
  \bibinfo {author} {\bibfnamefont {E.}~\bibnamefont {Olsen}}, \bibinfo
  {author} {\bibfnamefont {M.}~\bibnamefont {Bender}}, \ and\ \bibinfo {author}
  {\bibfnamefont {W.}~\bibnamefont {Ryssens}},\ }\href {\doibase
  10.1140/epja/s10050-021-00642-1} {\bibfield  {journal} {\bibinfo  {journal}
  {Eur. Phys. J. A}\ }\textbf {\bibinfo {volume} {57}},\ \bibinfo {pages} {333}
  (\bibinfo {year} {2021})}\BibitemShut {NoStop}%
\bibitem [{\citenamefont {Meng}\ and\ \citenamefont
  {Ring}(1996)}]{Meng1996PRL}%
  \BibitemOpen
  \bibfield  {author} {\bibinfo {author} {\bibfnamefont {J.}~\bibnamefont
  {Meng}}\ and\ \bibinfo {author} {\bibfnamefont {P.}~\bibnamefont {Ring}},\
  }\href {\doibase 10.1103/PhysRevLett.77.3963} {\bibfield  {journal} {\bibinfo
   {journal} {Phys. Rev. Lett.}\ }\textbf {\bibinfo {volume} {77}},\ \bibinfo
  {pages} {3963} (\bibinfo {year} {1996})}\BibitemShut {NoStop}%
\bibitem [{\citenamefont {Meng}(1998)}]{Meng1998NPA}%
  \BibitemOpen
  \bibfield  {author} {\bibinfo {author} {\bibfnamefont {J.}~\bibnamefont
  {Meng}},\ }\href {\doibase 10.1016/S0375-9474(98)00178-X} {\bibfield
  {journal} {\bibinfo  {journal} {Nucl. Phys. A}\ }\textbf {\bibinfo {volume}
  {635}},\ \bibinfo {pages} {3 } (\bibinfo {year} {1998})}\BibitemShut
  {NoStop}%
\bibitem [{\citenamefont {Meng}\ and\ \citenamefont
  {Ring}(1998)}]{Meng1998PRL}%
  \BibitemOpen
  \bibfield  {author} {\bibinfo {author} {\bibfnamefont {J.}~\bibnamefont
  {Meng}}\ and\ \bibinfo {author} {\bibfnamefont {P.}~\bibnamefont {Ring}},\
  }\href {\doibase 10.1103/PhysRevLett.80.460} {\bibfield  {journal} {\bibinfo
  {journal} {Phys. Rev. Lett.}\ }\textbf {\bibinfo {volume} {80}},\ \bibinfo
  {pages} {460} (\bibinfo {year} {1998})}\BibitemShut {NoStop}%
\bibitem [{\citenamefont {Meng}\ \emph
  {et~al.}(1998{\natexlab{b}})\citenamefont {Meng}, \citenamefont {Tanihata},\
  and\ \citenamefont {Yamaji}}]{Meng1998PLB}%
  \BibitemOpen
  \bibfield  {author} {\bibinfo {author} {\bibfnamefont {J.}~\bibnamefont
  {Meng}}, \bibinfo {author} {\bibfnamefont {I.}~\bibnamefont {Tanihata}}, \
  and\ \bibinfo {author} {\bibfnamefont {S.}~\bibnamefont {Yamaji}},\ }\href
  {\doibase 10.1016/S0370-2693(97)01386-5} {\bibfield  {journal} {\bibinfo
  {journal} {Phys. Lett. B}\ }\textbf {\bibinfo {volume} {419}},\ \bibinfo
  {pages} {1 } (\bibinfo {year} {1998}{\natexlab{b}})}\BibitemShut {NoStop}%
\bibitem [{\citenamefont {Meng}\ \emph
  {et~al.}(2002{\natexlab{a}})\citenamefont {Meng}, \citenamefont {Zhou},\ and\
  \citenamefont {Tanihata}}]{Meng2002PLB}%
  \BibitemOpen
  \bibfield  {author} {\bibinfo {author} {\bibfnamefont {J.}~\bibnamefont
  {Meng}}, \bibinfo {author} {\bibfnamefont {S.-G.}\ \bibnamefont {Zhou}}, \
  and\ \bibinfo {author} {\bibfnamefont {I.}~\bibnamefont {Tanihata}},\ }\href
  {\doibase 10.1016/S0370-2693(02)01574-5} {\bibfield  {journal} {\bibinfo
  {journal} {Phys. Lett. B}\ }\textbf {\bibinfo {volume} {532}},\ \bibinfo
  {pages} {209 } (\bibinfo {year} {2002}{\natexlab{a}})}\BibitemShut {NoStop}%
\bibitem [{\citenamefont {Meng}\ \emph
  {et~al.}(2002{\natexlab{b}})\citenamefont {Meng}, \citenamefont {Toki},
  \citenamefont {Zeng}, \citenamefont {Zhang},\ and\ \citenamefont
  {Zhou}}]{Meng2002PRC}%
  \BibitemOpen
  \bibfield  {author} {\bibinfo {author} {\bibfnamefont {J.}~\bibnamefont
  {Meng}}, \bibinfo {author} {\bibfnamefont {H.}~\bibnamefont {Toki}}, \bibinfo
  {author} {\bibfnamefont {J.~Y.}\ \bibnamefont {Zeng}}, \bibinfo {author}
  {\bibfnamefont {S.~Q.}\ \bibnamefont {Zhang}}, \ and\ \bibinfo {author}
  {\bibfnamefont {S.-G.}\ \bibnamefont {Zhou}},\ }\href {\doibase
  10.1103/PhysRevC.65.041302} {\bibfield  {journal} {\bibinfo  {journal} {Phys.
  Rev. C}\ }\textbf {\bibinfo {volume} {65}},\ \bibinfo {pages} {041302}
  (\bibinfo {year} {2002}{\natexlab{b}})}\BibitemShut {NoStop}%
\bibitem [{\citenamefont {Zhang}\ \emph
  {et~al.}(2002{\natexlab{a}})\citenamefont {Zhang}, \citenamefont {Meng},
  \citenamefont {Zhou},\ and\ \citenamefont {Zeng}}]{Zhang2002CPL}%
  \BibitemOpen
  \bibfield  {author} {\bibinfo {author} {\bibfnamefont {S.-Q.}\ \bibnamefont
  {Zhang}}, \bibinfo {author} {\bibfnamefont {J.}~\bibnamefont {Meng}},
  \bibinfo {author} {\bibfnamefont {S.-G.}\ \bibnamefont {Zhou}}, \ and\
  \bibinfo {author} {\bibfnamefont {J.-Y.}\ \bibnamefont {Zeng}},\ }\href
  {\doibase 10.1088/0256-307x/19/3/308} {\bibfield  {journal} {\bibinfo
  {journal} {Chin. Phys. Lett.}\ }\textbf {\bibinfo {volume} {19}},\ \bibinfo
  {pages} {312} (\bibinfo {year} {2002}{\natexlab{a}})}\BibitemShut {NoStop}%
\bibitem [{\citenamefont {L{\"u}}\ \emph {et~al.}(2003)\citenamefont {L{\"u}},
  \citenamefont {Meng}, \citenamefont {Zhang},\ and\ \citenamefont
  {Zhou}}]{Lv2003EPJA}%
  \BibitemOpen
  \bibfield  {author} {\bibinfo {author} {\bibfnamefont {H.~F.}\ \bibnamefont
  {L{\"u}}}, \bibinfo {author} {\bibfnamefont {J.}~\bibnamefont {Meng}},
  \bibinfo {author} {\bibfnamefont {S.~Q.}\ \bibnamefont {Zhang}}, \ and\
  \bibinfo {author} {\bibfnamefont {S.-G.}\ \bibnamefont {Zhou}},\ }\href
  {\doibase 10.1140/epja/i2002-10136-3} {\bibfield  {journal} {\bibinfo
  {journal} {Eur. Phys. J. A}\ }\textbf {\bibinfo {volume} {17}},\ \bibinfo
  {pages} {19} (\bibinfo {year} {2003})}\BibitemShut {NoStop}%
\bibitem [{\citenamefont {Zhang}\ \emph {et~al.}(2005)\citenamefont {Zhang},
  \citenamefont {Meng}, \citenamefont {Zhang}, \citenamefont {Geng},\ and\
  \citenamefont {Toki}}]{Zhang2005NPA}%
  \BibitemOpen
  \bibfield  {author} {\bibinfo {author} {\bibfnamefont {W.}~\bibnamefont
  {Zhang}}, \bibinfo {author} {\bibfnamefont {J.}~\bibnamefont {Meng}},
  \bibinfo {author} {\bibfnamefont {S.~Q.}\ \bibnamefont {Zhang}}, \bibinfo
  {author} {\bibfnamefont {L.~S.}\ \bibnamefont {Geng}}, \ and\ \bibinfo
  {author} {\bibfnamefont {H.}~\bibnamefont {Toki}},\ }\href {\doibase
  10.1016/j.nuclphysa.2005.02.086} {\bibfield  {journal} {\bibinfo  {journal}
  {Nucl. Phys. A}\ }\textbf {\bibinfo {volume} {753}},\ \bibinfo {pages} {106 }
  (\bibinfo {year} {2005})}\BibitemShut {NoStop}%
\bibitem [{\citenamefont {Lim}\ \emph {et~al.}(2016)\citenamefont {Lim},
  \citenamefont {Xia},\ and\ \citenamefont {Kim}}]{Lim2016PRC}%
  \BibitemOpen
  \bibfield  {author} {\bibinfo {author} {\bibfnamefont {Y.}~\bibnamefont
  {Lim}}, \bibinfo {author} {\bibfnamefont {X.}~\bibnamefont {Xia}}, \ and\
  \bibinfo {author} {\bibfnamefont {Y.}~\bibnamefont {Kim}},\ }\href {\doibase
  10.1103/PhysRevC.93.014314} {\bibfield  {journal} {\bibinfo  {journal} {Phys.
  Rev. C}\ }\textbf {\bibinfo {volume} {93}},\ \bibinfo {pages} {014314}
  (\bibinfo {year} {2016})}\BibitemShut {NoStop}%
\bibitem [{\citenamefont {Zhang}\ and\ \citenamefont
  {Xia}(2016)}]{Zhang2016CPC}%
  \BibitemOpen
  \bibfield  {author} {\bibinfo {author} {\bibfnamefont {L.-F.}\ \bibnamefont
  {Zhang}}\ and\ \bibinfo {author} {\bibfnamefont {X.-W.}\ \bibnamefont
  {Xia}},\ }\href {\doibase 10.1088/1674-1137/40/5/054102} {\bibfield
  {journal} {\bibinfo  {journal} {Chin. Phys. C}\ }\textbf {\bibinfo {volume}
  {40}},\ \bibinfo {pages} {054102} (\bibinfo {year} {2016})}\BibitemShut
  {NoStop}%
\bibitem [{\citenamefont {Kuang}\ \emph {et~al.}(2023)\citenamefont {Kuang},
  \citenamefont {Tu}, \citenamefont {Zhang}, \citenamefont {Zhang},\ and\
  \citenamefont {Li}}]{Kuang2023EPJA}%
  \BibitemOpen
  \bibfield  {author} {\bibinfo {author} {\bibfnamefont {Y.}~\bibnamefont
  {Kuang}}, \bibinfo {author} {\bibfnamefont {X.~L.}\ \bibnamefont {Tu}},
  \bibinfo {author} {\bibfnamefont {J.~T.}\ \bibnamefont {Zhang}}, \bibinfo
  {author} {\bibfnamefont {K.~Y.}\ \bibnamefont {Zhang}}, \ and\ \bibinfo
  {author} {\bibfnamefont {Z.~P.}\ \bibnamefont {Li}},\ }\href {\doibase
  10.1140/epja/s10050-023-01072-x} {\bibfield  {journal} {\bibinfo  {journal}
  {Eur. Phys. J. A}\ }\textbf {\bibinfo {volume} {59}},\ \bibinfo {pages} {160}
  (\bibinfo {year} {2023})}\BibitemShut {NoStop}%
\bibitem [{\citenamefont {Guo}\ \emph {et~al.}(2024{\natexlab{b}})\citenamefont
  {Guo}, \citenamefont {Yu}, \citenamefont {Wu}, \citenamefont {Pan},\ and\
  \citenamefont {Zhang}}]{Guo2024PRC}%
  \BibitemOpen
  \bibfield  {author} {\bibinfo {author} {\bibfnamefont {Y.~Y.}\ \bibnamefont
  {Guo}}, \bibinfo {author} {\bibfnamefont {T.}~\bibnamefont {Yu}}, \bibinfo
  {author} {\bibfnamefont {X.~H.}\ \bibnamefont {Wu}}, \bibinfo {author}
  {\bibfnamefont {C.}~\bibnamefont {Pan}}, \ and\ \bibinfo {author}
  {\bibfnamefont {K.~Y.}\ \bibnamefont {Zhang}},\ }\href {\doibase
  10.1103/PhysRevC.110.064310} {\bibfield  {journal} {\bibinfo  {journal}
  {Phys. Rev. C}\ }\textbf {\bibinfo {volume} {110}},\ \bibinfo {pages}
  {064310} (\bibinfo {year} {2024}{\natexlab{b}})}\BibitemShut {NoStop}%
\bibitem [{\citenamefont {Wu}\ \emph {et~al.}(2024)\citenamefont {Wu},
  \citenamefont {Pan}, \citenamefont {Zhang},\ and\ \citenamefont
  {Hu}}]{Wu2024PRC_KRR}%
  \BibitemOpen
  \bibfield  {author} {\bibinfo {author} {\bibfnamefont {X.~H.}\ \bibnamefont
  {Wu}}, \bibinfo {author} {\bibfnamefont {C.}~\bibnamefont {Pan}}, \bibinfo
  {author} {\bibfnamefont {K.~Y.}\ \bibnamefont {Zhang}}, \ and\ \bibinfo
  {author} {\bibfnamefont {J.}~\bibnamefont {Hu}},\ }\href {\doibase
  10.1103/PhysRevC.109.024310} {\bibfield  {journal} {\bibinfo  {journal}
  {Phys. Rev. C}\ }\textbf {\bibinfo {volume} {109}},\ \bibinfo {pages}
  {024310} (\bibinfo {year} {2024})}\BibitemShut {NoStop}%
\bibitem [{\citenamefont {Zhang}\ \emph
  {et~al.}(2002{\natexlab{b}})\citenamefont {Zhang}, \citenamefont {Meng},
  \citenamefont {Zhou},\ and\ \citenamefont {Zeng}}]{Zhang2002EPJA}%
  \BibitemOpen
  \bibfield  {author} {\bibinfo {author} {\bibfnamefont {S.~Q.}\ \bibnamefont
  {Zhang}}, \bibinfo {author} {\bibfnamefont {J.}~\bibnamefont {Meng}},
  \bibinfo {author} {\bibfnamefont {S.-G.}\ \bibnamefont {Zhou}}, \ and\
  \bibinfo {author} {\bibfnamefont {J.~Y.}\ \bibnamefont {Zeng}},\ }\href
  {\doibase 10.1007/s10050-002-8757-6} {\bibfield  {journal} {\bibinfo
  {journal} {Eur. Phys. J. A}\ }\textbf {\bibinfo {volume} {13}},\ \bibinfo
  {pages} {285} (\bibinfo {year} {2002}{\natexlab{b}})}\BibitemShut {NoStop}%
\bibitem [{\citenamefont {Zhou}\ \emph {et~al.}(2010)\citenamefont {Zhou},
  \citenamefont {Meng}, \citenamefont {Ring},\ and\ \citenamefont
  {Zhao}}]{Zhou2010PRC}%
  \BibitemOpen
  \bibfield  {author} {\bibinfo {author} {\bibfnamefont {S.-G.}\ \bibnamefont
  {Zhou}}, \bibinfo {author} {\bibfnamefont {J.}~\bibnamefont {Meng}}, \bibinfo
  {author} {\bibfnamefont {P.}~\bibnamefont {Ring}}, \ and\ \bibinfo {author}
  {\bibfnamefont {E.-G.}\ \bibnamefont {Zhao}},\ }\href {\doibase
  10.1103/PhysRevC.82.011301} {\bibfield  {journal} {\bibinfo  {journal} {Phys.
  Rev. C}\ }\textbf {\bibinfo {volume} {82}},\ \bibinfo {pages} {011301}
  (\bibinfo {year} {2010})}\BibitemShut {NoStop}%
\bibitem [{\citenamefont {Li}\ \emph {et~al.}(2012{\natexlab{a}})\citenamefont
  {Li}, \citenamefont {Meng}, \citenamefont {Ring}, \citenamefont {Zhao},\ and\
  \citenamefont {Zhou}}]{Li2012PRC}%
  \BibitemOpen
  \bibfield  {author} {\bibinfo {author} {\bibfnamefont {L.}~\bibnamefont
  {Li}}, \bibinfo {author} {\bibfnamefont {J.}~\bibnamefont {Meng}}, \bibinfo
  {author} {\bibfnamefont {P.}~\bibnamefont {Ring}}, \bibinfo {author}
  {\bibfnamefont {E.-G.}\ \bibnamefont {Zhao}}, \ and\ \bibinfo {author}
  {\bibfnamefont {S.-G.}\ \bibnamefont {Zhou}},\ }\href {\doibase
  10.1103/PhysRevC.85.024312} {\bibfield  {journal} {\bibinfo  {journal} {Phys.
  Rev. C}\ }\textbf {\bibinfo {volume} {85}},\ \bibinfo {pages} {024312}
  (\bibinfo {year} {2012}{\natexlab{a}})}\BibitemShut {NoStop}%
\bibitem [{\citenamefont {Yang}\ \emph {et~al.}(2021)\citenamefont {Yang},
  \citenamefont {Kubota}, \citenamefont {Corsi}, \citenamefont {Yoshida},
  \citenamefont {Sun}, \citenamefont {Li}, \citenamefont {Kimura},
  \citenamefont {Michel}, \citenamefont {Ogata}, \citenamefont {Yuan},
  \citenamefont {Yuan}, \citenamefont {Authelet}, \citenamefont {Baba},
  \citenamefont {Caesar}, \citenamefont {Calvet}, \citenamefont {Delbart},
  \citenamefont {Dozono}, \citenamefont {Feng}, \citenamefont {Flavigny},
  \citenamefont {Gheller}, \citenamefont {Gibelin}, \citenamefont {Giganon},
  \citenamefont {Gillibert}, \citenamefont {Hasegawa}, \citenamefont {Isobe},
  \citenamefont {Kanaya}, \citenamefont {Kawakami}, \citenamefont {Kim},
  \citenamefont {Kiyokawa}, \citenamefont {Kobayashi}, \citenamefont
  {Kobayashi}, \citenamefont {Kobayashi}, \citenamefont {Kondo}, \citenamefont
  {Korkulu}, \citenamefont {Koyama}, \citenamefont {Lapoux}, \citenamefont
  {Maeda}, \citenamefont {Marqu\'es}, \citenamefont {Motobayashi},
  \citenamefont {Miyazaki}, \citenamefont {Nakamura}, \citenamefont
  {Nakatsuka}, \citenamefont {Nishio}, \citenamefont {Obertelli}, \citenamefont
  {Ohkura}, \citenamefont {Orr}, \citenamefont {Ota}, \citenamefont {Otsu},
  \citenamefont {Ozaki}, \citenamefont {Panin}, \citenamefont {Paschalis},
  \citenamefont {Pollacco}, \citenamefont {Reichert}, \citenamefont {Rouss\'e},
  \citenamefont {Saito}, \citenamefont {Sakaguchi}, \citenamefont {Sako},
  \citenamefont {Santamaria}, \citenamefont {Sasano}, \citenamefont {Sato},
  \citenamefont {Shikata}, \citenamefont {Shimizu}, \citenamefont {Shindo},
  \citenamefont {Stuhl}, \citenamefont {Sumikama}, \citenamefont {Sun},
  \citenamefont {Tabata}, \citenamefont {Togano}, \citenamefont {Tsubota},
  \citenamefont {Xu}, \citenamefont {Yasuda}, \citenamefont {Yoneda},
  \citenamefont {Zenihiro}, \citenamefont {Zhou}, \citenamefont {Zuo},\ and\
  \citenamefont {Uesaka}}]{Yang2021PRL}%
  \BibitemOpen
  \bibfield  {author} {\bibinfo {author} {\bibfnamefont {Z.~H.}\ \bibnamefont
  {Yang}}, \bibinfo {author} {\bibfnamefont {Y.}~\bibnamefont {Kubota}},
  \bibinfo {author} {\bibfnamefont {A.}~\bibnamefont {Corsi}}, \bibinfo
  {author} {\bibfnamefont {K.}~\bibnamefont {Yoshida}}, \bibinfo {author}
  {\bibfnamefont {X.-X.}\ \bibnamefont {Sun}}, \bibinfo {author} {\bibfnamefont
  {J.~G.}\ \bibnamefont {Li}}, \bibinfo {author} {\bibfnamefont
  {M.}~\bibnamefont {Kimura}}, \bibinfo {author} {\bibfnamefont
  {N.}~\bibnamefont {Michel}}, \bibinfo {author} {\bibfnamefont
  {K.}~\bibnamefont {Ogata}}, \bibinfo {author} {\bibfnamefont {C.~X.}\
  \bibnamefont {Yuan}}, \bibinfo {author} {\bibfnamefont {Q.}~\bibnamefont
  {Yuan}}, \bibinfo {author} {\bibfnamefont {G.}~\bibnamefont {Authelet}},
  \bibinfo {author} {\bibfnamefont {H.}~\bibnamefont {Baba}}, \bibinfo {author}
  {\bibfnamefont {C.}~\bibnamefont {Caesar}}, \bibinfo {author} {\bibfnamefont
  {D.}~\bibnamefont {Calvet}}, \bibinfo {author} {\bibfnamefont
  {A.}~\bibnamefont {Delbart}}, \bibinfo {author} {\bibfnamefont
  {M.}~\bibnamefont {Dozono}}, \bibinfo {author} {\bibfnamefont
  {J.}~\bibnamefont {Feng}}, \bibinfo {author} {\bibfnamefont {F.}~\bibnamefont
  {Flavigny}}, \bibinfo {author} {\bibfnamefont {J.-M.}\ \bibnamefont
  {Gheller}}, \bibinfo {author} {\bibfnamefont {J.}~\bibnamefont {Gibelin}},
  \bibinfo {author} {\bibfnamefont {A.}~\bibnamefont {Giganon}}, \bibinfo
  {author} {\bibfnamefont {A.}~\bibnamefont {Gillibert}}, \bibinfo {author}
  {\bibfnamefont {K.}~\bibnamefont {Hasegawa}}, \bibinfo {author}
  {\bibfnamefont {T.}~\bibnamefont {Isobe}}, \bibinfo {author} {\bibfnamefont
  {Y.}~\bibnamefont {Kanaya}}, \bibinfo {author} {\bibfnamefont
  {S.}~\bibnamefont {Kawakami}}, \bibinfo {author} {\bibfnamefont
  {D.}~\bibnamefont {Kim}}, \bibinfo {author} {\bibfnamefont {Y.}~\bibnamefont
  {Kiyokawa}}, \bibinfo {author} {\bibfnamefont {M.}~\bibnamefont {Kobayashi}},
  \bibinfo {author} {\bibfnamefont {N.}~\bibnamefont {Kobayashi}}, \bibinfo
  {author} {\bibfnamefont {T.}~\bibnamefont {Kobayashi}}, \bibinfo {author}
  {\bibfnamefont {Y.}~\bibnamefont {Kondo}}, \bibinfo {author} {\bibfnamefont
  {Z.}~\bibnamefont {Korkulu}}, \bibinfo {author} {\bibfnamefont
  {S.}~\bibnamefont {Koyama}}, \bibinfo {author} {\bibfnamefont
  {V.}~\bibnamefont {Lapoux}}, \bibinfo {author} {\bibfnamefont
  {Y.}~\bibnamefont {Maeda}}, \bibinfo {author} {\bibfnamefont {F.~M.}\
  \bibnamefont {Marqu\'es}}, \bibinfo {author} {\bibfnamefont {T.}~\bibnamefont
  {Motobayashi}}, \bibinfo {author} {\bibfnamefont {T.}~\bibnamefont
  {Miyazaki}}, \bibinfo {author} {\bibfnamefont {T.}~\bibnamefont {Nakamura}},
  \bibinfo {author} {\bibfnamefont {N.}~\bibnamefont {Nakatsuka}}, \bibinfo
  {author} {\bibfnamefont {Y.}~\bibnamefont {Nishio}}, \bibinfo {author}
  {\bibfnamefont {A.}~\bibnamefont {Obertelli}}, \bibinfo {author}
  {\bibfnamefont {A.}~\bibnamefont {Ohkura}}, \bibinfo {author} {\bibfnamefont
  {N.~A.}\ \bibnamefont {Orr}}, \bibinfo {author} {\bibfnamefont
  {S.}~\bibnamefont {Ota}}, \bibinfo {author} {\bibfnamefont {H.}~\bibnamefont
  {Otsu}}, \bibinfo {author} {\bibfnamefont {T.}~\bibnamefont {Ozaki}},
  \bibinfo {author} {\bibfnamefont {V.}~\bibnamefont {Panin}}, \bibinfo
  {author} {\bibfnamefont {S.}~\bibnamefont {Paschalis}}, \bibinfo {author}
  {\bibfnamefont {E.~C.}\ \bibnamefont {Pollacco}}, \bibinfo {author}
  {\bibfnamefont {S.}~\bibnamefont {Reichert}}, \bibinfo {author}
  {\bibfnamefont {J.-Y.}\ \bibnamefont {Rouss\'e}}, \bibinfo {author}
  {\bibfnamefont {A.~T.}\ \bibnamefont {Saito}}, \bibinfo {author}
  {\bibfnamefont {S.}~\bibnamefont {Sakaguchi}}, \bibinfo {author}
  {\bibfnamefont {M.}~\bibnamefont {Sako}}, \bibinfo {author} {\bibfnamefont
  {C.}~\bibnamefont {Santamaria}}, \bibinfo {author} {\bibfnamefont
  {M.}~\bibnamefont {Sasano}}, \bibinfo {author} {\bibfnamefont
  {H.}~\bibnamefont {Sato}}, \bibinfo {author} {\bibfnamefont {M.}~\bibnamefont
  {Shikata}}, \bibinfo {author} {\bibfnamefont {Y.}~\bibnamefont {Shimizu}},
  \bibinfo {author} {\bibfnamefont {Y.}~\bibnamefont {Shindo}}, \bibinfo
  {author} {\bibfnamefont {L.}~\bibnamefont {Stuhl}}, \bibinfo {author}
  {\bibfnamefont {T.}~\bibnamefont {Sumikama}}, \bibinfo {author}
  {\bibfnamefont {Y.~L.}\ \bibnamefont {Sun}}, \bibinfo {author} {\bibfnamefont
  {M.}~\bibnamefont {Tabata}}, \bibinfo {author} {\bibfnamefont
  {Y.}~\bibnamefont {Togano}}, \bibinfo {author} {\bibfnamefont
  {J.}~\bibnamefont {Tsubota}}, \bibinfo {author} {\bibfnamefont {F.~R.}\
  \bibnamefont {Xu}}, \bibinfo {author} {\bibfnamefont {J.}~\bibnamefont
  {Yasuda}}, \bibinfo {author} {\bibfnamefont {K.}~\bibnamefont {Yoneda}},
  \bibinfo {author} {\bibfnamefont {J.}~\bibnamefont {Zenihiro}}, \bibinfo
  {author} {\bibfnamefont {S.-G.}\ \bibnamefont {Zhou}}, \bibinfo {author}
  {\bibfnamefont {W.}~\bibnamefont {Zuo}}, \ and\ \bibinfo {author}
  {\bibfnamefont {T.}~\bibnamefont {Uesaka}},\ }\href {\doibase
  10.1103/PhysRevLett.126.082501} {\bibfield  {journal} {\bibinfo  {journal}
  {Phys. Rev. Lett.}\ }\textbf {\bibinfo {volume} {126}},\ \bibinfo {pages}
  {082501} (\bibinfo {year} {2021})}\BibitemShut {NoStop}%
\bibitem [{\citenamefont {Sun}(2021)}]{Sun2021PRC}%
  \BibitemOpen
  \bibfield  {author} {\bibinfo {author} {\bibfnamefont {X.-X.}\ \bibnamefont
  {Sun}},\ }\href {\doibase 10.1103/PhysRevC.103.054315} {\bibfield  {journal}
  {\bibinfo  {journal} {Phys. Rev. C}\ }\textbf {\bibinfo {volume} {103}},\
  \bibinfo {pages} {054315} (\bibinfo {year} {2021})}\BibitemShut {NoStop}%
\bibitem [{\citenamefont {Sun}\ \emph {et~al.}(2018)\citenamefont {Sun},
  \citenamefont {Zhao},\ and\ \citenamefont {Zhou}}]{Sun2018PLB}%
  \BibitemOpen
  \bibfield  {author} {\bibinfo {author} {\bibfnamefont {X.-X.}\ \bibnamefont
  {Sun}}, \bibinfo {author} {\bibfnamefont {J.}~\bibnamefont {Zhao}}, \ and\
  \bibinfo {author} {\bibfnamefont {S.-G.}\ \bibnamefont {Zhou}},\ }\href
  {\doibase 10.1016/j.physletb.2018.08.071} {\bibfield  {journal} {\bibinfo
  {journal} {Phys. Lett. B}\ }\textbf {\bibinfo {volume} {785}},\ \bibinfo
  {pages} {530 } (\bibinfo {year} {2018})}\BibitemShut {NoStop}%
\bibitem [{\citenamefont {Sun}\ \emph {et~al.}(2020)\citenamefont {Sun},
  \citenamefont {Zhao},\ and\ \citenamefont {Zhou}}]{Sun2020NPA}%
  \BibitemOpen
  \bibfield  {author} {\bibinfo {author} {\bibfnamefont {X.-X.}\ \bibnamefont
  {Sun}}, \bibinfo {author} {\bibfnamefont {J.}~\bibnamefont {Zhao}}, \ and\
  \bibinfo {author} {\bibfnamefont {S.-G.}\ \bibnamefont {Zhou}},\ }\href
  {\doibase 10.1016/j.nuclphysa.2020.122011} {\bibfield  {journal} {\bibinfo
  {journal} {Nucl. Phys. A}\ }\textbf {\bibinfo {volume} {1003}},\ \bibinfo
  {pages} {122011} (\bibinfo {year} {2020})}\BibitemShut {NoStop}%
\bibitem [{\citenamefont {Wang}\ \emph {et~al.}(2024)\citenamefont {Wang},
  \citenamefont {Zhang}, \citenamefont {An},\ and\ \citenamefont
  {Zhang}}]{Wang2024EPJA}%
  \BibitemOpen
  \bibfield  {author} {\bibinfo {author} {\bibfnamefont {L.-Y.}\ \bibnamefont
  {Wang}}, \bibinfo {author} {\bibfnamefont {K.}~\bibnamefont {Zhang}},
  \bibinfo {author} {\bibfnamefont {J.-L.}\ \bibnamefont {An}}, \ and\ \bibinfo
  {author} {\bibfnamefont {S.-S.}\ \bibnamefont {Zhang}},\ }\href {\doibase
  10.1140/epja/s10050-024-01464-7} {\bibfield  {journal} {\bibinfo  {journal}
  {Eur. Phys. J. A}\ }\textbf {\bibinfo {volume} {60}},\ \bibinfo {pages} {251}
  (\bibinfo {year} {2024})}\BibitemShut {NoStop}%
\bibitem [{\citenamefont {Zhong}\ \emph {et~al.}(2022)\citenamefont {Zhong},
  \citenamefont {Zhang}, \citenamefont {Sun},\ and\ \citenamefont
  {Smith}}]{Zhong2022SCP}%
  \BibitemOpen
  \bibfield  {author} {\bibinfo {author} {\bibfnamefont {S.-Y.}\ \bibnamefont
  {Zhong}}, \bibinfo {author} {\bibfnamefont {S.-S.}\ \bibnamefont {Zhang}},
  \bibinfo {author} {\bibfnamefont {X.-X.}\ \bibnamefont {Sun}}, \ and\
  \bibinfo {author} {\bibfnamefont {M.~S.}\ \bibnamefont {Smith}},\ }\href
  {\doibase 10.1007/s11433-022-1894-6} {\bibfield  {journal} {\bibinfo
  {journal} {Sci. China Phys. Mech. Astron.}\ }\textbf {\bibinfo {volume}
  {65}},\ \bibinfo {pages} {262011} (\bibinfo {year} {2022})}\BibitemShut
  {NoStop}%
\bibitem [{\citenamefont {Pan}\ \emph {et~al.}(2024)\citenamefont {Pan},
  \citenamefont {Zhang},\ and\ \citenamefont {Zhang}}]{Pan2024PLB}%
  \BibitemOpen
  \bibfield  {author} {\bibinfo {author} {\bibfnamefont {C.}~\bibnamefont
  {Pan}}, \bibinfo {author} {\bibfnamefont {K.}~\bibnamefont {Zhang}}, \ and\
  \bibinfo {author} {\bibfnamefont {S.}~\bibnamefont {Zhang}},\ }\href
  {\doibase https://doi.org/10.1016/j.physletb.2024.138792} {\bibfield
  {journal} {\bibinfo  {journal} {Phys. Lett. B}\ }\textbf {\bibinfo {volume}
  {855}},\ \bibinfo {pages} {138792} (\bibinfo {year} {2024})}\BibitemShut
  {NoStop}%
\bibitem [{\citenamefont {Zhang}\ \emph
  {et~al.}(2023{\natexlab{a}})\citenamefont {Zhang}, \citenamefont
  {Papakonstantinou}, \citenamefont {Mun}, \citenamefont {Kim}, \citenamefont
  {Yan},\ and\ \citenamefont {Sun}}]{Zhang2023PRC_Na}%
  \BibitemOpen
  \bibfield  {author} {\bibinfo {author} {\bibfnamefont {K.~Y.}\ \bibnamefont
  {Zhang}}, \bibinfo {author} {\bibfnamefont {P.}~\bibnamefont
  {Papakonstantinou}}, \bibinfo {author} {\bibfnamefont {M.-H.}\ \bibnamefont
  {Mun}}, \bibinfo {author} {\bibfnamefont {Y.}~\bibnamefont {Kim}}, \bibinfo
  {author} {\bibfnamefont {H.}~\bibnamefont {Yan}}, \ and\ \bibinfo {author}
  {\bibfnamefont {X.-X.}\ \bibnamefont {Sun}},\ }\href {\doibase
  10.1103/PhysRevC.107.L041303} {\bibfield  {journal} {\bibinfo  {journal}
  {Phys. Rev. C}\ }\textbf {\bibinfo {volume} {107}},\ \bibinfo {pages}
  {L041303} (\bibinfo {year} {2023}{\natexlab{a}})}\BibitemShut {NoStop}%
\bibitem [{\citenamefont {Zhang}\ \emph
  {et~al.}(2023{\natexlab{b}})\citenamefont {Zhang}, \citenamefont {Yang},
  \citenamefont {An}, \citenamefont {Zhang}, \citenamefont {Papakonstantinou},
  \citenamefont {Mun}, \citenamefont {Kim},\ and\ \citenamefont
  {Yan}}]{Zhang2023PLB}%
  \BibitemOpen
  \bibfield  {author} {\bibinfo {author} {\bibfnamefont {K.~Y.}\ \bibnamefont
  {Zhang}}, \bibinfo {author} {\bibfnamefont {S.~Q.}\ \bibnamefont {Yang}},
  \bibinfo {author} {\bibfnamefont {J.~L.}\ \bibnamefont {An}}, \bibinfo
  {author} {\bibfnamefont {S.~S.}\ \bibnamefont {Zhang}}, \bibinfo {author}
  {\bibfnamefont {P.}~\bibnamefont {Papakonstantinou}}, \bibinfo {author}
  {\bibfnamefont {M.-H.}\ \bibnamefont {Mun}}, \bibinfo {author} {\bibfnamefont
  {Y.}~\bibnamefont {Kim}}, \ and\ \bibinfo {author} {\bibfnamefont
  {H.}~\bibnamefont {Yan}},\ }\href {\doibase
  https://doi.org/10.1016/j.physletb.2023.138112} {\bibfield  {journal}
  {\bibinfo  {journal} {Phys. Lett. B}\ }\textbf {\bibinfo {volume} {844}},\
  \bibinfo {pages} {138112} (\bibinfo {year} {2023}{\natexlab{b}})}\BibitemShut
  {NoStop}%
\bibitem [{\citenamefont {An}\ \emph {et~al.}(2024{\natexlab{b}})\citenamefont
  {An}, \citenamefont {Zhang}, \citenamefont {Lu}, \citenamefont {Zhong},\ and\
  \citenamefont {Zhang}}]{An2024PLB}%
  \BibitemOpen
  \bibfield  {author} {\bibinfo {author} {\bibfnamefont {J.-L.}\ \bibnamefont
  {An}}, \bibinfo {author} {\bibfnamefont {K.-Y.}\ \bibnamefont {Zhang}},
  \bibinfo {author} {\bibfnamefont {Q.}~\bibnamefont {Lu}}, \bibinfo {author}
  {\bibfnamefont {S.-Y.}\ \bibnamefont {Zhong}}, \ and\ \bibinfo {author}
  {\bibfnamefont {S.-S.}\ \bibnamefont {Zhang}},\ }\href {\doibase
  https://doi.org/10.1016/j.physletb.2023.138422} {\bibfield  {journal}
  {\bibinfo  {journal} {Phys. Lett. B}\ }\textbf {\bibinfo {volume} {849}},\
  \bibinfo {pages} {138422} (\bibinfo {year} {2024}{\natexlab{b}})}\BibitemShut
  {NoStop}%
\bibitem [{\citenamefont {Zhang}\ \emph
  {et~al.}(2024{\natexlab{a}})\citenamefont {Zhang}, \citenamefont {Pan},\ and\
  \citenamefont {Wang}}]{Zhang2024PRC_Al}%
  \BibitemOpen
  \bibfield  {author} {\bibinfo {author} {\bibfnamefont {K.~Y.}\ \bibnamefont
  {Zhang}}, \bibinfo {author} {\bibfnamefont {C.}~\bibnamefont {Pan}}, \ and\
  \bibinfo {author} {\bibfnamefont {S.}~\bibnamefont {Wang}},\ }\href {\doibase
  10.1103/PhysRevC.110.014320} {\bibfield  {journal} {\bibinfo  {journal}
  {Phys. Rev. C}\ }\textbf {\bibinfo {volume} {110}},\ \bibinfo {pages}
  {014320} (\bibinfo {year} {2024}{\natexlab{a}})}\BibitemShut {NoStop}%
\bibitem [{\citenamefont {In}\ \emph {et~al.}(2021)\citenamefont {In},
  \citenamefont {Papakonstantinou}, \citenamefont {Kim},\ and\ \citenamefont
  {Hong}}]{In2021IJMPE}%
  \BibitemOpen
  \bibfield  {author} {\bibinfo {author} {\bibfnamefont {E.~J.}\ \bibnamefont
  {In}}, \bibinfo {author} {\bibfnamefont {P.}~\bibnamefont
  {Papakonstantinou}}, \bibinfo {author} {\bibfnamefont {Y.}~\bibnamefont
  {Kim}}, \ and\ \bibinfo {author} {\bibfnamefont {S.-W.}\ \bibnamefont
  {Hong}},\ }\href {\doibase 10.1142/S0218301321500099} {\bibfield  {journal}
  {\bibinfo  {journal} {Int. J. Mod. Phys. E}\ }\textbf {\bibinfo {volume}
  {30}},\ \bibinfo {pages} {2150009} (\bibinfo {year} {2021})}\BibitemShut
  {NoStop}%
\bibitem [{\citenamefont {Zhang}\ \emph {et~al.}(2021)\citenamefont {Zhang},
  \citenamefont {He}, \citenamefont {Meng}, \citenamefont {Pan}, \citenamefont
  {Shen}, \citenamefont {Wang},\ and\ \citenamefont {Zhang}}]{Zhang2021PRC}%
  \BibitemOpen
  \bibfield  {author} {\bibinfo {author} {\bibfnamefont {K.}~\bibnamefont
  {Zhang}}, \bibinfo {author} {\bibfnamefont {X.}~\bibnamefont {He}}, \bibinfo
  {author} {\bibfnamefont {J.}~\bibnamefont {Meng}}, \bibinfo {author}
  {\bibfnamefont {C.}~\bibnamefont {Pan}}, \bibinfo {author} {\bibfnamefont
  {C.}~\bibnamefont {Shen}}, \bibinfo {author} {\bibfnamefont {C.}~\bibnamefont
  {Wang}}, \ and\ \bibinfo {author} {\bibfnamefont {S.}~\bibnamefont {Zhang}},\
  }\href {\doibase 10.1103/PhysRevC.104.L021301} {\bibfield  {journal}
  {\bibinfo  {journal} {Phys. Rev. C}\ }\textbf {\bibinfo {volume} {104}},\
  \bibinfo {pages} {L021301} (\bibinfo {year} {2021})}\BibitemShut {NoStop}%
\bibitem [{\citenamefont {Pan}\ \emph {et~al.}(2021)\citenamefont {Pan},
  \citenamefont {Zhang}, \citenamefont {Chong}, \citenamefont {Heo},
  \citenamefont {Ho}, \citenamefont {Lee}, \citenamefont {Li}, \citenamefont
  {Sun}, \citenamefont {Tam}, \citenamefont {Wong}, \citenamefont {Yeung},
  \citenamefont {Yiu},\ and\ \citenamefont {Zhang}}]{Pan2021PRC}%
  \BibitemOpen
  \bibfield  {author} {\bibinfo {author} {\bibfnamefont {C.}~\bibnamefont
  {Pan}}, \bibinfo {author} {\bibfnamefont {K.~Y.}\ \bibnamefont {Zhang}},
  \bibinfo {author} {\bibfnamefont {P.~S.}\ \bibnamefont {Chong}}, \bibinfo
  {author} {\bibfnamefont {C.}~\bibnamefont {Heo}}, \bibinfo {author}
  {\bibfnamefont {M.~C.}\ \bibnamefont {Ho}}, \bibinfo {author} {\bibfnamefont
  {J.}~\bibnamefont {Lee}}, \bibinfo {author} {\bibfnamefont {Z.~P.}\
  \bibnamefont {Li}}, \bibinfo {author} {\bibfnamefont {W.}~\bibnamefont
  {Sun}}, \bibinfo {author} {\bibfnamefont {C.~K.}\ \bibnamefont {Tam}},
  \bibinfo {author} {\bibfnamefont {S.~H.}\ \bibnamefont {Wong}}, \bibinfo
  {author} {\bibfnamefont {R.~W.-Y.}\ \bibnamefont {Yeung}}, \bibinfo {author}
  {\bibfnamefont {T.~C.}\ \bibnamefont {Yiu}}, \ and\ \bibinfo {author}
  {\bibfnamefont {S.~Q.}\ \bibnamefont {Zhang}},\ }\href {\doibase
  10.1103/PhysRevC.104.024331} {\bibfield  {journal} {\bibinfo  {journal}
  {Phys. Rev. C}\ }\textbf {\bibinfo {volume} {104}},\ \bibinfo {pages}
  {024331} (\bibinfo {year} {2021})}\BibitemShut {NoStop}%
\bibitem [{\citenamefont {He}\ \emph {et~al.}(2021)\citenamefont {He},
  \citenamefont {Wang}, \citenamefont {Zhang},\ and\ \citenamefont
  {Shen}}]{He2021CPC}%
  \BibitemOpen
  \bibfield  {author} {\bibinfo {author} {\bibfnamefont {X.}~\bibnamefont
  {He}}, \bibinfo {author} {\bibfnamefont {C.}~\bibnamefont {Wang}}, \bibinfo
  {author} {\bibfnamefont {K.}~\bibnamefont {Zhang}}, \ and\ \bibinfo {author}
  {\bibfnamefont {C.}~\bibnamefont {Shen}},\ }\href {\doibase
  10.1088/1674-1137/ac1b99} {\bibfield  {journal} {\bibinfo  {journal} {Chin.
  Phys. C}\ }\textbf {\bibinfo {volume} {45}},\ \bibinfo {pages} {101001}
  (\bibinfo {year} {2021})}\BibitemShut {NoStop}%
\bibitem [{\citenamefont {He}\ \emph {et~al.}(2024)\citenamefont {He},
  \citenamefont {Wu}, \citenamefont {Zhang},\ and\ \citenamefont
  {Shen}}]{He2024PRC}%
  \BibitemOpen
  \bibfield  {author} {\bibinfo {author} {\bibfnamefont {X.-T.}\ \bibnamefont
  {He}}, \bibinfo {author} {\bibfnamefont {J.-W.}\ \bibnamefont {Wu}}, \bibinfo
  {author} {\bibfnamefont {K.-Y.}\ \bibnamefont {Zhang}}, \ and\ \bibinfo
  {author} {\bibfnamefont {C.-W.}\ \bibnamefont {Shen}},\ }\href {\doibase
  10.1103/PhysRevC.110.014301} {\bibfield  {journal} {\bibinfo  {journal}
  {Phys. Rev. C}\ }\textbf {\bibinfo {volume} {110}},\ \bibinfo {pages}
  {014301} (\bibinfo {year} {2024})}\BibitemShut {NoStop}%
\bibitem [{\citenamefont {Pan}\ \emph {et~al.}(2019)\citenamefont {Pan},
  \citenamefont {Zhang},\ and\ \citenamefont {Zhang}}]{Pan2019IJMPE}%
  \BibitemOpen
  \bibfield  {author} {\bibinfo {author} {\bibfnamefont {C.}~\bibnamefont
  {Pan}}, \bibinfo {author} {\bibfnamefont {K.}~\bibnamefont {Zhang}}, \ and\
  \bibinfo {author} {\bibfnamefont {S.}~\bibnamefont {Zhang}},\ }\href
  {\doibase 10.1142/S0218301319500824} {\bibfield  {journal} {\bibinfo
  {journal} {Int. J. Mod. Phys. E}\ }\textbf {\bibinfo {volume} {28}},\
  \bibinfo {pages} {1950082} (\bibinfo {year} {2019})}\BibitemShut {NoStop}%
\bibitem [{\citenamefont {Choi}\ \emph {et~al.}(2022)\citenamefont {Choi},
  \citenamefont {Lee}, \citenamefont {Mun},\ and\ \citenamefont
  {Kim}}]{Choi2022PRC}%
  \BibitemOpen
  \bibfield  {author} {\bibinfo {author} {\bibfnamefont {Y.-B.}\ \bibnamefont
  {Choi}}, \bibinfo {author} {\bibfnamefont {C.-H.}\ \bibnamefont {Lee}},
  \bibinfo {author} {\bibfnamefont {M.-H.}\ \bibnamefont {Mun}}, \ and\
  \bibinfo {author} {\bibfnamefont {Y.}~\bibnamefont {Kim}},\ }\href {\doibase
  10.1103/PhysRevC.105.024306} {\bibfield  {journal} {\bibinfo  {journal}
  {Phys. Rev. C}\ }\textbf {\bibinfo {volume} {105}},\ \bibinfo {pages}
  {024306} (\bibinfo {year} {2022})}\BibitemShut {NoStop}%
\bibitem [{\citenamefont {Kim}\ \emph {et~al.}(2022)\citenamefont {Kim},
  \citenamefont {Mun}, \citenamefont {Cheoun},\ and\ \citenamefont
  {Ha}}]{Kim2022PRC}%
  \BibitemOpen
  \bibfield  {author} {\bibinfo {author} {\bibfnamefont {S.}~\bibnamefont
  {Kim}}, \bibinfo {author} {\bibfnamefont {M.-H.}\ \bibnamefont {Mun}},
  \bibinfo {author} {\bibfnamefont {M.-K.}\ \bibnamefont {Cheoun}}, \ and\
  \bibinfo {author} {\bibfnamefont {E.}~\bibnamefont {Ha}},\ }\href {\doibase
  10.1103/PhysRevC.105.034340} {\bibfield  {journal} {\bibinfo  {journal}
  {Phys. Rev. C}\ }\textbf {\bibinfo {volume} {105}},\ \bibinfo {pages}
  {034340} (\bibinfo {year} {2022})}\BibitemShut {NoStop}%
\bibitem [{\citenamefont {Guo}\ \emph {et~al.}(2023)\citenamefont {Guo},
  \citenamefont {Pan}, \citenamefont {Zhao}, \citenamefont {Du},\ and\
  \citenamefont {Zhang}}]{Guo2023PRC}%
  \BibitemOpen
  \bibfield  {author} {\bibinfo {author} {\bibfnamefont {P.}~\bibnamefont
  {Guo}}, \bibinfo {author} {\bibfnamefont {C.}~\bibnamefont {Pan}}, \bibinfo
  {author} {\bibfnamefont {Y.~C.}\ \bibnamefont {Zhao}}, \bibinfo {author}
  {\bibfnamefont {X.~K.}\ \bibnamefont {Du}}, \ and\ \bibinfo {author}
  {\bibfnamefont {S.~Q.}\ \bibnamefont {Zhang}},\ }\href {\doibase
  10.1103/PhysRevC.108.014319} {\bibfield  {journal} {\bibinfo  {journal}
  {Phys. Rev. C}\ }\textbf {\bibinfo {volume} {108}},\ \bibinfo {pages}
  {014319} (\bibinfo {year} {2023})}\BibitemShut {NoStop}%
\bibitem [{\citenamefont {Mun}\ \emph {et~al.}(2023)\citenamefont {Mun},
  \citenamefont {Kim}, \citenamefont {Cheoun}, \citenamefont {So},
  \citenamefont {Choi},\ and\ \citenamefont {Ha}}]{Mun2023PLB}%
  \BibitemOpen
  \bibfield  {author} {\bibinfo {author} {\bibfnamefont {M.-H.}\ \bibnamefont
  {Mun}}, \bibinfo {author} {\bibfnamefont {S.}~\bibnamefont {Kim}}, \bibinfo
  {author} {\bibfnamefont {M.-K.}\ \bibnamefont {Cheoun}}, \bibinfo {author}
  {\bibfnamefont {W.~Y.}\ \bibnamefont {So}}, \bibinfo {author} {\bibfnamefont
  {S.}~\bibnamefont {Choi}}, \ and\ \bibinfo {author} {\bibfnamefont
  {E.}~\bibnamefont {Ha}},\ }\href {\doibase
  https://doi.org/10.1016/j.physletb.2023.138298} {\bibfield  {journal}
  {\bibinfo  {journal} {Phys. Lett. B}\ }\textbf {\bibinfo {volume} {847}},\
  \bibinfo {pages} {138298} (\bibinfo {year} {2023})}\BibitemShut {NoStop}%
\bibitem [{\citenamefont {Xiao}\ \emph {et~al.}(2023)\citenamefont {Xiao},
  \citenamefont {Xu}, \citenamefont {Zheng}, \citenamefont {Sun}, \citenamefont
  {Geng},\ and\ \citenamefont {Zhang}}]{Xiao2023PLB}%
  \BibitemOpen
  \bibfield  {author} {\bibinfo {author} {\bibfnamefont {Y.}~\bibnamefont
  {Xiao}}, \bibinfo {author} {\bibfnamefont {S.-Z.}\ \bibnamefont {Xu}},
  \bibinfo {author} {\bibfnamefont {R.-Y.}\ \bibnamefont {Zheng}}, \bibinfo
  {author} {\bibfnamefont {X.-X.}\ \bibnamefont {Sun}}, \bibinfo {author}
  {\bibfnamefont {L.-S.}\ \bibnamefont {Geng}}, \ and\ \bibinfo {author}
  {\bibfnamefont {S.-S.}\ \bibnamefont {Zhang}},\ }\href {\doibase
  https://doi.org/10.1016/j.physletb.2023.138160} {\bibfield  {journal}
  {\bibinfo  {journal} {Phys. Lett. B}\ }\textbf {\bibinfo {volume} {845}},\
  \bibinfo {pages} {138160} (\bibinfo {year} {2023})}\BibitemShut {NoStop}%
\bibitem [{\citenamefont {Choi}\ \emph {et~al.}(2024)\citenamefont {Choi},
  \citenamefont {Lee}, \citenamefont {Mun}, \citenamefont {Choi},\ and\
  \citenamefont {Kim}}]{Choi2024PRC}%
  \BibitemOpen
  \bibfield  {author} {\bibinfo {author} {\bibfnamefont {Y.-B.}\ \bibnamefont
  {Choi}}, \bibinfo {author} {\bibfnamefont {C.-H.}\ \bibnamefont {Lee}},
  \bibinfo {author} {\bibfnamefont {M.-H.}\ \bibnamefont {Mun}}, \bibinfo
  {author} {\bibfnamefont {S.}~\bibnamefont {Choi}}, \ and\ \bibinfo {author}
  {\bibfnamefont {Y.}~\bibnamefont {Kim}},\ }\href {\doibase
  10.1103/PhysRevC.109.054310} {\bibfield  {journal} {\bibinfo  {journal}
  {Phys. Rev. C}\ }\textbf {\bibinfo {volume} {109}},\ \bibinfo {pages}
  {054310} (\bibinfo {year} {2024})}\BibitemShut {NoStop}%
\bibitem [{\citenamefont {Zhang}\ \emph
  {et~al.}(2024{\natexlab{b}})\citenamefont {Zhang}, \citenamefont {Huang},
  \citenamefont {Sun}, \citenamefont {Peng},\ and\ \citenamefont
  {Zhang}}]{Zhang2024CPC}%
  \BibitemOpen
  \bibfield  {author} {\bibinfo {author} {\bibfnamefont {W.}~\bibnamefont
  {Zhang}}, \bibinfo {author} {\bibfnamefont {J.-K.}\ \bibnamefont {Huang}},
  \bibinfo {author} {\bibfnamefont {T.-T.}\ \bibnamefont {Sun}}, \bibinfo
  {author} {\bibfnamefont {J.}~\bibnamefont {Peng}}, \ and\ \bibinfo {author}
  {\bibfnamefont {S.-Q.}\ \bibnamefont {Zhang}},\ }\href {\doibase
  10.1088/1674-1137/ad62dd} {\bibfield  {journal} {\bibinfo  {journal} {Chinese
  Physics C}\ }\textbf {\bibinfo {volume} {48}},\ \bibinfo {pages} {104105}
  (\bibinfo {year} {2024}{\natexlab{b}})}\BibitemShut {NoStop}%
\bibitem [{\citenamefont {Zhang}\ \emph {et~al.}(2019)\citenamefont {Zhang},
  \citenamefont {Wang},\ and\ \citenamefont {Zhang}}]{Zhang2019PRC}%
  \BibitemOpen
  \bibfield  {author} {\bibinfo {author} {\bibfnamefont {K.~Y.}\ \bibnamefont
  {Zhang}}, \bibinfo {author} {\bibfnamefont {D.~Y.}\ \bibnamefont {Wang}}, \
  and\ \bibinfo {author} {\bibfnamefont {S.~Q.}\ \bibnamefont {Zhang}},\ }\href
  {\doibase 10.1103/PhysRevC.100.034312} {\bibfield  {journal} {\bibinfo
  {journal} {Phys. Rev. C}\ }\textbf {\bibinfo {volume} {100}},\ \bibinfo
  {pages} {034312} (\bibinfo {year} {2019})}\BibitemShut {NoStop}%
\bibitem [{\citenamefont {Sun}\ and\ \citenamefont
  {Zhou}(2021{\natexlab{a}})}]{Sun2021SciB}%
  \BibitemOpen
  \bibfield  {author} {\bibinfo {author} {\bibfnamefont {X.-X.}\ \bibnamefont
  {Sun}}\ and\ \bibinfo {author} {\bibfnamefont {S.-G.}\ \bibnamefont {Zhou}},\
  }\href {\doibase 10.1016/j.scib.2021.07.005} {\bibfield  {journal} {\bibinfo
  {journal} {Sci. Bull.}\ }\textbf {\bibinfo {volume} {66}},\ \bibinfo {pages}
  {2072} (\bibinfo {year} {2021}{\natexlab{a}})}\BibitemShut {NoStop}%
\bibitem [{\citenamefont {Sun}\ and\ \citenamefont
  {Zhou}(2021{\natexlab{b}})}]{Sun2021PRC_AMP}%
  \BibitemOpen
  \bibfield  {author} {\bibinfo {author} {\bibfnamefont {X.-X.}\ \bibnamefont
  {Sun}}\ and\ \bibinfo {author} {\bibfnamefont {S.-G.}\ \bibnamefont {Zhou}},\
  }\href {\doibase 10.1103/PhysRevC.104.064319} {\bibfield  {journal} {\bibinfo
   {journal} {Phys. Rev. C}\ }\textbf {\bibinfo {volume} {104}},\ \bibinfo
  {pages} {064319} (\bibinfo {year} {2021}{\natexlab{b}})}\BibitemShut
  {NoStop}%
\bibitem [{\citenamefont {Sun}\ \emph {et~al.}(2022)\citenamefont {Sun},
  \citenamefont {Zhang}, \citenamefont {Pan}, \citenamefont {Fan},
  \citenamefont {Zhang},\ and\ \citenamefont {Li}}]{Sun2022CPC}%
  \BibitemOpen
  \bibfield  {author} {\bibinfo {author} {\bibfnamefont {W.}~\bibnamefont
  {Sun}}, \bibinfo {author} {\bibfnamefont {K.-Y.}\ \bibnamefont {Zhang}},
  \bibinfo {author} {\bibfnamefont {C.}~\bibnamefont {Pan}}, \bibinfo {author}
  {\bibfnamefont {X.-H.}\ \bibnamefont {Fan}}, \bibinfo {author} {\bibfnamefont
  {S.}~\bibnamefont {Zhang}}, \ and\ \bibinfo {author} {\bibfnamefont {Z.-P.}\
  \bibnamefont {Li}},\ }\href {\doibase 10.1088/1674-1137/ac53fa} {\bibfield
  {journal} {\bibinfo  {journal} {Chin. Phys. C}\ }\textbf {\bibinfo {volume}
  {46}},\ \bibinfo {pages} {064103} (\bibinfo {year} {2022})}\BibitemShut
  {NoStop}%
\bibitem [{\citenamefont {Sun}\ and\ \citenamefont {Meng}(2022)}]{Sun2022PRC}%
  \BibitemOpen
  \bibfield  {author} {\bibinfo {author} {\bibfnamefont {X.}~\bibnamefont
  {Sun}}\ and\ \bibinfo {author} {\bibfnamefont {J.}~\bibnamefont {Meng}},\
  }\href {\doibase 10.1103/PhysRevC.105.044312} {\bibfield  {journal} {\bibinfo
   {journal} {Phys. Rev. C}\ }\textbf {\bibinfo {volume} {105}},\ \bibinfo
  {pages} {044312} (\bibinfo {year} {2022})}\BibitemShut {NoStop}%
\bibitem [{\citenamefont {Zhang}\ \emph
  {et~al.}(2022{\natexlab{b}})\citenamefont {Zhang}, \citenamefont {Pan},\ and\
  \citenamefont {Zhang}}]{Zhang2022PRC}%
  \BibitemOpen
  \bibfield  {author} {\bibinfo {author} {\bibfnamefont {K.~Y.}\ \bibnamefont
  {Zhang}}, \bibinfo {author} {\bibfnamefont {C.}~\bibnamefont {Pan}}, \ and\
  \bibinfo {author} {\bibfnamefont {S.~Q.}\ \bibnamefont {Zhang}},\ }\href
  {\doibase 10.1103/PhysRevC.106.024302} {\bibfield  {journal} {\bibinfo
  {journal} {Phys. Rev. C}\ }\textbf {\bibinfo {volume} {106}},\ \bibinfo
  {pages} {024302} (\bibinfo {year} {2022}{\natexlab{b}})}\BibitemShut
  {NoStop}%
\bibitem [{\citenamefont {Zhang}\ \emph
  {et~al.}(2023{\natexlab{c}})\citenamefont {Zhang}, \citenamefont {Niu},
  \citenamefont {Sun},\ and\ \citenamefont {Xia}}]{Zhang2023PRC_2DCH}%
  \BibitemOpen
  \bibfield  {author} {\bibinfo {author} {\bibfnamefont {X.~Y.}\ \bibnamefont
  {Zhang}}, \bibinfo {author} {\bibfnamefont {Z.~M.}\ \bibnamefont {Niu}},
  \bibinfo {author} {\bibfnamefont {W.}~\bibnamefont {Sun}}, \ and\ \bibinfo
  {author} {\bibfnamefont {X.~W.}\ \bibnamefont {Xia}},\ }\href {\doibase
  10.1103/PhysRevC.108.024310} {\bibfield  {journal} {\bibinfo  {journal}
  {Phys. Rev. C}\ }\textbf {\bibinfo {volume} {108}},\ \bibinfo {pages}
  {024310} (\bibinfo {year} {2023}{\natexlab{c}})}\BibitemShut {NoStop}%
\bibitem [{\citenamefont {Zhao}\ \emph {et~al.}(2023)\citenamefont {Zhao},
  \citenamefont {Sun}, \citenamefont {Tanihata}, \citenamefont {Terashima},
  \citenamefont {Prochazka}, \citenamefont {Xu}, \citenamefont {Zhu},
  \citenamefont {Meng}, \citenamefont {Su}, \citenamefont {Zhang},
  \citenamefont {Geng}, \citenamefont {He}, \citenamefont {Liu}, \citenamefont
  {Li}, \citenamefont {Lu}, \citenamefont {Lin}, \citenamefont {Lin},
  \citenamefont {Liu}, \citenamefont {Ren}, \citenamefont {Sun}, \citenamefont
  {Wang}, \citenamefont {Wang}, \citenamefont {Wang}, \citenamefont {Wang},
  \citenamefont {Wei}, \citenamefont {Xu}, \citenamefont {Zhang}, \citenamefont
  {Zhang},\ and\ \citenamefont {Zhang}}]{Zhao2023PLB}%
  \BibitemOpen
  \bibfield  {author} {\bibinfo {author} {\bibfnamefont {J.~W.}\ \bibnamefont
  {Zhao}}, \bibinfo {author} {\bibfnamefont {B.-H.}\ \bibnamefont {Sun}},
  \bibinfo {author} {\bibfnamefont {I.}~\bibnamefont {Tanihata}}, \bibinfo
  {author} {\bibfnamefont {S.}~\bibnamefont {Terashima}}, \bibinfo {author}
  {\bibfnamefont {A.}~\bibnamefont {Prochazka}}, \bibinfo {author}
  {\bibfnamefont {J.~Y.}\ \bibnamefont {Xu}}, \bibinfo {author} {\bibfnamefont
  {L.~H.}\ \bibnamefont {Zhu}}, \bibinfo {author} {\bibfnamefont
  {J.}~\bibnamefont {Meng}}, \bibinfo {author} {\bibfnamefont {J.}~\bibnamefont
  {Su}}, \bibinfo {author} {\bibfnamefont {K.~Y.}\ \bibnamefont {Zhang}},
  \bibinfo {author} {\bibfnamefont {L.~S.}\ \bibnamefont {Geng}}, \bibinfo
  {author} {\bibfnamefont {L.~C.}\ \bibnamefont {He}}, \bibinfo {author}
  {\bibfnamefont {C.~Y.}\ \bibnamefont {Liu}}, \bibinfo {author} {\bibfnamefont
  {G.~S.}\ \bibnamefont {Li}}, \bibinfo {author} {\bibfnamefont {C.~G.}\
  \bibnamefont {Lu}}, \bibinfo {author} {\bibfnamefont {W.~J.}\ \bibnamefont
  {Lin}}, \bibinfo {author} {\bibfnamefont {W.~P.}\ \bibnamefont {Lin}},
  \bibinfo {author} {\bibfnamefont {Z.}~\bibnamefont {Liu}}, \bibinfo {author}
  {\bibfnamefont {P.~P.}\ \bibnamefont {Ren}}, \bibinfo {author} {\bibfnamefont
  {Z.~Y.}\ \bibnamefont {Sun}}, \bibinfo {author} {\bibfnamefont
  {F.}~\bibnamefont {Wang}}, \bibinfo {author} {\bibfnamefont {J.}~\bibnamefont
  {Wang}}, \bibinfo {author} {\bibfnamefont {M.}~\bibnamefont {Wang}}, \bibinfo
  {author} {\bibfnamefont {S.~T.}\ \bibnamefont {Wang}}, \bibinfo {author}
  {\bibfnamefont {X.~L.}\ \bibnamefont {Wei}}, \bibinfo {author} {\bibfnamefont
  {X.~D.}\ \bibnamefont {Xu}}, \bibinfo {author} {\bibfnamefont {J.~C.}\
  \bibnamefont {Zhang}}, \bibinfo {author} {\bibfnamefont {M.~X.}\ \bibnamefont
  {Zhang}}, \ and\ \bibinfo {author} {\bibfnamefont {X.~H.}\ \bibnamefont
  {Zhang}},\ }\href {\doibase https://doi.org/10.1016/j.physletb.2023.138269}
  {\bibfield  {journal} {\bibinfo  {journal} {Phys. Lett. B}\ }\textbf
  {\bibinfo {volume} {847}},\ \bibinfo {pages} {138269} (\bibinfo {year}
  {2023})}\BibitemShut {NoStop}%
\bibitem [{\citenamefont {Mun}\ \emph {et~al.}(2024)\citenamefont {Mun},
  \citenamefont {Cheoun}, \citenamefont {Ha}, \citenamefont {Sagawa},\ and\
  \citenamefont {Col\`o}}]{Mun2024PRC}%
  \BibitemOpen
  \bibfield  {author} {\bibinfo {author} {\bibfnamefont {M.-H.}\ \bibnamefont
  {Mun}}, \bibinfo {author} {\bibfnamefont {M.-K.}\ \bibnamefont {Cheoun}},
  \bibinfo {author} {\bibfnamefont {E.}~\bibnamefont {Ha}}, \bibinfo {author}
  {\bibfnamefont {H.}~\bibnamefont {Sagawa}}, \ and\ \bibinfo {author}
  {\bibfnamefont {G.}~\bibnamefont {Col\`o}},\ }\href {\doibase
  10.1103/PhysRevC.110.014314} {\bibfield  {journal} {\bibinfo  {journal}
  {Phys. Rev. C}\ }\textbf {\bibinfo {volume} {110}},\ \bibinfo {pages}
  {014314} (\bibinfo {year} {2024})}\BibitemShut {NoStop}%
\bibitem [{\citenamefont {Zhang}\ \emph
  {et~al.}(2024{\natexlab{c}})\citenamefont {Zhang}, \citenamefont {Liu},
  \citenamefont {Zhang},\ and\ \citenamefont {Yao}}]{Zhang2024PRC_shell}%
  \BibitemOpen
  \bibfield  {author} {\bibinfo {author} {\bibfnamefont {Y.~X.}\ \bibnamefont
  {Zhang}}, \bibinfo {author} {\bibfnamefont {B.~R.}\ \bibnamefont {Liu}},
  \bibinfo {author} {\bibfnamefont {K.~Y.}\ \bibnamefont {Zhang}}, \ and\
  \bibinfo {author} {\bibfnamefont {J.~M.}\ \bibnamefont {Yao}},\ }\href
  {\doibase 10.1103/PhysRevC.110.024302} {\bibfield  {journal} {\bibinfo
  {journal} {Phys. Rev. C}\ }\textbf {\bibinfo {volume} {110}},\ \bibinfo
  {pages} {024302} (\bibinfo {year} {2024}{\natexlab{c}})}\BibitemShut
  {NoStop}%
\bibitem [{\citenamefont {Pan}\ \emph {et~al.}(2025)\citenamefont {Pan},
  \citenamefont {Yang}, \citenamefont {Jiang},\ and\ \citenamefont
  {Wu}}]{Pan2025arXiv}%
  \BibitemOpen
  \bibfield  {author} {\bibinfo {author} {\bibfnamefont {C.}~\bibnamefont
  {Pan}}, \bibinfo {author} {\bibfnamefont {Y.~C.}\ \bibnamefont {Yang}},
  \bibinfo {author} {\bibfnamefont {X.~F.}\ \bibnamefont {Jiang}}, \ and\
  \bibinfo {author} {\bibfnamefont {X.~H.}\ \bibnamefont {Wu}},\ }\href
  {https://arxiv.org/abs/2503.09324} {\bibfield  {journal} {\bibinfo  {journal}
  {arXiv}\ } (\bibinfo {year} {2025})},\ \Eprint
  {http://arxiv.org/abs/2503.09324} {arXiv:2503.09324 [nucl-th]} \BibitemShut
  {NoStop}%
\bibitem [{\citenamefont {Zhao}\ \emph {et~al.}(2010)\citenamefont {Zhao},
  \citenamefont {Li}, \citenamefont {Yao},\ and\ \citenamefont
  {Meng}}]{Zhao2010PRC}%
  \BibitemOpen
  \bibfield  {author} {\bibinfo {author} {\bibfnamefont {P.~W.}\ \bibnamefont
  {Zhao}}, \bibinfo {author} {\bibfnamefont {Z.~P.}\ \bibnamefont {Li}},
  \bibinfo {author} {\bibfnamefont {J.~M.}\ \bibnamefont {Yao}}, \ and\
  \bibinfo {author} {\bibfnamefont {J.}~\bibnamefont {Meng}},\ }\href {\doibase
  10.1103/PhysRevC.82.054319} {\bibfield  {journal} {\bibinfo  {journal} {Phys.
  Rev. C}\ }\textbf {\bibinfo {volume} {82}},\ \bibinfo {pages} {054319}
  (\bibinfo {year} {2010})}\BibitemShut {NoStop}%
\bibitem [{\citenamefont {Zhang}\ \emph {et~al.}(2020)\citenamefont {Zhang},
  \citenamefont {Cheoun}, \citenamefont {Choi}, \citenamefont {Chong},
  \citenamefont {Dong}, \citenamefont {Geng}, \citenamefont {Ha}, \citenamefont
  {He}, \citenamefont {Heo}, \citenamefont {Ho}, \citenamefont {In},
  \citenamefont {Kim}, \citenamefont {Kim}, \citenamefont {Lee}, \citenamefont
  {Lee}, \citenamefont {Li}, \citenamefont {Luo}, \citenamefont {Meng},
  \citenamefont {Mun}, \citenamefont {Niu}, \citenamefont {Pan}, \citenamefont
  {Papakonstantinou}, \citenamefont {Shang}, \citenamefont {Shen},
  \citenamefont {Shen}, \citenamefont {Sun}, \citenamefont {Sun}, \citenamefont
  {Tam}, \citenamefont {Thaivayongnou}, \citenamefont {Wang}, \citenamefont
  {Wong}, \citenamefont {Xia}, \citenamefont {Yan}, \citenamefont {Yeung},
  \citenamefont {Yiu}, \citenamefont {Zhang}, \citenamefont {Zhang},\ and\
  \citenamefont {Zhou}}]{Zhang2020PRC}%
  \BibitemOpen
  \bibfield  {author} {\bibinfo {author} {\bibfnamefont {K.}~\bibnamefont
  {Zhang}}, \bibinfo {author} {\bibfnamefont {M.-K.}\ \bibnamefont {Cheoun}},
  \bibinfo {author} {\bibfnamefont {Y.-B.}\ \bibnamefont {Choi}}, \bibinfo
  {author} {\bibfnamefont {P.~S.}\ \bibnamefont {Chong}}, \bibinfo {author}
  {\bibfnamefont {J.}~\bibnamefont {Dong}}, \bibinfo {author} {\bibfnamefont
  {L.}~\bibnamefont {Geng}}, \bibinfo {author} {\bibfnamefont {E.}~\bibnamefont
  {Ha}}, \bibinfo {author} {\bibfnamefont {X.}~\bibnamefont {He}}, \bibinfo
  {author} {\bibfnamefont {C.}~\bibnamefont {Heo}}, \bibinfo {author}
  {\bibfnamefont {M.~C.}\ \bibnamefont {Ho}}, \bibinfo {author} {\bibfnamefont
  {E.~J.}\ \bibnamefont {In}}, \bibinfo {author} {\bibfnamefont
  {S.}~\bibnamefont {Kim}}, \bibinfo {author} {\bibfnamefont {Y.}~\bibnamefont
  {Kim}}, \bibinfo {author} {\bibfnamefont {C.-H.}\ \bibnamefont {Lee}},
  \bibinfo {author} {\bibfnamefont {J.}~\bibnamefont {Lee}}, \bibinfo {author}
  {\bibfnamefont {Z.}~\bibnamefont {Li}}, \bibinfo {author} {\bibfnamefont
  {T.}~\bibnamefont {Luo}}, \bibinfo {author} {\bibfnamefont {J.}~\bibnamefont
  {Meng}}, \bibinfo {author} {\bibfnamefont {M.-H.}\ \bibnamefont {Mun}},
  \bibinfo {author} {\bibfnamefont {Z.}~\bibnamefont {Niu}}, \bibinfo {author}
  {\bibfnamefont {C.}~\bibnamefont {Pan}}, \bibinfo {author} {\bibfnamefont
  {P.}~\bibnamefont {Papakonstantinou}}, \bibinfo {author} {\bibfnamefont
  {X.}~\bibnamefont {Shang}}, \bibinfo {author} {\bibfnamefont
  {C.}~\bibnamefont {Shen}}, \bibinfo {author} {\bibfnamefont {G.}~\bibnamefont
  {Shen}}, \bibinfo {author} {\bibfnamefont {W.}~\bibnamefont {Sun}}, \bibinfo
  {author} {\bibfnamefont {X.-X.}\ \bibnamefont {Sun}}, \bibinfo {author}
  {\bibfnamefont {C.~K.}\ \bibnamefont {Tam}}, \bibinfo {author} {\bibnamefont
  {Thaivayongnou}}, \bibinfo {author} {\bibfnamefont {C.}~\bibnamefont {Wang}},
  \bibinfo {author} {\bibfnamefont {S.~H.}\ \bibnamefont {Wong}}, \bibinfo
  {author} {\bibfnamefont {X.}~\bibnamefont {Xia}}, \bibinfo {author}
  {\bibfnamefont {Y.}~\bibnamefont {Yan}}, \bibinfo {author} {\bibfnamefont
  {R.~W.-Y.}\ \bibnamefont {Yeung}}, \bibinfo {author} {\bibfnamefont {T.~C.}\
  \bibnamefont {Yiu}}, \bibinfo {author} {\bibfnamefont {S.}~\bibnamefont
  {Zhang}}, \bibinfo {author} {\bibfnamefont {W.}~\bibnamefont {Zhang}}, \ and\
  \bibinfo {author} {\bibfnamefont {S.-G.}\ \bibnamefont {Zhou}} (\bibinfo
  {collaboration} {DRHBc Mass Table Collaboration}),\ }\href {\doibase
  10.1103/PhysRevC.102.024314} {\bibfield  {journal} {\bibinfo  {journal}
  {Phys. Rev. C}\ }\textbf {\bibinfo {volume} {102}},\ \bibinfo {pages}
  {024314} (\bibinfo {year} {2020})}\BibitemShut {NoStop}%
\bibitem [{\citenamefont {Pan}\ \emph {et~al.}(2022)\citenamefont {Pan},
  \citenamefont {Cheoun}, \citenamefont {Choi}, \citenamefont {Dong},
  \citenamefont {Du}, \citenamefont {Fan}, \citenamefont {Gao}, \citenamefont
  {Geng}, \citenamefont {Ha}, \citenamefont {He}, \citenamefont {Huang},
  \citenamefont {Huang}, \citenamefont {Kim}, \citenamefont {Kim},
  \citenamefont {Lee}, \citenamefont {Lee}, \citenamefont {Li}, \citenamefont
  {Liu}, \citenamefont {Ma}, \citenamefont {Meng}, \citenamefont {Mun},
  \citenamefont {Niu}, \citenamefont {Papakonstantinou}, \citenamefont {Shang},
  \citenamefont {Shen}, \citenamefont {Shen}, \citenamefont {Sun},
  \citenamefont {Sun}, \citenamefont {Wu}, \citenamefont {Wu}, \citenamefont
  {Xia}, \citenamefont {Yan}, \citenamefont {Yiu}, \citenamefont {Zhang},
  \citenamefont {Zhang}, \citenamefont {Zhang}, \citenamefont {Zhang},
  \citenamefont {Zhao}, \citenamefont {Zheng},\ and\ \citenamefont
  {Zhou}}]{Pan2022PRC}%
  \BibitemOpen
  \bibfield  {author} {\bibinfo {author} {\bibfnamefont {C.}~\bibnamefont
  {Pan}}, \bibinfo {author} {\bibfnamefont {M.-K.}\ \bibnamefont {Cheoun}},
  \bibinfo {author} {\bibfnamefont {Y.-B.}\ \bibnamefont {Choi}}, \bibinfo
  {author} {\bibfnamefont {J.}~\bibnamefont {Dong}}, \bibinfo {author}
  {\bibfnamefont {X.}~\bibnamefont {Du}}, \bibinfo {author} {\bibfnamefont
  {X.-H.}\ \bibnamefont {Fan}}, \bibinfo {author} {\bibfnamefont
  {W.}~\bibnamefont {Gao}}, \bibinfo {author} {\bibfnamefont {L.}~\bibnamefont
  {Geng}}, \bibinfo {author} {\bibfnamefont {E.}~\bibnamefont {Ha}}, \bibinfo
  {author} {\bibfnamefont {X.-T.}\ \bibnamefont {He}}, \bibinfo {author}
  {\bibfnamefont {J.}~\bibnamefont {Huang}}, \bibinfo {author} {\bibfnamefont
  {K.}~\bibnamefont {Huang}}, \bibinfo {author} {\bibfnamefont
  {S.}~\bibnamefont {Kim}}, \bibinfo {author} {\bibfnamefont {Y.}~\bibnamefont
  {Kim}}, \bibinfo {author} {\bibfnamefont {C.-H.}\ \bibnamefont {Lee}},
  \bibinfo {author} {\bibfnamefont {J.}~\bibnamefont {Lee}}, \bibinfo {author}
  {\bibfnamefont {Z.}~\bibnamefont {Li}}, \bibinfo {author} {\bibfnamefont
  {Z.-R.}\ \bibnamefont {Liu}}, \bibinfo {author} {\bibfnamefont
  {Y.}~\bibnamefont {Ma}}, \bibinfo {author} {\bibfnamefont {J.}~\bibnamefont
  {Meng}}, \bibinfo {author} {\bibfnamefont {M.-H.}\ \bibnamefont {Mun}},
  \bibinfo {author} {\bibfnamefont {Z.}~\bibnamefont {Niu}}, \bibinfo {author}
  {\bibfnamefont {P.}~\bibnamefont {Papakonstantinou}}, \bibinfo {author}
  {\bibfnamefont {X.}~\bibnamefont {Shang}}, \bibinfo {author} {\bibfnamefont
  {C.}~\bibnamefont {Shen}}, \bibinfo {author} {\bibfnamefont {G.}~\bibnamefont
  {Shen}}, \bibinfo {author} {\bibfnamefont {W.}~\bibnamefont {Sun}}, \bibinfo
  {author} {\bibfnamefont {X.-X.}\ \bibnamefont {Sun}}, \bibinfo {author}
  {\bibfnamefont {J.}~\bibnamefont {Wu}}, \bibinfo {author} {\bibfnamefont
  {X.}~\bibnamefont {Wu}}, \bibinfo {author} {\bibfnamefont {X.}~\bibnamefont
  {Xia}}, \bibinfo {author} {\bibfnamefont {Y.}~\bibnamefont {Yan}}, \bibinfo
  {author} {\bibfnamefont {T.~C.}\ \bibnamefont {Yiu}}, \bibinfo {author}
  {\bibfnamefont {K.}~\bibnamefont {Zhang}}, \bibinfo {author} {\bibfnamefont
  {S.}~\bibnamefont {Zhang}}, \bibinfo {author} {\bibfnamefont
  {W.}~\bibnamefont {Zhang}}, \bibinfo {author} {\bibfnamefont
  {X.}~\bibnamefont {Zhang}}, \bibinfo {author} {\bibfnamefont
  {Q.}~\bibnamefont {Zhao}}, \bibinfo {author} {\bibfnamefont {R.}~\bibnamefont
  {Zheng}}, \ and\ \bibinfo {author} {\bibfnamefont {S.-G.}\ \bibnamefont
  {Zhou}} (\bibinfo {collaboration} {DRHBc Mass Table Collaboration}),\ }\href
  {\doibase 10.1103/PhysRevC.106.014316} {\bibfield  {journal} {\bibinfo
  {journal} {Phys. Rev. C}\ }\textbf {\bibinfo {volume} {106}},\ \bibinfo
  {pages} {014316} (\bibinfo {year} {2022})}\BibitemShut {NoStop}%
\bibitem [{\citenamefont {Zhou}\ \emph
  {et~al.}(2003{\natexlab{b}})\citenamefont {Zhou}, \citenamefont {Meng},\ and\
  \citenamefont {Ring}}]{Zhou2003PRC}%
  \BibitemOpen
  \bibfield  {author} {\bibinfo {author} {\bibfnamefont {S.-G.}\ \bibnamefont
  {Zhou}}, \bibinfo {author} {\bibfnamefont {J.}~\bibnamefont {Meng}}, \ and\
  \bibinfo {author} {\bibfnamefont {P.}~\bibnamefont {Ring}},\ }\href {\doibase
  10.1103/PhysRevC.68.034323} {\bibfield  {journal} {\bibinfo  {journal} {Phys.
  Rev. C}\ }\textbf {\bibinfo {volume} {68}},\ \bibinfo {pages} {034323}
  (\bibinfo {year} {2003}{\natexlab{b}})}\BibitemShut {NoStop}%
\bibitem [{\citenamefont {Ring}\ and\ \citenamefont
  {Schuck}(1980)}]{Ring1980NMBP}%
  \BibitemOpen
  \bibfield  {author} {\bibinfo {author} {\bibfnamefont {P.}~\bibnamefont
  {Ring}}\ and\ \bibinfo {author} {\bibfnamefont {P.}~\bibnamefont {Schuck}},\
  }\href {https://www.springer.com/us/book/9783540212065} {\emph {\bibinfo
  {title} {{The Nuclear Many-Body Problem}}}}\ (\bibinfo  {publisher}
  {Springer-Verlag},\ \bibinfo {year} {1980})\BibitemShut {NoStop}%
\bibitem [{\citenamefont {Li}\ \emph {et~al.}(2012{\natexlab{b}})\citenamefont
  {Li}, \citenamefont {Meng}, \citenamefont {Ring}, \citenamefont {Zhao},\ and\
  \citenamefont {Zhou}}]{Li2012CPL}%
  \BibitemOpen
  \bibfield  {author} {\bibinfo {author} {\bibfnamefont {L.-L.}\ \bibnamefont
  {Li}}, \bibinfo {author} {\bibfnamefont {J.}~\bibnamefont {Meng}}, \bibinfo
  {author} {\bibfnamefont {P.}~\bibnamefont {Ring}}, \bibinfo {author}
  {\bibfnamefont {E.-G.}\ \bibnamefont {Zhao}}, \ and\ \bibinfo {author}
  {\bibfnamefont {S.-G.}\ \bibnamefont {Zhou}},\ }\href {\doibase
  10.1088/0256-307x/29/4/042101} {\bibfield  {journal} {\bibinfo  {journal}
  {Chin. Phys. Lett.}\ }\textbf {\bibinfo {volume} {29}},\ \bibinfo {pages}
  {042101} (\bibinfo {year} {2012}{\natexlab{b}})}\BibitemShut {NoStop}%
\bibitem [{\citenamefont {Ekstr\"om}\ \emph {et~al.}(2015)\citenamefont
  {Ekstr\"om}, \citenamefont {Jansen}, \citenamefont {Wendt}, \citenamefont
  {Hagen}, \citenamefont {Papenbrock}, \citenamefont {Carlsson}, \citenamefont
  {Forss\'en}, \citenamefont {Hjorth-Jensen}, \citenamefont {Navr\'atil},\ and\
  \citenamefont {Nazarewicz}}]{Ekstrom2015PRC}%
  \BibitemOpen
  \bibfield  {author} {\bibinfo {author} {\bibfnamefont {A.}~\bibnamefont
  {Ekstr\"om}}, \bibinfo {author} {\bibfnamefont {G.~R.}\ \bibnamefont
  {Jansen}}, \bibinfo {author} {\bibfnamefont {K.~A.}\ \bibnamefont {Wendt}},
  \bibinfo {author} {\bibfnamefont {G.}~\bibnamefont {Hagen}}, \bibinfo
  {author} {\bibfnamefont {T.}~\bibnamefont {Papenbrock}}, \bibinfo {author}
  {\bibfnamefont {B.~D.}\ \bibnamefont {Carlsson}}, \bibinfo {author}
  {\bibfnamefont {C.}~\bibnamefont {Forss\'en}}, \bibinfo {author}
  {\bibfnamefont {M.}~\bibnamefont {Hjorth-Jensen}}, \bibinfo {author}
  {\bibfnamefont {P.}~\bibnamefont {Navr\'atil}}, \ and\ \bibinfo {author}
  {\bibfnamefont {W.}~\bibnamefont {Nazarewicz}},\ }\href {\doibase
  10.1103/PhysRevC.91.051301} {\bibfield  {journal} {\bibinfo  {journal} {Phys.
  Rev. C}\ }\textbf {\bibinfo {volume} {91}},\ \bibinfo {pages} {051301}
  (\bibinfo {year} {2015})}\BibitemShut {NoStop}%
\bibitem [{\citenamefont {Pritychenko}\ \emph {et~al.}(2016)\citenamefont
  {Pritychenko}, \citenamefont {Birch}, \citenamefont {Singh},\ and\
  \citenamefont {Horoi}}]{Pritychenko2016ADNDT}%
  \BibitemOpen
  \bibfield  {author} {\bibinfo {author} {\bibfnamefont {B.}~\bibnamefont
  {Pritychenko}}, \bibinfo {author} {\bibfnamefont {M.}~\bibnamefont {Birch}},
  \bibinfo {author} {\bibfnamefont {B.}~\bibnamefont {Singh}}, \ and\ \bibinfo
  {author} {\bibfnamefont {M.}~\bibnamefont {Horoi}},\ }\href {\doibase
  10.1016/j.adt.2015.10.001} {\bibfield  {journal} {\bibinfo  {journal} {At.
  Data Nucl. Data Tables}\ }\textbf {\bibinfo {volume} {107}},\ \bibinfo
  {pages} {1} (\bibinfo {year} {2016})}\BibitemShut {NoStop}%
\bibitem [{\citenamefont {Maass}\ \emph {et~al.}(2025)\citenamefont {Maass},
  \citenamefont {Ryssens}, \citenamefont {Bender}, \citenamefont {Burdette},
  \citenamefont {Clark}, \citenamefont {Dockery}, \citenamefont {Grams},
  \citenamefont {Horst}, \citenamefont {Imgram}, \citenamefont {K\"{o}nig},
  \citenamefont {Minamisono}, \citenamefont {M\"{u}ller}, \citenamefont
  {M\"{u}ller}, \citenamefont {N\"{a}ortersh\"{a}user}, \citenamefont {Pineda},
  \citenamefont {Rausch}, \citenamefont {Renth}, \citenamefont {Rickey},
  \citenamefont {Santiago-Gonzalez}, \citenamefont {Savard}, \citenamefont
  {Sommer},\ and\ \citenamefont {Valverde}}]{Maass2025arXiv}%
  \BibitemOpen
  \bibfield  {author} {\bibinfo {author} {\bibfnamefont {B.}~\bibnamefont
  {Maass}}, \bibinfo {author} {\bibfnamefont {W.}~\bibnamefont {Ryssens}},
  \bibinfo {author} {\bibfnamefont {M.}~\bibnamefont {Bender}}, \bibinfo
  {author} {\bibfnamefont {D.~P.}\ \bibnamefont {Burdette}}, \bibinfo {author}
  {\bibfnamefont {J.}~\bibnamefont {Clark}}, \bibinfo {author} {\bibfnamefont
  {A.}~\bibnamefont {Dockery}}, \bibinfo {author} {\bibfnamefont
  {G.}~\bibnamefont {Grams}}, \bibinfo {author} {\bibfnamefont
  {M.}~\bibnamefont {Horst}}, \bibinfo {author} {\bibfnamefont
  {P.}~\bibnamefont {Imgram}}, \bibinfo {author} {\bibfnamefont
  {K.}~\bibnamefont {K\"{o}nig}}, \bibinfo {author} {\bibfnamefont
  {K.}~\bibnamefont {Minamisono}}, \bibinfo {author} {\bibfnamefont
  {P.}~\bibnamefont {M\"{u}ller}}, \bibinfo {author} {\bibfnamefont
  {P.}~\bibnamefont {M\"{u}ller}}, \bibinfo {author} {\bibfnamefont
  {W.}~\bibnamefont {N\"{a}ortersh\"{a}user}}, \bibinfo {author} {\bibfnamefont
  {S.~V.}\ \bibnamefont {Pineda}}, \bibinfo {author} {\bibfnamefont
  {S.}~\bibnamefont {Rausch}}, \bibinfo {author} {\bibfnamefont
  {L.}~\bibnamefont {Renth}}, \bibinfo {author} {\bibfnamefont
  {B.}~\bibnamefont {Rickey}}, \bibinfo {author} {\bibfnamefont
  {D.}~\bibnamefont {Santiago-Gonzalez}}, \bibinfo {author} {\bibfnamefont
  {G.}~\bibnamefont {Savard}}, \bibinfo {author} {\bibfnamefont
  {F.}~\bibnamefont {Sommer}}, \ and\ \bibinfo {author} {\bibfnamefont {A.~A.}\
  \bibnamefont {Valverde}},\ }\href {https://arxiv.org/abs/2503.07841}
  {\bibfield  {journal} {\bibinfo  {journal} {arXiv}\ ,\ \bibinfo {pages}
  {2503.07841}} (\bibinfo {year} {2025})}\BibitemShut {NoStop}%
\bibitem [{\citenamefont {Zhang}\ \emph
  {et~al.}(2023{\natexlab{d}})\citenamefont {Zhang}, \citenamefont {Zhang},\
  and\ \citenamefont {Meng}}]{Zhang2023PRC_TRHBc}%
  \BibitemOpen
  \bibfield  {author} {\bibinfo {author} {\bibfnamefont {K.~Y.}\ \bibnamefont
  {Zhang}}, \bibinfo {author} {\bibfnamefont {S.~Q.}\ \bibnamefont {Zhang}}, \
  and\ \bibinfo {author} {\bibfnamefont {J.}~\bibnamefont {Meng}},\ }\href
  {\doibase 10.1103/PhysRevC.108.L041301} {\bibfield  {journal} {\bibinfo
  {journal} {Phys. Rev. C}\ }\textbf {\bibinfo {volume} {108}},\ \bibinfo
  {pages} {L041301} (\bibinfo {year} {2023}{\natexlab{d}})}\BibitemShut
  {NoStop}%
\end{thebibliography}


\end{document}